\documentclass[aps,prx,twocolumn,superscriptaddress,longbibliography]{revtex4-2}

\usepackage{amssymb}
\usepackage{amsmath}
\usepackage{bm}
\usepackage{graphicx}
\usepackage{amsfonts}
\usepackage{braket}
\usepackage{subfigure}
\usepackage{hyperref,color}
\usepackage[version=3]{mhchem}

\begin{document}
\title{Effective-Hamiltonian reconstruction through Bloch-wave interferometry in bulk GaAs driven by strong THz fields}

\author{Qile Wu}
\affiliation{Physics Department, University of California, Santa Barbara, California 93106, USA}
\affiliation{Institute for Terahertz Science and Technology, University of California, Santa Barbara, California 93106, USA}

\author{Seamus D. O'Hara}
\affiliation{Department of Physics and Astronomy, University of Pennsylvania, Philadelphia, Pennsylvania 19104, USA}

\author{Joseph B. Costello}
\affiliation{Electrical Engineering and Computer Science, Exponent, Menlo Park, California 94025, USA}

\author{Loren N. Pfeiffer}
\affiliation{Electrical Engineering Department, Princeton University, Princeton, New Jersey 08544, USA}

\author{Ken W. West}
\affiliation{Electrical Engineering Department, Princeton University, Princeton, New Jersey 08544, USA}

\author{Mark S. Sherwin}
\affiliation{Physics Department, University of California, Santa Barbara, California 93106, USA}
\affiliation{Institute for Terahertz Science and Technology, University of California, Santa Barbara, California 93106, USA}

\date{\today}

\begin{abstract}
Reconstructing effective Hamiltonians of condensed matter systems directly from experimental data is challenging because of the intricate relationship between Hamiltonian parameters and observables. Here, we reconstruct an effective three-band electron-hole Hamiltonian in bulk GaAs based on high-order sideband generation (HSG) induced by quasi-continuous near-infrared (NIR) and terahertz (THz) lasers. We perform polarimetry of high-order sidebands while varying the wavelength and polarization of the NIR laser, as well as the strength of the THz field (27 to 64\,kV/cm at 447\,GHz). An analytic model is derived to incorporate the effects of both dephasing and quantum fluctuations around the semiclassical electron-hole recollision pathways. Surprisingly, the contribution of quantum fluctuations to the decay of sideband intensity with increasing sideband order is comparable to the contribution of dephasing. Assuming that the exciton reduced mass and the single parameter that defines the hole Bloch wavefunctions in bulk GaAs are known from previous experiments, we simultaneously and unambiguously determine through Bloch-wave interferometry the following: the effective Hamiltonian parameter that determines the electron-hole reduced masses, $\xi=0.125\pm0.011$; the bandgap of GaAs at 30\,K, $E_{\rm g}=1.530\pm0.001$\,eV; and two dephasing constants associated with two electron-hole species, $\Gamma_{\rm E-HH}=10.4\pm0.2$\,meV and $\Gamma_{\rm E-LH}=7.6\pm0.2$\,meV. We demonstrate that full Hamiltonian reconstruction can be achieved by combining HSG measurements with absorbance spectroscopy. Unexpectedly, we find that the extracted bandgap of GaAs is about 10\,meV larger than the value inferred from previous absorbance measurements. Quantum-kinetic analysis suggests that, in the HSG experiments, the electron-hole energy may be renormalized through Fr\"ohlich interaction that is modulated by the strong THz fields. We also show that the energy threshold for optical-phonon emission can be suppressed by applying a strong THz field, leading to nearly constant dephasing rates. Our work provides an opportunity to explore possible modification of polaronic effects under strong THz fields. We show that polarimetry of high-order sidebands, together with interband absorption measurements, symmetry analysis, and analytic theory, can potentially enable reconstruction of effective Hamiltonians for other direct-gap insulators and semiconductors, whose bulk is not directly accessible to surface-sensitive techniques such as angle-resolved photoemission spectroscopy (ARPES).
\end{abstract}

\maketitle

\section{Introduction}\label{SEC:introduction}
Condensed-matter physicists focus on understanding the emergent phenomena arising from interactions among an enormous number of electrons and atomic nuclei. Although the underlying many-body Hamiltonians can, in principle, be written down based on our knowledge of few-body systems such as isolated atoms, their connection with the emergent phenomena is far from transparent, and direct calculation of the associated quantum wavefunctions remains a formidable task~\cite{laughlin2000theory}. Effective Hamiltonians are powerful tools for reducing the many-body complexity by focusing on the degrees of freedom most relevant to explaining experimental observations, with the effects of other degrees of freedom encoded in the Hamiltonian parameters or treated as perturbations~\cite{powell2009introduction}.

The development of our understanding of crystalline solids has been inextricably linked to the construction of effective Hamiltonians~\cite{ashcroft1976solid}. By focusing on small distortions of crystal lattices from their equilibrium configurations, effective phonon Hamiltonians derived from the Born-Oppenheimer approximation~\cite{born1927zur} have long been used to study the thermal and elastic properties of solid materials. The degrees of freedom associated with the valence electrons are incorporated into the force constants that determine the phonon dispersion, as well as into the coefficients in the anharmonic terms describing phonon-phonon interactions. To investigate the electronic and optical properties of a crystalline solid, one of the most successful starting points has been the band theory. In this theoretical framework, by assuming the crystal lattice to be perfectly periodic with the phonon degrees of freedom integrated out, various effective electron Hamiltonians are developed under the independent-electron approximation, in which electron-electron interactions are averaged into a mean electrostatic potential. Crucially, these effective electron Hamiltonians need not be built upon Hilbert spaces consisting of the actual microscopic wavefunctions in order to explain experimental data. For example, in the empirical pseudopotential method, by projecting out the rapidly oscillating components of the valence-electron wavefunctions, effective Hamiltonians are constructed in a basis of the so-called pseudo-wavefunctions that retain only the smooth components of the true wavefunctions~\cite{cohen1970fitting}. In the $\bf k\cdot\bf p$ method, effective Hamiltonians can be constructed even when only the symmetry properties of the basis states are known~\cite{willatzen2009kp}. When the band theory alone is insufficient to account for experimental observations, various interaction effects can be introduced, often leading to more intriguing physics. For instance, electron-electron interactions are incorporated in the Hubbard model~\cite{hubbard1963electron,arovas2022hubbard}, which, despite its simple form, has been instrumental in exploring correlated phenomena ranging from the Mott metal-insulator transition~\cite{mott1968metal,imada1998metal} to high-temperature superconductivity~\cite{lee2006doping,fradkin2015colloquium}.

Reconstruction of an effective Hamiltonian generally begins with a trial Hamiltonian, whose form is iteratively modified until the experimental data are reasonably reproduced. In each iteration, the effective Hamiltonian, as an operator on a specified Hilbert space, can always be expanded as a linear combination of linearly independent operators. A key step in the Hamiltonian-reconstruction process is determining the effective-Hamiltonian parameters, which serve as the coefficients for the operator combinations, by minimizing the deviation between theory and experiment. If significant theory-experiment deviation remains after the Hamiltonian parameters are optimized, a redefinition of the Hilbert space is necessary. While calculating measurable quantities with a known effective Hamiltonian is straightforward when the underlying Hilbert space is numerically manageable, extracting effective-Hamiltonian parameters directly from experimental data is usually challenging. In the absence of experimental inputs, \textit{ab initio} calculations have been used to determine effective Hamiltonians including their parameters, such as in $\bf k\cdot \bf p$ models~\cite{jocic2020ab}. Nevertheless, experimental measurements are ultimately required to benchmark their accuracy.

Here, we focus on the reconstruction of effective electron Hamiltonians within the framework of the band theory as a general starting point for understanding crystalline solids. According to Bloch's theorem~\cite{bloch1929quantenmechanik}, an effective electron Hamiltonian in the band theory can always be written as $\hat{H}_{\rm eff}=\sum_{N,\bf k}E_{N,\bf k}|\Psi_{N,\bf k}\rangle\langle\Psi_{N,\bf k}|$, where $E_{N,\bf k}$ and $|\Psi_{N,\bf k}\rangle$ are respectively the dispersion relation and the associated Bloch wavefunction labeled by the band index $N$ and wavevector $\bf k$. Reconstruction of such a single-electron effective Hamiltonian is therefore equivalent to reconstructing the band structure and Bloch wavefunctions. Angle-resolved photoemission spectroscopy (ARPES) has been a powerful tool for measuring band structures~\cite{sobota2021angle,zhang2022angle}. Reconstruction of Bloch wavefunctions from ARPES is also possible by using ionizing radiation with a tunable linear-polarization angle, provided that the Wannier functions associated with the photoexcited electrons can be expressed as linear combinations of known atomic orbitals~\cite{schuler2022polarization}. However, due to the finite mean free path of the photoelectrons, ARPES is sensitive primarily to electronic properties near material surfaces. The recent development of strong laser fields has enabled bulk-sensitive techniques for probing these fundamental quantities based on highly nonlinear and nonequilibium processes such as high-order harmonic generation (HHG)~\cite{ghimire2011observation,schubert2014sub,hohenleutner2015real,liu2017high} and high-order sideband generation (HSG)~\cite{liu2007high,zaks2012experimental}.

In HHG, a single laser field drives both interband transition and intraband acceleration, which are generally intertwined with each other~\cite{golde2008high,golde2009microscopic,avetissian2022high}. By isolating the contribution to HHG from interband electron-hole recombination, Vampa et. al. employed a weak second-harmonic field to modulate the intensities of the resulting even-order harmonics and retrieved the electron-hole band-energy difference in zinc oxide (ZnO) with two fitting parameters based on simulations of the semiconductor Bloch equations (SBEs)~\cite{lindberg1988effective} in one-dimensional quasi-momentum space~\cite{vampa2015all}. Reconstruction of band structure based on saddle-point analysis~\cite{vampa2015semiclassical} of the interband HHG has also been proposed~\cite{li2020determination,chen2021reconstruction}. In all these works based on interband HHG, the energy dispersion relations of the electron-hole pairs were expanded as linear combinations of cosine functions with the coefficients treated as fitting parameters, a single conduction band and a single valence band were assumed, and the energy bandgaps were assumed to be known. With similar Fourier expansions for the dispersion relations, the contribution to HHG from intraband acceleration has been used to extract band energies by considering semiclassical motion of electrons in single-band models~\cite{luu2015extreme,lanin2017mapping,lv2021high}. In real insulators, two-band models, as used in all of the works based on interband HHG~\cite{vampa2015all,li2020determination,chen2021reconstruction}, are rarely accurate---there are almost always multiple conduction or valence bands that must be considered. More recently, reconstruction of three-dimensional multi-band effective Hamiltonians without distinguishing the interband and intraband contributions was proposed by solving the SBEs with empirical tight-binding models as inputs~\cite{parks2025full}. In the theoretical demonstration, two HHG intensity spectra were generated as ``experimental data" by solving the SBEs with a preassigned target Hamiltonian. It was then shown that, starting from a reasonably good initial guess for the Hamiltonian parameters, and tuning them to fit the ``data" by solving the same SBEs, one could obtain a Hamiltonian that yields energy bands close to those associated with the target Hamiltonian. It is yet to be tested in real HHG experiments whether the effective-Hamiltonian parameters can be unambiguously determined.

Reconstruction of effective Hamiltonians based on HSG is also at the demonstration stage. In contrast to HHG, interband transition and intraband acceleration in HSG are disentangled and separately controlled by two different laser fields, resulting in comparatively simple physical pictures~\cite{zaks2012experimental,zaks2013high,banks2013terahertz,langer2016lightwave,banks2017dynamical,valovcin2018optical,langer2018lightwave,borsch2020super,nagai2020dynamical,costello2021reconstruction,freudenstein2022attosecond,liu2024dephasing,o2024bloch,costello2023breaking}. Based on simulations of SBEs including up to four-point correlations, the HSG intensity spectra in monolayer tungsten diselenide have been linked to the band structure by considering the maximum electron-hole momentum attainable from an oscillating electric field~\cite{borsch2020super}. 
Progress has also been made toward reconstructing an effective Hamiltonian based on HSG in bulk gallium arsenide (GaAs). By near-resonantly exciting bulk GaAs with a weak near-infrared (NIR) laser while simultaneously applying a linearly polarized, strong terahertz (THz) field, reconstruction of the Bloch wavefunctions of holes in bulk GaAs has been achieved through a simple algebraic equation derived from a three-band model~\cite{costello2021reconstruction}. In this three-band model, the conduction-band electrons (Es) are described by a parabolic band with dispersion relation $H_c=E_{\rm g}+{\hbar^2k^2}/{(2m_c)}$, where $E_{\rm g}$ is the bandgap, $\hbar$ is the reduced Planck's constant, and $m_c$ is the conduction-band effective mass. For the valence bands, there are two species of holes called heavy holes (HHs) and light holes (LHs), which are described by the Luttinger Hamiltonian~\cite{luttinger1955motion}
\begin{align}
H_v
=
&
-\frac{\hbar^2}{2m_0}[
(\gamma_1+\frac{5}{2}\gamma_2)k^2{\bf 1}_4
-
2\gamma_3({\bf k}\cdot {\bf J})^2\notag\\
&+
2(\gamma_3-\gamma_2)(k_X^2J_X^2+k_Y^2J_Y^2+k_Z^2J_Z^2)
],
	\label{EQ:Luttinger}
\end{align}
where $m_0$ is the electron rest mass, $\gamma_1$, $\gamma_2$, and $\gamma_3$ are three Luttinger parameters, ${\bf 1}_4$ is the identity matrix of order four, and the components of $\bf J$, $J_X$, $J_Y$, and $J_Z$, are spin-3/2 matrices. Here, the $X$, $Y$, and $Z$ axes are defined by the crystal axes [100], [010], and [001], respectively. Reconstruction of the associated effective electron-hole Hamiltonian requires extraction of four parameters: the bandgap $E_{\rm g}$; the combined parameter $\mu_{\rm ex}/m_0\equiv (m_0/m_c+\gamma_1)^{-1}$, which enters the diagonal matrix elements; and the two Luttinger parameters $\gamma_2$ and $\gamma_3$. The parameter $\mu_{\rm ex}/m_0$ can be determined from the 1s-exciton binding energy~\cite{willatzen2009kp}, which has been extracted from absorbance spectra of bulk GaAs at 2\,K~\cite{sell1972resolved}. The capability of using HSG to extract the ratio $\gamma_3/\gamma_2$ has been demonstrated through reconstruction of the hole Bloch wavefunctions~\cite{costello2021reconstruction}. All of these Hamiltonian parameters determine the band energies and have been shown to be encoded in the E-HH and E-LH propagators, which govern acceleration of the electron-hole pairs under strong THz fields in HSG~\cite{costello2021reconstruction}. Although the electron-hole propagators can be determined up to a constant factor by measuring the polarization states of high-order sidebands~\cite{costello2021reconstruction,o2024bloch,costello2023breaking}, inverting the propagators to obtain the Hamiltonian parameters is challeging, because there are generally infinitely many quantum trajectories associated with an electron-hole propagator even in a parabolic two-band model~\cite{wu2023explicit}. Because of excitonic effects, the bandgap of a semiconductor is often difficult to determine precisely through traditional optical techniques such as linear absorption spectroscopy, except at very low temperatures~\cite{sell1972resolved}, and it generally varies with temperature~\cite{varshni1967temperature}. Surface-sensitive techniques such as ARPES~\cite{sobota2021angle,zhang2022angle} and scanning tunneling microscopy~\cite{feenstra2001recent} can be used to measure free electron-hole bandgaps, but the bandgaps near the material surfaces may differ from those in the bulk. To date, extraction of semiconductor bandgaps based on HHG or HSG has not been demonstrated.

By leveraging the theoretical work on tailoring an electron-hole propagator into contributions associated with a single electron-hole trajectory~\cite{wu2023explicit}, it was experimentally demonstrated that HSG from bulk GaAs that is near-resonantly excited by a weak NIR laser and simultaneously driven by a linearly polarized, strong terahertz (THz) field can be viewed as a Michelson-like interferometer for Bloch waves based on the three-band model discussed above~\cite{o2024bloch}. The polarizations of the sidebands, which serve as interferograms of the Bloch-wave interferometer, were reasonably reproduced by using an analytic model of the electron-hole propagators based on a classical description of electron-hole recollisions in a THz electric field. By neglecting the detuning of the NIR laser with respect to the bandgap $E_{\rm g}$ and taking the remaining Hamiltonian parameters from the literature, the Bloch-wave interferograms were used to extract an average dephasing constant for the two species of electron-hole pairs, the E-HH and E-LH pairs~\cite{o2024bloch}. Under the same assumption, it has also been shown that experimentally distinguishing the dephasing rates of the two electron-hole species is possible by studying the temperature dependences of the electron-hole propagators~\cite{costello2023breaking}.

In this paper, we demonstrate reconstruction of the effective electron-hole Hamiltonian in bulk GaAs based on the understanding of HSG in terms of the Bloch-wave interferometry~\cite{o2024bloch}. In our experiment, the GaAs sample exhibits a small exciton-peak splitting in the absorbance spectrum corresponding to a separation of the two valence bands near the band edge, possibly due to a strain induced by the substrate. We assign two different bandgaps to the two electron-hole species and assume that the curvatures of the energy bands remain the same as in unstrained samples. We simultaneously extract the two bandgaps at 30\,K, two dephasing constants associated respectively with the E-HH and E-LH pairs, and the combined parameter $\gamma_2\mu_{\rm ex}/m_0$. Different from previous works~\cite{o2024bloch,costello2023breaking}, information about the electron-hole propagators not only from the polarizations of the sidebands but also from the sideband intensity spectra, including their dependences on the THz-field strength and the NIR-laser frequency, was systematically collected from HSG experiments and compared with the theoretical results to unambiguously determine the dephasing constants and Hamiltonian parameters. Instead of using a classical picture of electron-hole recollsions, we employ here a more sophisticated analytic model based on saddle-point analysis to incorporate corrections from quantum fluctuations. Since the parameter $\mu_{\rm ex}/m_0$ can be determined from low-temperature absorbance spectra~\cite{sell1972resolved}, and the ratio $\gamma_3/\gamma_2$ can be extracted from HSG measurements~\cite{costello2021reconstruction}, we thus show that reconstruction of the three-band electron-hole Hamiltonian in bulk GaAs can be achieved by combining HSG experiments with absorbance spectroscopy. Interestingly, the bandgaps we extract are about 10\,meV larger than the values inferred from the 1s-exciton binding energy, which was determined by low-temperature absorbance measurements~\cite{sell1972resolved}. Through quantum-kinetic analysis for a generic electron-phonon system, we show that the electron-hole bandgaps in HSG may be renormalized through Fr\"ohlich interaction~\cite{frohlich1950xx} that is modulated by the strong THz field. We also show that the energy threshold for emission of optical phonons can be suppressed by applying a strong THz field, leading to nearly constant dephasing rates. It has long been known that an electron moving in a polar crystal can be dressed by optical phonons to form a polaron, a quasiparticle whose energy dispersion differs from that of a bare electron~\cite{lee1952motion}. Our work offers an opportunity to explore possible modification of polaronic effects under strong THz fields.

\section{Dynamical Jones matrices from high-order sideband polarimetry}\label{SEC:dyanmical_jones_mat}

To establish the connection between the effective electron-hole Hamiltonian of bulk GaAs and high-order sideband generation (HSG), we begin with a general discussion of the measurable quantities in HSG experiments. To simplify the analysis, we consider here HSG induced by quasi-continuous near-infrared (NIR) and terahertz (THz) waves. Because a photon possesses only two helicity components, the electric field of a sideband or of the NIR laser can always be represented by a two-component vector called a Jones vector. In this paper, the incident NIR laser or a sideband propagating in the air is treated as a monochromatic plane wave with a specific wavevector $q$ and an angular frequency $\omega$. Multiplying the associated Jones vector by the exponential factor $\exp{[i(qz-\omega t)]}$ gives the complex representation of the electric field propagating along the $z$ axis defined by the [001] crystal axis of GaAs. Because of the linearity of HSG with respect to the NIR-laser field, the Jones vector of each sideband can be connected with the Jones vector of the NIR laser through a two-by-two matrix, which is generally complex. For convenience in theoretical treatment, we use here the circular basis vectors $\hat{\sigma}_{\pm}=\pm({\hat{X}\pm i{\hat{Y}}})/{\sqrt{2}}$, where $\hat{X}$ and $\hat{Y}$ are the unit vectors along the [100] and [010] crystal axes of GaAs, respectively. In this basis, we can write
\begin{align}
	 \begin{pmatrix}
	    {E}_{+,n}\\
	    {E}_{-,n}
          \end{pmatrix} 	      
=
 	 \begin{pmatrix}
	T_{++,n}     & 	T_{+-,n}\\
	T_{-+,n}     & 	T_{--,n}
          \end{pmatrix} 	      
 	 \begin{pmatrix}
	    {E}_{+,\rm NIR}\\
	    {E}_{-,\rm NIR}
          \end{pmatrix}, 	      
 	\label{EQ:dynamical_Jones_matrix}
\end{align}
where $({E}_{+,n},{E}_{-,n})^{T}$ and $({E}_{+,\rm NIR},{E}_{-,\rm NIR})^{T}$ are the Jones vectors of the $n$th-order sideband and the NIR-laser field, respectively, with $\pm$ labeling the two helicity components. The two-by-two matrix with elements $T_{\pm\pm,n}$ is called a dynamical Jones matrix~\cite{banks2017dynamical}, in analogy to a Jones matrix for describing polarization-transforming optical components. All information accessible from HSG signals is therefore compactly encoded in the dynamical Jones matrices.

\begin{figure*}
	\includegraphics[width=0.94\textwidth]{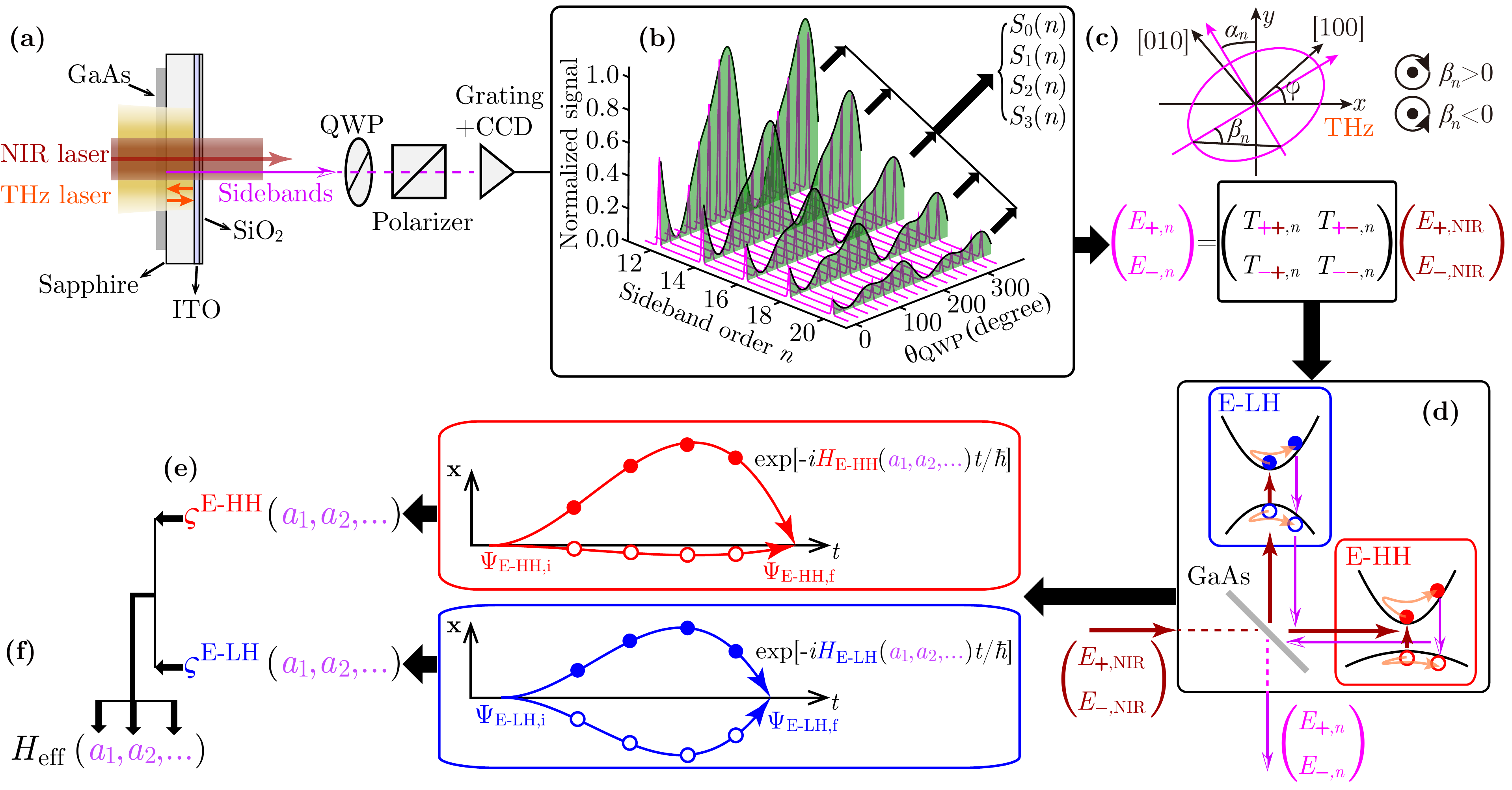}
	\caption{Effective-Hamiltonian reconstruction through Bloch-wave interferometry in bulk gallium arsenide (GaAs). (a) Experimental setup. A near-infrared (NIR) laser and a terahertz (THz) laser are focused collinearly onto a GaAs epilayer mounted on a sapphire substrate. An indium-tin-oxide (ITO) film on the opposite side of the substrate reflects the THz field to enhance the THz-field strength at the GaAs epilayer through constructive interference. A silicon-dioxide ($\rm SiO_2$) layer deposited on top of the ITO film acts as an anti-reflection coating for the NIR laser and the sidebands. Polarimetry of high-order sidebands is performed by passing the sideband fields through a quarter-wave plate (QWP) and a linear polarizer. A diffraction grating and a charge-coupled device (CCD) are combined to measure the intensities of a series of sidebands simultaneously. (b) The QWP is rotated by $360^\circ$ in $22.5^\circ$ steps. At each QWP rotation angle $\theta_{\rm QWP}$, an intensity spectrum is measured and plotted as a function of the sideband order $n$ (magenta curves), which is defined as the offset of the sideband frequency with respect to the NIR-laser frequency in units of the THz-laser frequency, with the laser linewidths ignored. For each sideband order, the total intensity is calculated as the area under the corresponding sideband peak in an intensity spectrum; its dependence on $\theta_{\rm QWP}$ (green shaded areas) yields the associated Stokes parameters, $S_0(n)$, $S_1(n)$, $S_2(n)$, and $S_3(n)$. (c) The polarization of each sideband is characterized by an orientation angle $\alpha_n$ and an ellipticity angle $\beta_n$, which are defined with respect to the THz electric field that makes an angle $\varphi$ with the [100] crystal axis. The sign of $\beta_n$ is positive (negative) when the sideband electric field rotates clockwise (counterclockwise) as it propagates away from the observer. In the linear regime with respect to the NIR laser, each sideband electric field with two helicity components $E_{\pm,n}$ and the NIR-laser electric field with two helicity components $E_{\pm,\rm NIR}$ are connected through a two-by-two matrix called a dynamical Jones matrix, which contains four complex elements $T_{\pm\pm,n}$. Each dynamical Jones matrix can be determined up to an overall phase factor by the measured Stokes parameters. (d) High-order sideband generation (HSG) in bulk GaAs that is near-resonantly excited by a NIR laser and simultaneously driven by a sufficiently strong linearly polarized THz field can be viewed as a Michelson-like interferometer for Bloch waves. First, the NIR laser is incident on the GaAs, creating an electron-hole Bloch wave. Second, the GaAs acts like a beam splitter, ``splitting" the electron-hole Bloch wave, which is a superposition of electron-heavy hole (E-HH) and electron-light hole (E-LH) Bloch waves, into two ``arms", one for each electron-hole species (closed circles for the electrons and open circles for the holes). Third, the THz field drives the E-HH and E-LH Bloch waves along different trajectories in their respective energy bands. Fourth, upon sideband emission, the E-HH and E-LH Bloch waves ``merge" at the ``beam splitter" (GaAs) and interfere with each other. Fifth, the sideband electric field as a function of sideband order $n$ is recorded as a Bloch-wave interferogram. (e) Based on the description of HSG in bulk GaAs in terms of a Bloch-wave interferometer, the measured dynamical Jones matrices are decoded into physical information including the electron-hole propagators. For each sideband, the E-HH (E-LH) propagator $\varsigma^{\rm E-HH}$ ($\varsigma^{\rm E-LH}$) describes a recollision process governed by an effective Hamiltonian $H_{\rm E-HH}$ ($H_{\rm E-LH}$), which contains the parameters of the total effective Hamiltonian $H_{\rm eff}$ for bulk GaAs, $a_1,a_2,\,...$. Here, $\Psi_{\rm E-HH,i}$ ($\Psi_{\rm E-LH,i}$) and $\Psi_{\rm E-HH,f}$ ($\Psi_{\rm E-LH,f}$) represent the initial state and final state respectively for the E-HH (E-LH) pair. (f) By inverting the propagators $\varsigma_{\rm E-HH}$ and $\varsigma_{\rm E-LH}$, the parameters $a_1,a_2,\,...$ are obtained and the effective Hamiltonian $H_{\rm eff}$ is reconstructed.}
	\label{FIG:bloch_wave_interferometry}
\end{figure*}

Polarimetry experiments were performed to measure the Jones vectors of the sidebands to determine the dynamical Jones matrices. A 100-mW NIR laser and a linearly polarized THz laser were focused collinearly onto the same spot of a 500-nm-thick gallium arsenide (GaAs) epilayer and propagated normal to the epilayer surface [Fig.~\ref{FIG:bloch_wave_interferometry} (a)]. The NIR laser was generated from an M Squared SolTiS titanium:sapphire laser with a tunable wavelength that was monitored in real time by a WS6-600 wavemeter. The THz radiation, consisting of 40-ns pulses at 0.447\,THz, was generated by the University of California, Santa Barbara (UCSB) Millimeter-Wave Free Electron Laser (FEL). The linewidth of the NIR laser is less than 5\,MHz, while the linewidth of the FEL is on the order of 1 GHz. The GaAs epilayer was grown along the [001] crystal axis and then transferred onto a sapphire substrate, which was approximately 488\,$\mu$m thick, through van der Waals bonding. A 250-nm-thick layer of indium tin oxide (ITO), which is transmissive to the NIR laser and reflective to the THz field, was grown on the opposite side of the sapphire substrate. Constructive interference between the incident THz field and the THz field reflected from the ITO led to an enhancement factor of 1.314 in the strength of the THz field at the GaAs epilayer. A 150-nm-thick silicon dioxide ($\rm SiO_2$) anti-reflection coating was deposited on the ITO film to minimize the NIR reflection of the sample and to avoid NIR Fabry-Perot oscillations in the HSG spectra. The sample was placed in a cryogenic chamber that was maintained at 30\,K during the HSG experiments. The generated sidebands propagated through a quarter-wave plate (QWP) and a linear polarizer before going into the detector. The intensities of a series of sidebands were recorded simultaneously by combining a diffraction grating with a charge-coupled device (CCD). We used the same sample and experimental setup as in Refs.~\cite{costello2021reconstruction} and~\cite{o2024bloch}, where more details about the sample preparation and the optics are provided. The variance in each sideband intensity spectrum was established from four repeated CCD scans. The QWP was rotated by $360^{\circ}$ in $22.5^{\circ}$ steps, and the polarization of each sideband was retrieved from the sideband intensity as a function of the QWP rotation angle $\theta_{\rm QWP}$. For each $\theta_{\rm QWP}$, sideband peaks were detected at frequencies $f_{\rm SB,n}=f_{\rm NIR}+nf_{\rm THz}$, where $f_{\rm NIR}$ is the frequency of the NIR laser, $f_{\rm THz}= 0.447$\,THz is the frequency of the THz field, and $n$ is an integer called the sideband order. Because of the reflection symmetry of the (001) GaAs crystal planes, only even-order sidebands appeared. 

The polarimetry experiment was repeated for 36 sets of laser parameters by using three NIR-laser wavelengths (819.5, 818, and 815\,nm), four NIR-laser polarizations including two circular polarizations with opposite helicities and two linear polarizations that were mutually orthogonal and oriented at $45^\circ$ to the THz electric field, and three THz-field strengths ranging from 27 to 64\,kV/cm for each NIR-laser setting. The polarization of the NIR-laser beam was set with a quarter-wave plate and a half-wave plate and was measured by a Thorlabs PAX polarimeter. Two wire-grid polarizers were used to attenuate the THz field while maintaining the polarization of the THz field in the GaAs epilayer (see Appendix~\ref{APP:thz_field_strength} for details about the THz-field strengths).

Figure~\ref{FIG:bloch_wave_interferometry} (b) shows an example of experimental sideband peaks varying with the QWP rotation angle $\theta_{\rm QWP}$. For each sideband order, the intensity $I(n,\theta_{\rm QWP})$ at $\theta_{\rm QWP}$ is calculated as the area under the corresponding sideband peak in an intensity spectrum. Owing to the narrow linewidths of both the NIR laser and the FEL, the sideband peaks are well separated. In the experiments, the transmission axis of the linear polarizer was aligned with the THz electric field, which defines the $x$ axis throughout this paper. The values of $\theta_{\rm QWP}$ were measured relative to the direction of the THz electric field. The intensity $I(n,\theta_{\rm QWP})$ recorded by the CCD can be written as
\begin{align}
&I(n,\theta_{\rm QWP})
= 
\frac{S_0(n)}{2}+\frac{S_1(n)}{4}-\frac{S_3(n)}{2}\sin(2\theta_{\rm QWP})\notag\\
&+\frac{S_1(n)}{4}\cos(4\theta_{\rm QWP})+\frac{S_2(n)}{4}\sin(4\theta_{\rm QWP}),
\label{EQ:intensiry_stokes}
\end{align}
where $S_0(n)={\mathcal I}_n$, $S_1(n)={\mathcal I}_n{p}_n\cos(2\alpha_n)\cos(2\beta_n)$, $S_2(n)={\mathcal I}_n{p}_n\sin(2\alpha_n)\cos(2\beta_n)$, and $S_3(n)={\mathcal I}_n{p}_n\sin(2\beta_n)$ are the Stokes parameters that define the intensity and polarization of the $n$th-order sideband. Here, $\alpha_n\in[-\pi/2,\pi/2]$ and $\beta_n\in[-\pi/4,\pi/4]$ are respectively the polarization angle and ellipticity angle defined with respect to the THz electric field [Fig.~\ref{FIG:bloch_wave_interferometry} (c)], $p_n$ is the degree of polarization, and ${\mathcal I}_n$ is the total intensity of the $n$th-order sideband. The Stokes parameters can be extracted from the Fourier transform:
\begin{align}
{\mathcal F}_{l}(n)
=
\int_0^{2\pi}
\frac{d\theta_{\rm QWP}}{2\pi}
I(n,\theta_{\rm QWP})e^{-il\theta_{\rm QWP}},
\label{EQ:fourier_stokes}
\end{align}
which yields $S_0(n)=2{\mathcal F}_{0}(n)-4{\rm Re}[{\mathcal F}_{0}(n)]$, $S_1(n)=8{\rm Re}[{\mathcal F}_{4}(n)]$, $S_2(n)=-8{\rm Im}[{\mathcal F}_{4}(n)]$, and $S_3(n)=4{\rm Im}[{\mathcal F}_{2}(n)]$. In the calculation, cubic-spline interpolation is used to generate a smooth $I$-$\theta_{\rm QWP}$ curve for each sideband [black curves in Fig.~\ref{FIG:bloch_wave_interferometry} (b)] on an equidistant grid of $\theta_{\rm QWP}$ with spacing $\Delta\theta=\pi/100$, and the integral in Eq.~(\ref{EQ:fourier_stokes}) is computed with the trapezoidal rule. The Jones vector $({E}_{+,n},{E}_{-,n})^{T}$ of a sideband field with orientation angle $\alpha_n$ and ellipticity angle $\beta_n$ takes the form
\begin{align}
\begin{pmatrix}
E_{+,n}\\
E_{-,n}
\end{pmatrix}
\propto
\begin{pmatrix}
e^{i(\varphi-\alpha_n)}(\cos\beta_n+\sin\beta_n)\\
-e^{-i(\varphi-\alpha_n)}(\cos\beta_n-\sin\beta_n)
\end{pmatrix},
\label{EQ:Jones_angles}
\end{align}
where $\varphi=43^\circ$ is the angle between the THz electric field and the [100] crystal axis [Fig.~\ref{FIG:bloch_wave_interferometry} (c)]. Using Eq.~(\ref{EQ:Jones_angles}), we see that the Jones vector $({E}_{+,n},{E}_{-,n})^{T}$ can be determined up to an overall phase factor from the Stokes parameters through the compact equations
\begin{align}
&\frac{n_{\rm Air}c\varepsilon_0}{2}(|E_{+,n}|^2+|E_{-,n}|^2)
=
{\mathcal I}^{\rm pol}_n,
\label{EQ:Stokes_to_Jones_amp}
\\
&\frac{E_{+,n}}{E_{-,n}}
=
-e^{2i\varphi}
\frac{1+\tilde{S}_3(n)}{\tilde{S}_1(n)+i\tilde{S}_2(n)},
\label{EQ:Stokes_to_Jones_phase}
\end{align}
where $n_{\rm Air}$ is the refractive index of the air, $c$ is the speed of light, $\varepsilon_0$ is the vacuum permittivity, ${\mathcal I}^{\rm pol}_n=\sqrt{S_1(n)^2+S_2(n)^2+S_3(n)^2}$ is the intensity associated with the polarized sideband signal, and $\tilde{S}_j(n)\equiv S_j(n)/{\mathcal I}^{\rm pol}_n$ ($j=1,2,3$) are normalized Stokes parameters.

For a given Jones vector of the incident NIR laser, $({E}_{+,\rm NIR},{E}_{-,\rm NIR})^{T}$, Eqs.~(\ref{EQ:Stokes_to_Jones_amp}) and~(\ref{EQ:Stokes_to_Jones_phase}) provide two equations that relate the measured sideband Stokes parameters to the dynamical Jones matrices:
\begin{align}
&|T_{++,n}{\tilde E}_{+,\rm NIR} + T_{+-,n}{\tilde E}_{-,\rm NIR}|^2\notag\\
&+
|T_{-+,n}{\tilde E}_{+,\rm NIR} + T_{--,n}{\tilde E}_{-,\rm NIR}|^2
=
\frac{{\mathcal I}^{\rm pol}_n}{{\mathcal I}_{\rm NIR}},
\label{EQ:Stokes_to_Jonesmat_amp}
\\
&\frac{
T_{++,n}{\tilde E}_{+,\rm NIR} + T_{+-,n}{\tilde E}_{-,\rm NIR}
}
{
T_{-+,n}{\tilde E}_{+,\rm NIR} + T_{--,n}{\tilde E}_{-,\rm NIR}
}=
\frac{-e^{2i\varphi}[1+\tilde{S}_3(n)]}{\tilde{S}_1(n)+i\tilde{S}_2(n)},
\label{EQ:Stokes_to_Jonesmat_phase}
\end{align}
where $({\tilde E}_{+,\rm NIR},{\tilde E}_{-,\rm NIR})^{T}\equiv{\bf E}_{\rm NIR}/F_{\rm NIR}$ is the NIR-laser Jones vector normalized by the field amplitude $F_{\rm NIR}$, and ${\mathcal I}_{\rm NIR}=n_{\rm Air}c\varepsilon_0|F_{\rm NIR}|^2/2$ is the intensity of the NIR laser.
In principle, repeating the polarimetry experiment for three or more different NIR-laser polarizations, one can extract the ratios between the matrix elements $T_{\pm\pm,n}$ by using Eq.~(\ref{EQ:Stokes_to_Jonesmat_phase}), and then determine each dynamical Jones matrix up to an overall phase factor with the aid of Eq.~(\ref{EQ:Stokes_to_Jonesmat_amp})~\cite{banks2017dynamical}. Here, we adopt a different approach based on the structure of the dynamical Jones matrices revealed in an earlier HSG experiment with the same setup~\cite{costello2021reconstruction}:
\begin{align}
&T_{++,n}=T_{--,n},
\label{EQ:Tmat_diagonal}
\\
&
\frac
{T_{+-,n}}
{T_{-+,n}}
=
\frac
{\sin(2\theta)-i(\gamma_3/\gamma_2)\cos(2\theta)
}
{
\sin(2\theta)+i(\gamma_3/\gamma_2)\cos(2\theta)
},
\label{EQ:Tmat_offdiagonal}
\end{align}
where $\theta=\varphi+\pi/4$ is the angle between the THz electric field and the [110] crystal axis, and $\gamma_3/\gamma_2$ is the ratio of two Luttinger parameters. Equation~(\ref{EQ:Tmat_offdiagonal}) has previously been used to extract the ratio $\gamma_3/\gamma_2$, which determines the Bloch wavefunctions of holes in bulk GaAs~\cite{costello2021reconstruction}. Using Eqs.~(\ref{EQ:Stokes_to_Jonesmat_amp}),~(\ref{EQ:Stokes_to_Jonesmat_phase}),~(\ref{EQ:Tmat_diagonal}), and~(\ref{EQ:Tmat_offdiagonal}), we can determine each dynamical Jones matrix up to a phase factor individually for each of the 36 polarimetry experiments with different laser parameters (see Appendix~\ref{APP:cal_Jones_mat} for details about the calculation). In theory, the dynamical Jones matrices should not depend on the NIR-laser polarization as HSG is a linear response with respect to the NIR-laser field. This expectation is consistent with our experimental observations, as discussed below.

\section{Electron-hole propagators from dynamical Jone matrices}\label{SEC:eh_propagators}

Besides Eq.~(\ref{EQ:Tmat_offdiagonal}), additional relations between the dynamical Jones matrices and the effective-Hamiltonian parameters can be explored by examining the absolute value of $T_{++,n}$ and the ratio ${T_{-+,n}}/{T_{++,n}}$. These quantities were previously shown to be connected with the electron-hole propagators that govern the electron-hole recollision processes in HSG based on the three-band model discussed in Sec.~\ref{SEC:introduction}~\cite{costello2021reconstruction,o2024bloch,costello2023breaking}. Within this three-band model, generally, the HSG signal should include contributions from electron-hole pairs created at any wavevector $\bf k$ by the NIR laser and then accelerated along a straight line in the Brillouin zone under the linearly polarized THz field. The spin of an electron-hole pair moving along a straight line that does not contain the ${\bf k}={\bf 0}$ point in the Brillouin zone is not constant because of the coupling between the four spin-3/2 hole states described by the Luttinger Hamiltonian [Eq.~(\ref{EQ:Luttinger})]~\cite{costello2021reconstruction}. Nevertheless, previous experiments indicate that, in bulk GaAs near-resonantly excited by a NIR laser, HSG is dominantly described by electron-hole recollision pathways starting from ${\bf k}={\bf 0}$~\cite{costello2021reconstruction,o2024bloch}. By restricting the dynamics to such recollision pathways, along which the spins of the electron-hole pairs remain constant, a three-step model of HSG in bulk GaAs was developed by decomposing each accelerating electron-hole Bloch wave into two interfering components, the electron-heavy hole (E-HH) and electron-light hole (E-LH) Bloch waves~\cite{costello2021reconstruction}. First, the NIR laser creates an electron-hole Bloch wave, which is a superposition of E-HH and E-LH Bloch waves. Second, the THz field drives the E-HH and E-LH Bloch waves along different $k$-space trajectories in their respective energy bands. Third, upon sideband emission, the E-HH and E-LH Bloch waves interfere as two components of the same electron-hole Bloch wave, imprinting information about the electronic structure onto the sideband electric fields. Based on this physical picture, the Fourier component of the interband polarization corresponding to the $n$th-order sideband, ${\mathbb P}_{n}$, was connected with the Jones vector of the incident NIR laser, ${\bf E}_{\rm NIR}$, through~\cite{costello2021reconstruction}
\begin{align}
{\mathbb P}_{n}
=
\frac{1}{|d|^2}
\sum_{s} 
    \begin{pmatrix}
    {\bf D}^{\rm E-HH}_{s} \\
    {\bf D}^{\rm E-LH}_{s}
    \end{pmatrix}^{\dagger}
    \begin{pmatrix}
    {{\mathbb Q}^{{\rm E-HH}}_n} & 0 \\
    0 & {{\mathbb Q}^{\rm E-LH}_n}
    \end{pmatrix}\notag\\
    \times
    \begin{pmatrix}
    {\bf D}^{\rm E-HH}_{s}\cdot {\bf E}_{\rm NIR} \\
    {\bf D}^{\rm E-LH}_{s}\cdot {\bf E}_{\rm NIR}
    \end{pmatrix},
    \label{EQ:Jonesmat_theory_P}
\end{align}
where $s$ labels the two-fold spin degeneracy of the electron-hole pairs, $d$ is a constant that determines the magnitude of the dipole moment ${\bf D}^{\rm E-HH (E-LH)}_{s}$ associated with the E-HH (E-LH) pair, and ${\mathbb Q}^{\rm E-HH(E-LH)}_{n}$ is the E-HH (E-LH) propagator that describes the E-HH (E-LH) acceleration under the strong THz field. In the derivation of Eq.~(\ref{EQ:Jonesmat_theory_P}), the electric field of the NIR laser is assumed to be constant in the GaAs epilayer. In reality, the NIR-laser field decays slightly and acquires a $z$-dependent phase as it propagates through the GaAs epilayer. The interband polarization field in the GaAs epilayer acts as a source of the sideband radiation that propagates through the multilayer structure of the sample before reaching the detector. The Jones vector ${\bf E}_{n}={E}_{+,n}\hat{\sigma}_{+}+{E}_{-,n}\hat{\sigma}_{-}$ associated with the detected sideband electric field differs from the Fourier component ${\mathbb P}_{n}$ by a proportionality factor ${\mathcal T}_n$, which depends on the dielectric functions of the materials in the sample at the NIR-laser and sideband frequencies, as well as on the material thicknesses (see Appendix~\ref{APP:propagation_sidebands} for more details about the sideband propagation). Therefore, the Jones vectors ${\bf E}_{n}$ and ${\bf E}_{\rm NIR}$ satisfy a similar relation:
\begin{align}
{\bf E}_{n}
=
\frac{1}{|d|^2}
\sum_{s} 
    \begin{pmatrix}
    {\bf D}^{\rm E-HH}_{s} \\
    {\bf D}^{\rm E-LH}_{s}
    \end{pmatrix}^{\dagger}
    \begin{pmatrix}
    \varsigma^{\rm E-HH}_n & 0 \\
    0 &  \varsigma^{\rm E-LH}_n
    \end{pmatrix}\notag\\
    \times
    \begin{pmatrix}
    {\bf D}^{\rm E-HH}_{s}\cdot {\bf E}_{\rm NIR} \\
    {\bf D}^{\rm E-LH}_{s}\cdot {\bf E}_{\rm NIR}
    \end{pmatrix},
    \label{EQ:Jonesmat_theory_E}
\end{align}
where $\varsigma^{\rm E-HH(E-LH)}_n\equiv{\mathcal T}_n{\mathbb Q}^{\rm E-HH (E-LH)}_n$ is the electron-hole propagator incorporating the aforementioned propagation effects. 

By using the explicit forms of the dipole vectors, which are constant along the electron-hole recollision pathways containing ${\bf k}={\bf 0}$ and are determined by the eigenfunctions of the Luttinger Hamiltonian, a comparison of Eqs.~(\ref{EQ:dynamical_Jones_matrix}) and~(\ref{EQ:Jonesmat_theory_E}) yields the following relations between the dynamical Jones matrices and the electron-hole propagators~\cite{costello2021reconstruction}:
\begin{align}
        T_{++,n} = T_{--,n} &= \frac{2+n _Z}{3} \varsigma_{n}^{\rm E-HH} +  \frac{2 -n_Z}{3} \varsigma_{n}^{\rm E-LH}, 
        \label{EQ:tppmm_propagator}\\
               T_{-+,n}&= \frac{n_X - in_Y}{\sqrt{3}} (\varsigma_{n}^{\rm E-HH} - \varsigma_{n}^{\rm E-LH}),
        \label{EQ:tmp_propagator}\\
        T_{+-,n}&= \frac{n_X + in_Y}{\sqrt{3}} (\varsigma_{n}^{\rm E-HH} - \varsigma_{n}^{\rm E-LH}), 
        \label{EQ:tpm_propagator}
\end{align}
where $\hat{n}=(n_X,n_Y,n_Z)$ is a unit vector along the vector $((\sqrt{3}/2)\sin2\theta,-({\sqrt{3}\gamma_3}/{2\gamma_2})\cos2\theta,-{1}/{2})$ that is defined by the angle $\theta$ and the Luttinger-parameter ratio $\gamma_3/\gamma_2$. Using Eqs.~(\ref{EQ:tppmm_propagator}) and (~\ref{EQ:tmp_propagator}), we can calculate the electron-hole propagators from the measured dynamical Jones matrices as:
\begin{align}
 \varsigma^{\rm E-HH}_n
= 
\frac{3}{4}
{T_{n,++}}
[
1
+
\frac{2-n_Z}{ \sqrt{3}
        (n_{X}-in_{Y}) }
\frac{T_{-+,n}}{T_{++,n}}
],
        \label{EQ:ehh_propagator_tppmp}\\
 \varsigma^{\rm E-LH}_n
= 
\frac{3}{4}
{T_{n,++}}
[
1
-
\frac{2-n_Z}{ \sqrt{3}
        (n_{X}-in_{Y}) }
\frac{T_{-+,n}}{T_{++,n}}
]                             
        \label{EQ:elh_propagator_tppmp}.
\end{align}

As discussed in Sec.~\ref{SEC:dyanmical_jones_mat}, the ratio ${T_{-+,n}}/{T_{++,n}}$ can be determined by solving Eqs.~(\ref{EQ:Stokes_to_Jonesmat_phase}),~(\ref{EQ:Tmat_diagonal}), and~(\ref{EQ:Tmat_offdiagonal}), while the value of $T_{++,n}$ can be determined up to a phase factor by using Eq.~(\ref{EQ:Stokes_to_Jonesmat_amp}). Therefore, the ratio $\varsigma^{\rm E-HH}_n/ \varsigma^{\rm E-LH}_n$, which contains the relative phase between the E-HH and E-LH propagators $\varsigma^{\rm E-HH}_n$ and $\varsigma^{\rm E-LH}_n$, can be fully determined from the polarimetry experiments, whereas each propagator individually can only be determined up to an overall phase factor.

\section{Theoretical model of the electron-hole propagators}\label{SEC:theory}

The question now is how to extract the Hamiltonian parameters from the experimentally determined propagators $\varsigma^{\rm E-HH}_n$ and  $\varsigma^{\rm E-LH}_n$. Under the electric field of the linearly polarized THz field, ${\bf E}_{\rm THz}(t)$, in general, these propagators contain contributions from infinitely many $k$-space trajectories ${\bf k}(t)={\bf P}+(e/\hbar){\bf A}_{\rm THz}(t)$,
where $\hbar\bf P$ is the canonical momentum, $e$ is the elementary charge, and ${\bf A}_{\rm THz}$ is the vector potential of the THz electric field satisfying $-\dot{\bf A}_{\rm THz}={\bf E}_{\rm THz}$. To tackle this problem, theoretical consideration of HSG based on the saddle-point analysis has resulted in a way to tailor a two-band electron-hole propagator into an explicit form under the condition of a sufficiently large THz-field strength and sufficiently fast dephasing relative to the THz-field oscillation~\cite{wu2023explicit}. The explicit formula contains contributions only from the $k$-space trajectories in the vicinity the shortest electron-hole recollision pathways. The mapping of HSG polarimetry spectra onto the output of a Michelson-like interferometer for Bloch waves~\cite{o2024bloch}, as detailed in Fig.~\ref{FIG:bloch_wave_interferometry} (d), results when one considers only the shortest electron-hole recollision pathways starting from ${\bf k}={\bf 0}$. The bulk GaAs acts like a beam splitter, ``splitting" an electron-hole Bloch wave created by the NIR laser into two ``arms", one for each electron-hole species. In each ``arm", an electron-hole pair driven by the THz field accumulates a dynamic phase determined by the effective electron-hole Hamiltonian and suffers from dephasing, in analogy to optical light waves propagating in a lossy arm in a Michelson interferometer. Upon sideband emission, the two electron-hole components ``merge" at the ``beam splitter" and imprint the material information that is encoded in the electron-hole propagators onto the sideband polarizations as the interferograms. The main features in the interferograms associated with the sidebands emitted from the same GaAs epilayer used here were reproduced by using a simple analytic model of the electron-hole propagators~\cite{o2024bloch}. In this simple model, the electron-hole recollision pathways are calculated according to Newton's equations of motion, and the quantum fluctuations around the classical electron-hole recollision pathways are ignored~\cite{o2024bloch,costello2023breaking}.

Quantum fluctuations, however, have been shown to play an essential role in determining the absolute magnitudes and phases of the sideband polarizations in a parabolic two-band model~\cite{wu2023explicit}. To extract the effective-Hamiltonian parameter based a more quantitative understanding of the Bloch-wave interferometer, we model the electron-hole propagators by considering all electron-hole $k$-space trajectories passing ${\bf k}={\bf 0}$ as well as the trajectories nearby. Following Ref.~\cite{costello2021reconstruction}, under the free electron-hole approximation, when all these $k$-space trajectories are included, the Fourier component of the interband polarization corresponding to the $n$th-order sideband, ${\mathbb P}_n$, can be connected with the Jones vector of the incident NIR laser, ${\bf E}_{\rm NIR}$, through
\begin{widetext}
\begin{align}
{\mathbb P}_n
= 
&
\frac{i}{\hbar}
\sum_{s}
\int\frac{d^3{\bf P}}{(2\pi)^3}
\frac{1}{T_{\rm THz}}
\int_{0}^{T_{\rm THz}} dt 
e^{i(\omega_{\rm NIR}+n\omega_{\rm THz} )t} 
    \begin{pmatrix}
    {\bf D}^{\rm E-HH}_{{\bf k}(t),s} \\
    {\bf D}^{\rm E-LH}_{{\bf k}(t),s}
    \end{pmatrix}^{\dagger}
\notag\\
&
\times
\int_{-\infty}^tdt'
    \hat{T}
    \exp\{
    -\frac{i}{\hbar}
    \int_{t'}^t dt''
    [ e{\bf E}_{\rm THz}(t'')\cdot{\mathcal A}^{*}_{{\bf k}(t''),s}
    +
    \begin{pmatrix}
    E^{\rm E-HH}_{{\bf k}(t'')}-i\Gamma_{\rm E-HH} & 0 \\
    0 & E^{\rm E-LH}_{{\bf k}(t'')}-i\Gamma_{\rm E-LH}
    \end{pmatrix}
    ]
    \}
\notag\\
&
   \times
    \begin{pmatrix}
    {\bf D}^{\rm E-HH}_{{\bf k}(t'),s}\cdot {\bf E}_{\rm NIR} \\
    {\bf D}^{\rm E-LH}_{{\bf k}(t'),s}\cdot {\bf E}_{\rm NIR}
    \end{pmatrix}
    e^{-i\omega_{\rm NIR} t'},
        \label{EQ:interband_polarization_general}
\end{align}
\end{widetext}
which describes HSG in bulk GaAs as a more general three-step process. In the first step, an electron-hole pair is created by the NIR laser through the coupling between the dipole moments at ${\bf k}(t')$, ${\bf D}^{\rm E-HH}_{{\bf k}(t'),s}$ and ${\bf D}^{\rm E-LH}_{{\bf k}(t'),s}$ associated respectively with the E-HH and E-LH components, and the NIR-laser electric field described by the Jones vector ${\bf E}_{\rm NIR}$. In the second step, the electron-hole pair accumulates dynamic phases determined by the E-HH and E-LH energies, $E^{\rm E-HH}_{\bf k}$ and $E^{\rm E-LH}_{\bf k}$, as well as a non-Abelian Berry phase determined by the two-by-two Berry connection matrix ${\mathcal A}_{{\bf k},s}$ associated with the two valence bands. The electron-hole pair also suffers from dephasing described by the two dephasing constants $\Gamma_{\rm E-HH}$ and $\Gamma_{\rm E-LH}$, which are associated respectively with the E-HH and E-LH pairs. In the third step, the electron and hole recombine and emit sidebands through the dipole moments at ${\bf k}(t)$. Here, $s$ labels the two-fold spin degeneracy of the electron-hole pairs, $T_{\rm THz}=1/f_{\rm THz}$ is the period of the THz field, $\omega_{\rm NIR}=2\pi f_{\rm NIR}$ and $\omega_{\rm THz}=2\pi f_{\rm THz}$ are respectively the angular frequencies of the NIR and THz lasers, and $\hat{T}$ is the time-ordering operator. As in the discussion of Eq.~(\ref{EQ:Jonesmat_theory_P}), a constant NIR-laser electric field in the GaAs epilayer is assumed in the derivation of Eq.~(\ref{EQ:interband_polarization_general}). Because the thickness of the GaAs epilayer is much smaller than the THz-field wavelength, we use the $z$-independent continuous wave form ${\bf E}_{\rm THz}(t)=\hat{x}F_{\rm THz}\cos(\omega_{\rm THz}t)$ for the THz electric field with a field strength $F_{\rm THz}$. Since the electron-hole pairs are driven along the $x$ axis [Fig.~\ref{FIG:bloch_wave_interferometry}(c)], to evaluate the effects of quantum fluctuations in the electron-hole energies, we expand the electron-hole energies $E^{\rm E-HH}_{\bf k}$ and $E^{\rm E-LH}_{\bf k}$ to second order in the wavevector components $k_y\equiv k_X\sin\varphi+k_Y\cos\varphi$ and $k_z\equiv k_Z$ for finite $k_x\equiv k_X\cos\varphi-k_Y\sin\varphi$:
\begin{align}
E^{\nu, (2)}_{\bf k}
=
E_{\rm g}
+
\frac{\hbar^2}{2}
[\frac{k_x^2}{\mu^{\nu}_{xx}}
+
\frac{k_xk_y}{\mu^{\nu}_{xy}}
+
\frac{k_y^2}{{\mu}^{\nu}_{yy}}
+
\frac{k_z^2}{\mu^{\nu}_{zz}}
],
        \label{EQ:eh_energy}
\end{align}
where $\nu=\rm E-HH,E-LH$, and the reduced mass tensor $\mu^{\nu}_{jl}$ ($j,l=x,y,z$) is defined by
\begin{align}
(\frac{\mu^{\nu}_{xx}}{\mu_{\rm ex}})^{-1}
=
&
1
+
2\eta_\nu\xi f(\varphi),
        \label{EQ:mu_xx}
        \\
(\frac{{\mu}^{\nu}_{yy}}{\mu_{\rm ex}})^{-1}
=&
1
+
2\eta_\nu\xi\frac{\gamma_{32}[2-3\sin^2(2\varphi)]+1}{f(\varphi)},
        \label{EQ:mu_yy}
        \\
(\frac{\mu^{\nu}_{zz}}{\mu_{\rm ex}})^{-1}
 =
 &
1+2\eta_\nu\xi\frac{2\gamma_{32}+1}{f(\varphi)},
         \label{EQ:mu_zz}
         \\
 (\frac{\mu^{\nu}_{xy}}{\mu_{\rm ex}})^{-1}
=
&
-2\eta_\nu\xi\frac{\gamma_{32}\sin(4\varphi)}{f(\varphi)},
        \label{EQ:mu_xy}
\end{align}
with $\eta_{\rm E-HH}=-1$, and $\eta_{\rm E-LH}=+1$. Here, we have introduced two combined parameters, $\xi\equiv\gamma_2\mu_{\rm ex}/m_0$ and $\gamma_{32}\equiv 3[({\gamma_3}/{\gamma_2})^2-1]/4$, and a function of the crystal orientation angle $\varphi$, $f(\varphi)\equiv\sqrt{1+\gamma_{32}\sin^2(2\varphi)}$. We further ignore the variation of the hole spins for small $k_y$ and $k_z$, i.e., we take the Berry connection matrix ${\mathcal A}_{{\bf k},s}$ to be zero and the dipole moment ${\bf D}^{\rm E-HH (E-LH)}_{{\bf k},s}$ to be a constant vector ${\bf D}^{\rm E-HH (E-LH)}_{s}$ defined by the Bloch wavefunctions along the $k$-space trajectories containing ${\bf k}={\bf 0}$. Under these assumptions, the Fourier component ${\mathbb P}_n$ is still described by  Eq.~(\ref{EQ:Jonesmat_theory_P}), with the electron-hole propagator ${\mathbb Q}^{\nu}_n$ ($\nu=\rm E-HH,E-LH$) taking the following form:
\begin{align}
{\mathbb Q}^{\nu}_n
=
&
\frac{i}{\hbar}|d|^2
\int\frac{d^3{\bf P}}{(2\pi)^3}
\frac{1}{T_{\rm THz}}
\int_{0}^{T_{\rm THz}} dt 
e^{in\omega_{\rm THz} t} 
\int_{-\infty}^tdt'
\notag\\
&
\times
    \exp\{
    -\frac{i}{\hbar}
    \int_{t'}^t dt''
[    E^{\nu,(2)}_{{\bf k}(t'')}-\hbar\omega_{\rm NIR}-i\Gamma_{\nu}    ]
    \}.
        \label{EQ:propagator_forms}
\end{align}
By redefining the canonical momentum $\bf P$ in the integral through the transformations $P_x\rightarrow P_x-[{\mu^{\nu}_{xx}}/({2\mu^{\nu}_{xy}})]P_y$,  $P_y\rightarrow \sqrt{{{\mu}^{\nu}_{yy}}/{\mu^{\nu}_{xx}}}P_y$ with ${\tilde\mu}^{\nu}_{yy}\equiv [({{\mu}^{\nu}_{yy}})^{-1}-{\mu^{\nu}_{xx}}/{(2\mu^{\nu}_{xy})^2}]^{-1}$, and $P_z\rightarrow \sqrt{{\mu^{\nu}_{zz}}/{\mu^{\nu}_{xx}}}P_z$, Eq.~(\ref{EQ:propagator_forms}) can be written in a form of the Feynman path integrals that have been studied in the description of HSG processes in parabolic two-band models~\cite{liu2007high,yan2008theory,xie2013effects,wu2023explicit}:
\begin{align}
{\mathbb Q}^{\nu}_n
=
&
\frac{ \sqrt{ {\tilde\mu}^{\nu}_{yy}\mu^{\nu}_{zz} } }{\mu^{\nu}_{xx}}
\frac{i}{\hbar}
|d|^2
\frac{1}{T_{\rm THz}}
\int_{0}^{T_{\rm THz}} dt 
\int\frac{d^3{\bf P}}{(2\pi)^3}
\notag\\
&
\times
\int_{-\infty}^tdt'
    \exp\{
    \frac{i}{\hbar}
    S^{\nu}_n({\bf P},t',t)
    \},
        \label{EQ:propagator_isotropic}
\end{align}
with an action
\begin{align}
S^{\nu}_n({\bf P},t',t)
=
    & n\hbar\omega_{\rm THz} t 
    -
    \int_{t'}^t dt''\frac{\hbar^2}{2\mu^{\nu}_{xx}}[{\bf P}+\frac{e}{\hbar}{\bf A}_{\rm THz}(t'')]^2
    \notag\\
    &+i(\Gamma_{\nu}-i\Delta_{\rm NIR})(t-t'),
        \label{EQ:propagator_action}
\end{align}
where $\Delta_{\rm NIR}=\hbar\omega_{\rm NIR}-E_{\rm g}$ is the detuning of the NIR laser with respect to the bandgap $E_{\rm g}$. Equation~(\ref{EQ:propagator_isotropic}) contains all effective electron-hole Hamiltonian parameters including the bandgap $E_{\rm g}$, the combined parameter $\mu_{\rm ex}$, and the two Luttinger parameters $\gamma_2$ and $\gamma_3$. When the Hamiltonian parameters are known, numerical calculation of similar Feynman path integrals has been shown to be straightforward~\cite{liu2007high,yan2008theory,xie2013effects,wu2023explicit}. However, it is not clear whether the Hamiltonian parameters can be uniquely determined by HSG signals based on numerical integration.

Since the original prediction of HSG~\cite{liu2007high}, the saddle-point analysis has provided a way of simplifying the Feynman path integrals in Eq.~(\ref{EQ:propagator_isotropic}) into a sum over countably many electron-hole recollision pathways~\cite{yan2008theory,xie2013effects,wu2023explicit}. In each electron-hole recollision pathway associated with the $n$th-order sideband, an electron and a hole are created at time $t^{\prime\nu}_n$ at an initial wavevector ${\bf k}^{\nu}_n(t^{\prime\nu}_n)={\bf P}^{\nu}_n+(e/\hbar){\bf A}_{\rm THz}(t^{\prime\nu}_n)$, and they recombine at time $t^{\nu}_n$, with the complex saddle point $({\bf P}^{\nu}_n, t^{\prime\nu}_n, t^{\nu}_n)$ satisfying the saddle-point equations:
\begin{align}
&\int_{t^{\prime\nu}_n}^{t^{\nu}_n} dt''\frac{\hbar{\bf k}^{\nu}_n(t'')}{\mu^{\nu}_{xx}}
={\bf 0},
\label{EQ:saddle_point1}
\\
&\frac{\hbar^2}{2\mu^{\nu}_{xx}}[{\bf k}^{\nu}_n(t^{\prime\nu}_n)]^2
=i\Gamma_{\nu}+\Delta_{\rm NIR},
\label{EQ:saddle_point2}
\\
&
\frac{\hbar^2}{2\mu^{\nu}_{xx}}[{\bf k}^{\nu}_n(t^{\nu}_n)]^2
=n\hbar\omega_{\rm THz}+i\Gamma_{\nu}+\Delta_{\rm NIR}.
        \label{EQ:saddle_point3}
\end{align}
The first saddle-point equation corresponds to the condition that the electron and hole recollide at the place where they are created. The other two saddle-point equations describe generalized conditions of energy conservation at the electron-hole creation time $t^{\prime\nu}_n$ and recollision time $t^{\nu}_n$.
Since the THz field is periodic in time, in principle, there are infinitely many saddle points corresponding to infinitely many electron-hole recollision pathways with different acceleration times $t^{\nu}_n-t^{\prime\nu}_n$. Further simplification of the Feynman path integrals can be achieved by considering the regime of sufficiently large THz-field strength and sufficiently strong dephasing~\cite{wu2023explicit}. By using a sufficiently strong THz field, the ponderomotive energy $U^{\nu}_{\rm p}\equiv{e^2F_{\rm THz}^2}/{(4\mu^{\nu}_{xx}\omega_{\rm THz}^2)}$, which defines the kinetic energy gain of an electron-hole pair with a reduced mass $\mu^{\nu}_{xx}$ in a THz period $T_{\rm THz}$, can be much larger than the sideband energy offset $n\hbar\omega_{\rm THz}$, the dephasing constant $\Gamma_{\nu}$, and the NIR-laser detuning $\Delta_{\rm NIR}$. Under the condition $U^{\nu}_{\rm p}\gg\Gamma_{\nu},\Delta_{\rm NIR}$, the complex saddle points are close to their counterparts in the absence of dephasing and detuning, and the electron-hole recollisions can be approximately considered as governed by ordinary classical mechanics. Based on classical mechanics, when $U^{\nu}_{\rm p}\gg n\hbar\omega_{\rm THz}$, the shortest electron-hole recollision pathways associated with the $n$th-order sideband should lie around the nodes of the THz electric field, where the THz electric field is almost linear in time. When dephasing of the electron-hole pairs is sufficiently faster than the THz-field oscillations ($\Gamma_{\nu}/\hbar\gg f_{\rm THz}$), the shortest electron-hole recollision pathways should dominate. Following Ref.~\cite{wu2023explicit}, by taking the THz electric field as almost linear in time and including only the shortest recollision pathways together with the associated Gaussian quantum fluctuations, the electron-hole propagator ${\mathbb Q}^{\nu}_n$ can be approximated as
\begin{align}
{\mathbb Q}^{\nu}_n
\approx 
&
2i^n e^{i{\pi }/{4}}\frac{|d|^2}{\hbar\omega_{\rm THz}}  (\frac{\omega_{\rm THz}}{2\pi\hbar})^{3/2}\sqrt{ \mu^{\nu}_{xx}{\tilde\mu}^{\nu}_{yy}\mu^{\nu}_{zz} }
\notag\\
&
\times({\tilde U}^{\nu}_{\rm p})^{1/8}
\frac{\exp[-i\arg[q^{\nu}_{0}(n,i{\tilde\Gamma}_{\nu}+{\tilde\Delta}_{\rm NIR})]/2]}
{\sqrt{|q^{\nu}_{0}(n,i{\tilde\Gamma}_{\nu}+{\tilde\Delta}_{\rm NIR})|}}
\notag\\
&
\times
\exp
\{
i[
q^{\nu}_{1/4}(n,i{\tilde\Gamma}_{\nu}+{\tilde\Delta}_{\rm NIR})({\tilde U}^{\nu}_{\rm p})^{-1/4}
\notag\\
&
+
q^{\nu}_{3/4}(n,i{\tilde\Gamma}_{\nu}+{\tilde\Delta}_{\rm NIR})({\tilde U}^{\nu}_{\rm p})^{-3/4}
]
\},
\label{EQ:propagator_algebraic_form}
\end{align}
where ${\tilde U}^{\nu}_{\rm p}\equiv U^{\nu}_{\rm p}/\hbar\omega_{\rm THz}$, ${\tilde\Delta}_{\rm NIR}\equiv\Delta_{\rm NIR}/\hbar\omega_{\rm THz}$ and ${\tilde\Gamma}_{\nu}\equiv\Gamma_{\nu}/\hbar\omega_{\rm THz}$ are respectively the ponderomotive energy $U^{\nu}_{\rm p}$, the NIR-laser detuning $\Delta_{\rm NIR}$, and the dephasing constant $\Gamma_{\nu}$ in units of the THz photon energy $\hbar\omega_{\rm THz}$, and $q^{\nu}_{0}$, $q^{\nu}_{1/4}$, and $q^{\nu}_{3/4}$ are functions of the sideband order $n$ and the complex quantity $i{\tilde\Gamma}_{\nu}+{\tilde\Delta}_{\rm NIR}$ in the following forms:
\begin{align}
q^{\nu}_{0}(n, & i{\tilde\Gamma}_{\nu}+{\tilde\Delta}_{\rm NIR})
=
-\sqrt{32(3\sqrt{2})^3}\zeta_{\nu,0}\notag\\
&\times\zeta_{\nu,n}(\zeta_{\nu,n}-\zeta_{\nu,0})^{\frac{5}{2}},
\label{EQ:definition_q0}
\\
q^{\nu}_{1/4}(n, & i{\tilde\Gamma}_{\nu}+{\tilde\Delta}_{\rm NIR})
=
(\frac{2}{9})^{1/4}
\frac{4\sqrt{\zeta_{\nu,n} - \zeta_{\nu,0}}}{5}
\notag\\
&
\times(2\zeta_{\nu,0}^2+\zeta_{\nu,0}\zeta_{\nu,n}+2\zeta_{\nu,n}^2),
\label{EQ:definition_q1o4}
\\
q^{\nu}_{3/4}(n, & i{\tilde\Gamma}_{\nu}+{\tilde\Delta}_{\rm NIR})
=
(\frac{1}{18})^{1/4}
\frac{1}{1260\sqrt{\zeta_{\nu,n} - \zeta_{\nu,0}}}
\notag\\
&\times[103(\zeta_{\nu,n}^2-\zeta_{\nu,0}^2)^2+232\zeta_{\nu,0}\zeta_n(\zeta^2_{\nu,0}+\zeta^2_{\nu,n})
\notag\\
&-184\zeta_{\nu,0}^2\zeta^2_{\nu,n}],
\label{EQ:definition_q3o4}
\end{align}
with $\zeta_{\nu,n}\equiv\sqrt{i{\tilde\Gamma}_{\nu}+{\tilde\Delta}_{\rm NIR}+n}$. Here, a square root of a complex number is defined to have a nonnegative real part. The first two lines of Eq.~(\ref{EQ:propagator_algebraic_form}) incorporate the effects of quantum fluctuations by summing over the $k$-space trajectories in the vicinity the shortest electron-hole recolission pathways associated with the saddle points $({\bf P}^{\nu}_n, t^{\prime\nu}_n, t^{\nu}_n)$, while the other two lines describe the quantum mechanical phase as well as the dephasing of an electron-hole pair moving along a shortest recollision pathway. If the THz electric field is perfectly linear in time, the exponential factor will not include the term $q^{\nu}_{3/4}({\tilde U}^{\nu}_{\rm p})^{-3/4}$. The term $q^{\nu}_{3/4}({\tilde U}^{\nu}_{\rm p})^{-3/4}$ gives a correction to the linear-in-time approximation of the THz electric field, making the formula valid for a broader range of experimental parameters.

The analytic model of the electron-hole propagator $\varsigma^{\nu}_n$ used in Refs.~\cite{o2024bloch} and~\cite{costello2023breaking} can be considered as a limiting case of Eq.~(\ref{EQ:propagator_algebraic_form}) for sufficiently high sideband orders, with the quantum fluctuations and the correction term $q^{\nu}_{3/4}({\tilde U}^{\nu}_{\rm p})^{-3/4}$ ignored. For sideband orders satisfying $n>|i{\tilde\Gamma}_{\nu}+{\tilde\Delta}_{\rm NIR}|$, the function $q^{\nu}_{1/4}$ can be expanded as the following Taylor series:
\begin{align}
q^{\nu}_{1/4}(n,& i{\tilde\Gamma}_{\nu}+{\tilde\Delta}_{\rm NIR})
=
(18n)^{1/4}
[
\frac{8}{15}n
+
(i{\tilde\Gamma}_{\nu}+{\tilde\Delta}_{\rm NIR})
\notag\\
&
\times
(1
-
\frac{1}{3}\sqrt{\frac{i{\tilde\Gamma}_{\nu}+{\tilde\Delta}_{\rm NIR}}{n}}
+...
)
].
\label{EQ:taylor_q1o4}
\end{align}
Apart from a constant factor, the analytic model of the electron-hole propagator $\varsigma^{\nu}_n$ used in Refs.~\cite{o2024bloch} and~\cite{costello2023breaking} is just the exponential function $\exp[iq^{\nu,(1)}_{1/4}({\tilde U}^{\nu}_{\rm p})^{-1/4}]$, where $q^{\nu,(1)}_{1/4}=(18n)^{1/4}[(8/15)n+i{\tilde\Gamma}_{\nu}+{\tilde\Delta}_{\rm NIR}]$ is the Taylor expansion of $q^{\nu}_{1/4}$ [Eq.~(\ref{EQ:taylor_q1o4})] retained up to the first-order term in $i{\tilde\Gamma}_{\nu}+{\tilde\Delta}_{\rm NIR}$.

By using Eq.~(\ref{EQ:propagator_algebraic_form}), the experimentally measured electron-hole propagator $\varsigma^{\nu}_n\equiv{\mathcal T}_n{\mathbb Q}^{\nu}_n$ ($\nu=\rm E-HH,E-LH$) is associated with a shortest electron-hole recollision pathway and becomes an explicit function of the effective electron-hole Hamiltonian parameters [Fig.~\ref{FIG:bloch_wave_interferometry} (e)]. We will show how to extract the Hamiltonian parameters and the dephasing constants by employing the explicit functional forms of the electron-hole propagators and achieve the reconstruction of the three-band effective Hamiltonian for bulk GaAs [Fig.~\ref{FIG:bloch_wave_interferometry} (f)]. 

\begin{figure}
	\includegraphics[width=0.47\textwidth]{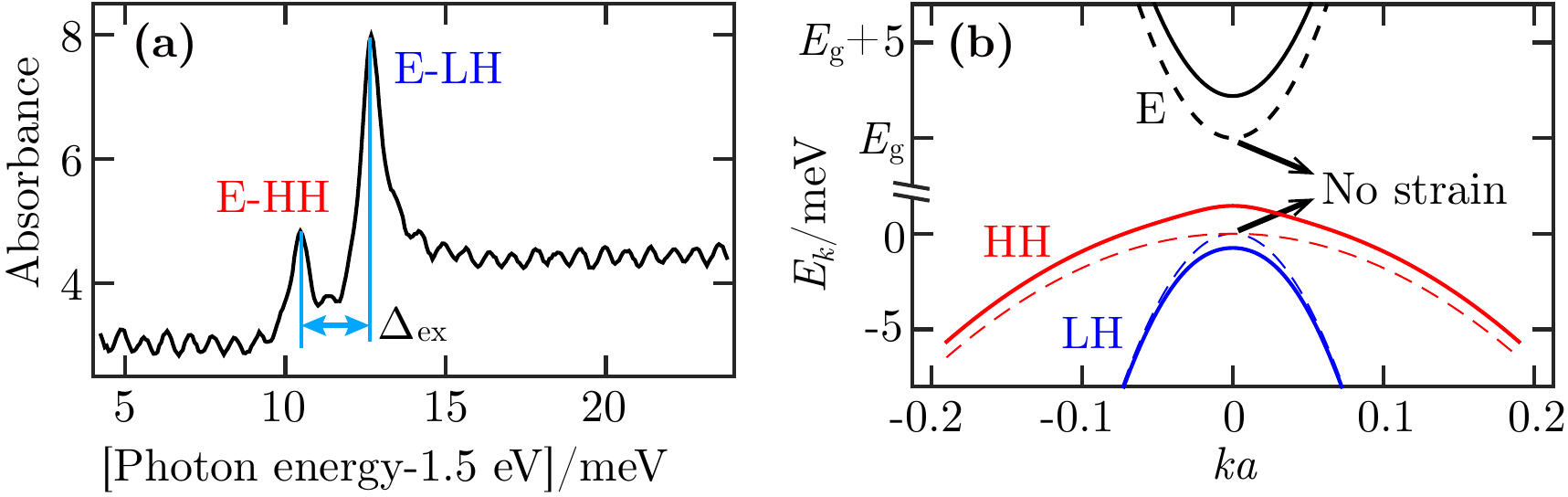}
	\caption{Different bandgaps for two electron-hole species. (a) An absorbance spectrum for the GaAs epilayer at 30\,K. Two exciton peaks with an energy splitting $\Delta_{\rm ex}\approx2.2$\,meV are observed. These peaks are associated with the E-HH and E-LH pairs, respectively. (b) Band structure of bulk GaAs including a lowest conduction band (E band) and two highest valence bands (HH and LH bands). The solid lines represent the energy bands calculated by including a biaxial strain that induces a 2.2-meV splitting between the HH and LH bands at ${\bf k}={\bf 0}$. The dashed lines represent the energy bands with no strain effects included. Here, the dimensionless wavevector $ka$ is used with $a=5.65$\,\AA\, being the lattice constant of GaAs~\cite{soma1982thermal,driscoll1975precision}, and $E_{\rm g}$ labels the bandgap.}
	\label{FIG:exciton_splitting}
\end{figure}
To ensure that the GaAs was near-resonantly excited by the NIR laser in the HSG experiments, an absorbance spectrum of the GaAs sample was measured by using a white-light source to locate the optical excitation gap. The powers of the white light transmitted through the cryogenic chamber with and without the sample, $W$ and $W_0$, were recorded, and the absorbance was calculated as $-10\log_{10}(W/W_0)$. In the absorbance spectrum [Fig.~\ref{FIG:exciton_splitting} (a)], the GaAs epilayer exhibited an exciton-peak splitting $\Delta_{\rm ex}$ of approximately 2.2\,meV, indicating that the degeneracy of the HH and LH bands was lifted. This splitting may arise from strain induced by the sapphire substrate, whose thermal expansion coefficient differs from that of GaAs. For example, as shown in Fig.~\ref{FIG:exciton_splitting} (b), the valence-band degeneracy can be lifted by a biaxial strain, which can also induce constant energy shifts in both the conduction and valence bands~\cite{willatzen2009kp} (see Appendix~\ref{APP:biaxial_strain} for more details about the band structure calculation). Since the exciton-peak splitting is small, without worrying about the details of the strain, we focus on the demonstration of Hamiltonian reconstruction by simply assuming that the E-HH and E-LH energies are still described by Eq.~(\ref{EQ:eh_energy}) but with two bandgaps for the slightly strained GaAs, $E_{\rm g,E-HH}$ and $E_{\rm g,E-LH}$, which are associated with the two electron-hole species. Accordingly, we will distinguish the NIR-laser detunings for the E-HH and E-LH pairs by defining $\Delta^{\nu}_{\rm NIR}\equiv\hbar\omega_{\rm NIR}-E_{\rm g,\nu}$ ($\nu=\rm E-HH,E-LH$). Based on the absorbance spectrum, we impose the constraint $E_{\rm g,E-LH}-E_{\rm g,E-HH}=\Delta_{\rm ex}$ without increasing the number of Hamiltonian parameters.

In our HSG experiments, the mean values of the ponderomotive energies $U^{\rm E-HH}_{\rm p}$ and $U^{\rm E-LH}_{\rm p}$ are estimated to lie in the ranges $[3.7\times10^2,2.0\times10^3]\hbar\omega_{\rm THz}$ and $[6.2\times10^2,3.4\times10^3]\hbar\omega_{\rm THz}$, respectively, by using the literature values of the conduction-band effective mass $m_c=0.067m_0$~\cite{ahmed1992far} and the Luttinger parameters $\gamma_1=6.98$, $\gamma_2=2.2$, and $\gamma_3=2.9$~\cite{skolnick1976investigation}. Accordingly, we focus on a small range of sideband orders, $12\le n\le 38$, to ensure that $U^{\nu}_{\rm p}\gg n\hbar\omega_{\rm THz}$. We will also show that the extracted dephasing constant $\Gamma_{\nu}$ and NIR-laser detuning $\Delta^{\nu}_{\rm NIR}$ indeed satisfy $U^{\nu}_{\rm p}\gg\Gamma_{\nu},\Delta^{\nu}_{\rm NIR}$ and $\Gamma_{\nu}/\hbar\gg f_{\rm THz}$.

\section{Collecting information for Hamiltonian reconstruction}\label{SEC:information}

Armed with an explicit formula for the theoretical electron-hole propagators, ${\mathbb Q}^{\rm E-HH}_n$ and ${\mathbb Q}^{\rm E-LH}_n$ [Eq.~(\ref{EQ:propagator_algebraic_form})], we are now in a position to collect information for Hamiltonian reconstruction from the experimentally measured electron-hole propagators, ${\varsigma}^{\rm E-HH}_n$ and ${\varsigma}^{\rm E-LH}_n$. As discussed in Secs.~\ref{SEC:dyanmical_jones_mat} and~\ref{SEC:eh_propagators}, in each of the 36 repeated polarimetry experiments, the propagator ratio ${\varsigma}^{\rm E-HH}_n/{\varsigma}^{\rm E-LH}_n$ was fully determined, whereas each of the electron-hole propagators ${\varsigma}^{\rm E-HH}_n$ and ${\varsigma}^{\rm E-LH}_n$, was only determined up to an overall phase factor. The propagator ratio ${\varsigma}^{\rm E-HH}_n/{\varsigma}^{\rm E-LH}_n$ contains the full information encoded in the matrix-element ratio ${T_{-+,n}}/{T_{++,n}}$ [see Eqs.~(\ref{EQ:ehh_propagator_tppmp}) and (\ref{EQ:elh_propagator_tppmp})], while the absolute values of the propagators, $|{\varsigma}^{\rm E-HH}_n|$ and $|{\varsigma}^{\rm E-LH}_n|$, contain additional information encoded in the absolute value of the matrix element $T_{++,n}$. 

\begin{figure*}
	\includegraphics[width=0.7\textwidth]{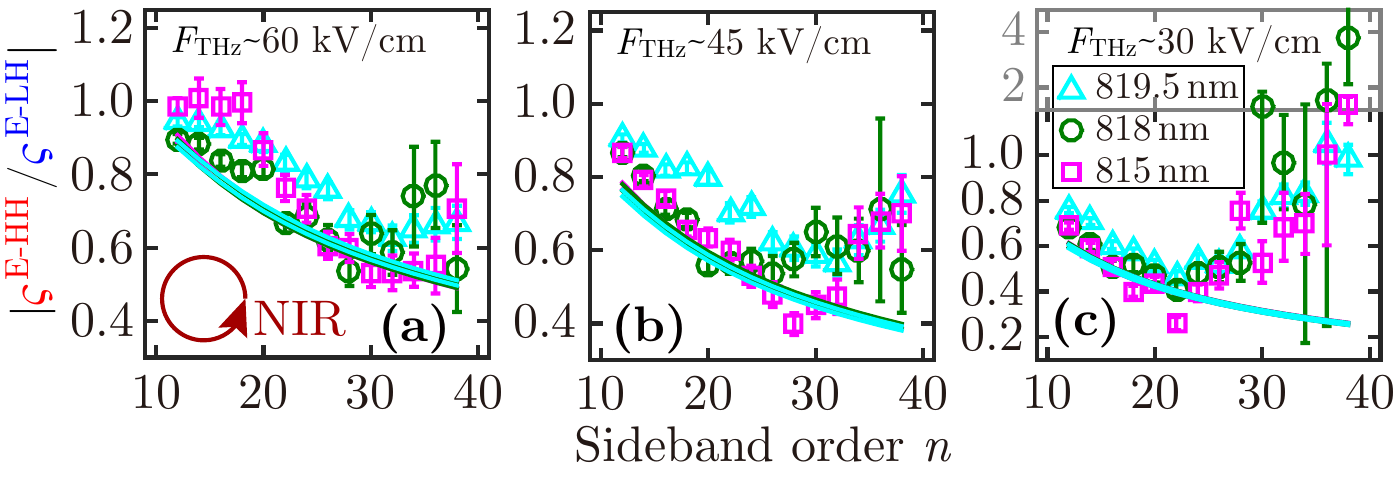}
	\caption{Absolute values of the propagator ratio $\varsigma^{\rm E-HH}/\varsigma^{\rm E-LH}$. The data were obtained by using a left-handed circular polarization (helicity -1) for the NIR laser. Panels (a), (b), and (c) show the data collected at three different THz-field strength levels: around 60\,kV/cm, 45\,kV/cm, and 30\,kV/cm, respectively (see Fig.~\ref{FIG:thz_field_strength} in Appendix~\ref{APP:thz_field_strength} for the exact THz-field strengths). In each panel, cyan triangles, dark green circles, and magenta squares represent the data obtained at three different NIR-laser wavelengths: 819.5\,nm, 818\,nm, and 815\,nm, respectively. The cyan, dark green, and magenta solid lines represent the corresponding theoretical results. For each set of laser parameters, two solid lines of the same color indicate the one-standard-deviation range resulting from uncertainties in the THz field strengths and the Hamiltonian parameters. The theoretical curves in each panel largely overlap. In (c), a larger $y$ scale is used for the data within the grey box.}
	\label{FIG:abs_propagator_ratio}
\end{figure*}
\begin{figure}
	\includegraphics[width=0.47\textwidth]{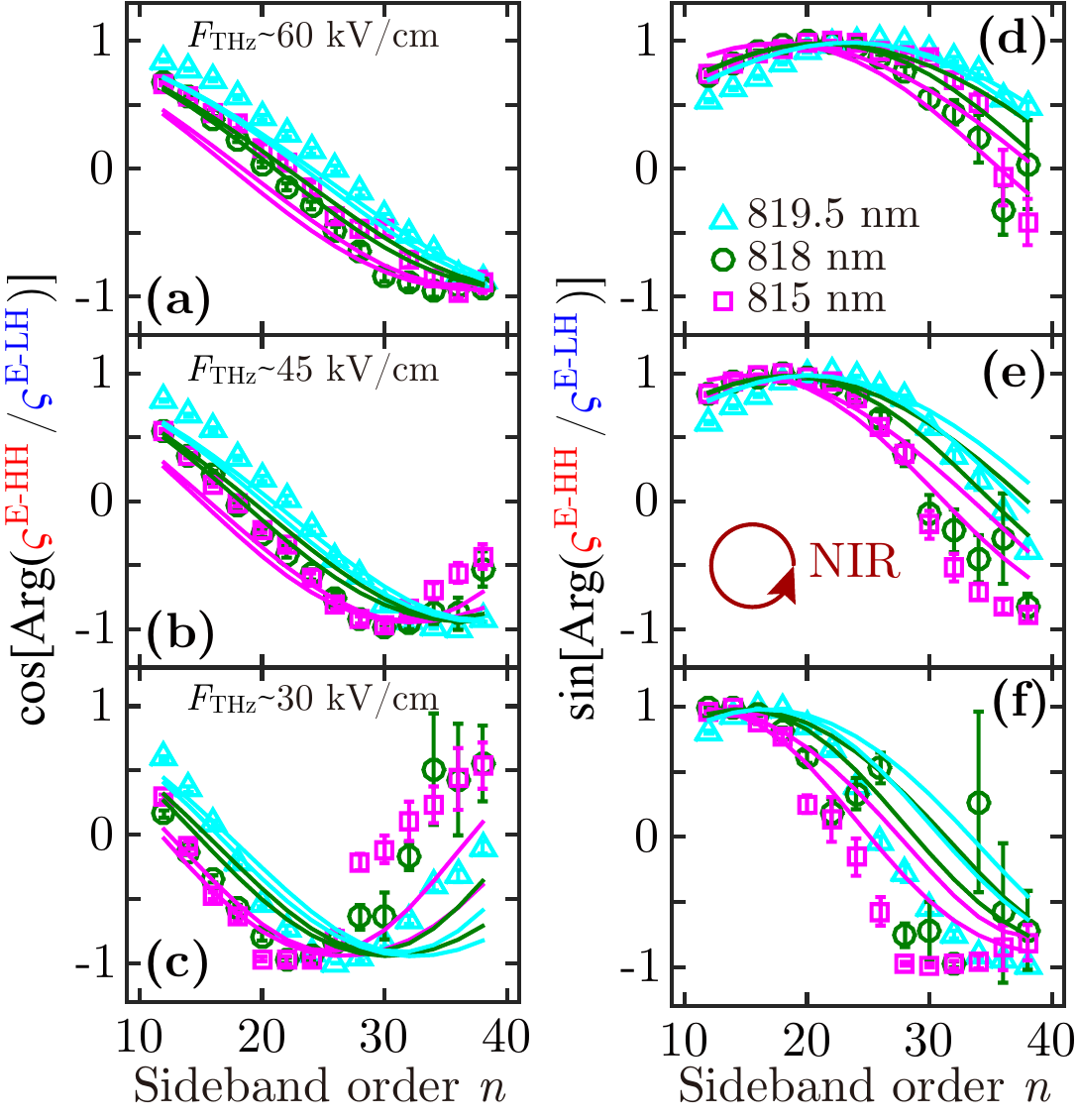}
	\caption{Phases of the propagator ratio $\varsigma^{\rm E-HH}/\varsigma^{\rm E-LH}$ expressed in terms of cosine and sine functions. The data were obtained by using a left-handed circular polarization (helicity -1) for the NIR laser. The first, second, and third rows show the data collected at three different THz-field strength levels: around 60\,kV/cm, 45\,kV/cm, and 30\,kV/cm, respectively (see Fig.~\ref{FIG:thz_field_strength} in Appendix~\ref{APP:thz_field_strength} for the exact THz-field strengths). In each panel, cyan triangles, dark green circles, and magenta squares represent the data obtained at three different NIR-laser wavelengths: 819.5\,nm, 818\,nm, and 815\,nm, respectively. The cyan, dark green, and magenta solid lines represent the corresponding theoretical results. For each set of laser parameters, two solid lines of the same color indicate the one-standard-deviation range resulting from uncertainties in the THz field strengths and the Hamiltonian parameters.}
	\label{FIG:arg_propagator}
\end{figure}
According to Eqs.~(\ref{EQ:Stokes_to_Jones_phase}),~(\ref{EQ:Stokes_to_Jonesmat_phase}),~(\ref{EQ:Tmat_diagonal}), and~(\ref{EQ:Tmat_offdiagonal}), the polarization of the $n$th-order sideband is determined by the ratio ${T_{-+,n}}/{T_{++,n}}$ and thus by the propagator ratio ${\varsigma}^{\rm E-HH}_n/{\varsigma}^{\rm E-LH}_n$. In previous work on Bloch-wave interferometry in bulk GaAs~\cite{o2024bloch}, only the polarization states of the sidebands, but not their amplitudes, were discussed, i.e., only the information contained in the propagator ratio ${\varsigma}^{\rm E-HH}_n/{\varsigma}^{\rm E-LH}_n$ was explored. Figures~\ref{FIG:abs_propagator_ratio} and~\ref{FIG:arg_propagator} show respectively the absolute values and phases of the propagator ratio ${\varsigma}^{\rm E-HH}_n/{\varsigma}^{\rm E-LH}_n$ measured in nine of the 36 polarimetry experiments by using a left-handed circular polarization (helicity -1) for the NIR laser (see Figs.~\ref{FIG:abs_ratio_all} and~\ref{FIG:arg_ratio_all} in Appendix~\ref{APP:sup_data} for data from all 36 polarimetry experiments and calculation of the error bars). Qualitatively, the data are consistent with the analytic model of the electron-hole propagator proposed in Ref.~\cite{o2024bloch} based on classical electron-hole recollisions. As discussed in Sec.~\ref{SEC:theory}, in the analytic model, the electron-hole propagator ${\varsigma}^{\nu}_n$ is an exponential function of the form ($\nu={\rm E-HH,E-LH}$):
\begin{align}
{\varsigma}^{\nu}_n\propto\exp[i(\frac{8}{15}n+\frac{i{\Gamma}_{\nu}+{\Delta}^{\nu}_{\rm NIR}}{\hbar\omega_{\rm THz}})(\frac{18n\hbar\omega_{\rm THz}}{{U}^{\nu}_{\rm p}})^{1/4}],
\label{EQ:propagator_analytic}
\end{align}
which is parametrized by the dephasing constant ${\Gamma}_{\nu}$, the NIR-laser detuning ${\Delta}^{\nu}_{\rm NIR}$, and the ponderomotive energy ${U}^{\nu}_{\rm p}$ in units of the THz photon energy $\hbar\omega_{\rm THz}$. The quantity $\tau^{\nu}_n\equiv[{(18n\hbar\omega_{\rm THz})}/{{U}^{\nu}_{\rm p}}]^{1/4}/\omega_{\rm THz}$ represents the acceleration time of the shortest classical electron-hole recollision pathways associated with the $n$th-order sideband, and the factor $({8}/{15})n\omega_{\rm THz}\tau^{\nu}_n$ is related to the quantum mechanical phase acquired by the electron-hole pairs~\cite{o2024bloch}. Based on this analytic model, if the exciton-peak splitting is neglected ($\Delta^{\nu}_{\rm NIR}\equiv\Delta_{\rm NIR}$), and the two electron–hole species are assumed to share a common dephasing constant $\bar{\Gamma}$, the propagator ratio ${\varsigma}^{\rm E-HH}_n/{\varsigma}^{\rm E-LH}_n$ can be written as
\begin{align}
\frac{{\varsigma}^{\rm E-HH}_n}{{\varsigma}^{\rm E-LH}_n}
=
\exp\{i&(\frac{8}{15}n+\frac{i{\bar\Gamma}+\Delta_{\rm NIR}}{\hbar\omega_{\rm THz}})
\notag\\
&
\times[\omega_{\rm THz}(\tau^{\rm E-HH}_n-\tau^{\rm E-LH}_n)]\}.
\label{EQ:propagator_ratio_analytic}
\end{align}
Because the E-HH pairs have a larger reduced mass $\mu^{\nu}_{xx}$ and therefore a smaller ponderomotive energy than the E-LH pairs, the acceleration time $\tau^{\nu}_n$ is longer for the E-HH pairs. Moreover, the difference in the acceleration times scales with the sideband order and THz-field strength as $\tau^{\rm E-HH}_n-\tau^{\rm E-LH}_n\propto n^{1/4}/\sqrt{F_{\rm THz}}$. Equation~(\ref{EQ:propagator_ratio_analytic}) therefore predicts that the absolute value of ${\varsigma}^{\rm E-HH}_n/{\varsigma}^{\rm E-LH}_n$ should be less than one and should decrease with increasing sideband order and decreasing THz-field strength. It also predicts that the absolute value of ${\varsigma}^{\rm E-HH}_n/{\varsigma}^{\rm E-LH}_n$ is insensitive to the NIR-laser wavelength. These trends are consistent with the data shown in Fig.~\ref{FIG:abs_propagator_ratio}, where the absolute values of ${\varsigma}^{\rm E-HH}_n/{\varsigma}^{\rm E-LH}_n$ mostly lie in the range from 0.4 to 1.0. For the phases of ${\varsigma}^{\rm E-HH}_n/{\varsigma}^{\rm E-LH}_n$, Eq.~(\ref{EQ:propagator_ratio_analytic}) implies that $\cos[{\rm Arg}({\varsigma}^{\rm E-HH}_n/{\varsigma}^{\rm E-LH}_n)]$ ($\sin[{\rm Arg}({\varsigma}^{\rm E-HH}_n/{\varsigma}^{\rm E-LH}_n)]$) should oscillate as a function of the sideband order, similar to the standard cosine (sine) function, with a shorter period for a weaker THz field. As shown in Fig.~\ref{FIG:arg_propagator}, approximately half of an oscillation cycle is observed in $\cos[{\rm Arg}({\varsigma}^{\rm E-HH}_n/{\varsigma}^{\rm E-LH}_n)]$ or $\sin[{\rm Arg}({\varsigma}^{\rm E-HH}_n/{\varsigma}^{\rm E-LH}_n)]$ for the lowest THz-field strength level [Fig.~\ref{FIG:arg_propagator} (c) and (f)]. Equation~(\ref{EQ:propagator_ratio_analytic}) also implies that ${\rm Arg}({\varsigma}^{\rm E-HH}_n/{\varsigma}^{\rm E-LH}_n)$ approaches zero in the limit of vanishing sideband order and infinitely strong THz field, consistent with the data shown in Fig.~\ref{FIG:arg_propagator}, where $\cos[{\rm Arg}({\varsigma}^{\rm E-HH}_n/{\varsigma}^{\rm E-LH}_n)]$ ($\sin[{\rm Arg}({\varsigma}^{\rm E-HH}_n/{\varsigma}^{\rm E-LH}_n)]$) is closer to one (zero) for stronger THz fields at the lowest recorded sideband order. In each panel of Fig.~\ref{FIG:arg_propagator}, the data corresponding to the same THz-field strength level indicate that the longest NIR-laser wavelength, for which $\Delta_{\rm NIR}$ is the smallest, corresponds to the smallest ${\rm Arg}({\varsigma}^{\rm E-HH}_n/{\varsigma}^{\rm E-LH}_n)$, in agreement with Eq.~(\ref{EQ:propagator_ratio_analytic}). Note that the THz-field strength varies slightly when the NIR laser was tuned to a different frequency due to fluctuations in the FEL output power (see Fig.~\ref{FIG:thz_field_strength} in Appendix~\ref{APP:thz_field_strength} for the exact THz-field strengths). When the NIR-laser wavelength was tuned to 819.5\,nm, the mean FEL output power was measured to be slightly higher, resulting in a smaller difference in the acceleration time, $\tau^{\rm E-HH}_n-\tau^{\rm E-LH}_n$, and therefore even smaller ${\rm Arg}({\varsigma}^{\rm E-HH}_n/{\varsigma}^{\rm E-LH}_n)$. We also see that, in each panel of Fig.~\ref{FIG:abs_propagator_ratio}, the absolute value of ${\varsigma}^{\rm E-HH}_n/{\varsigma}^{\rm E-LH}_n$ is overall slightly larger for the data corresponding to the 819.5-nm NIR-laser wavelength, as expected from Eq.~(\ref{EQ:propagator_ratio_analytic}).

\begin{figure}
	\includegraphics[width=0.47\textwidth]{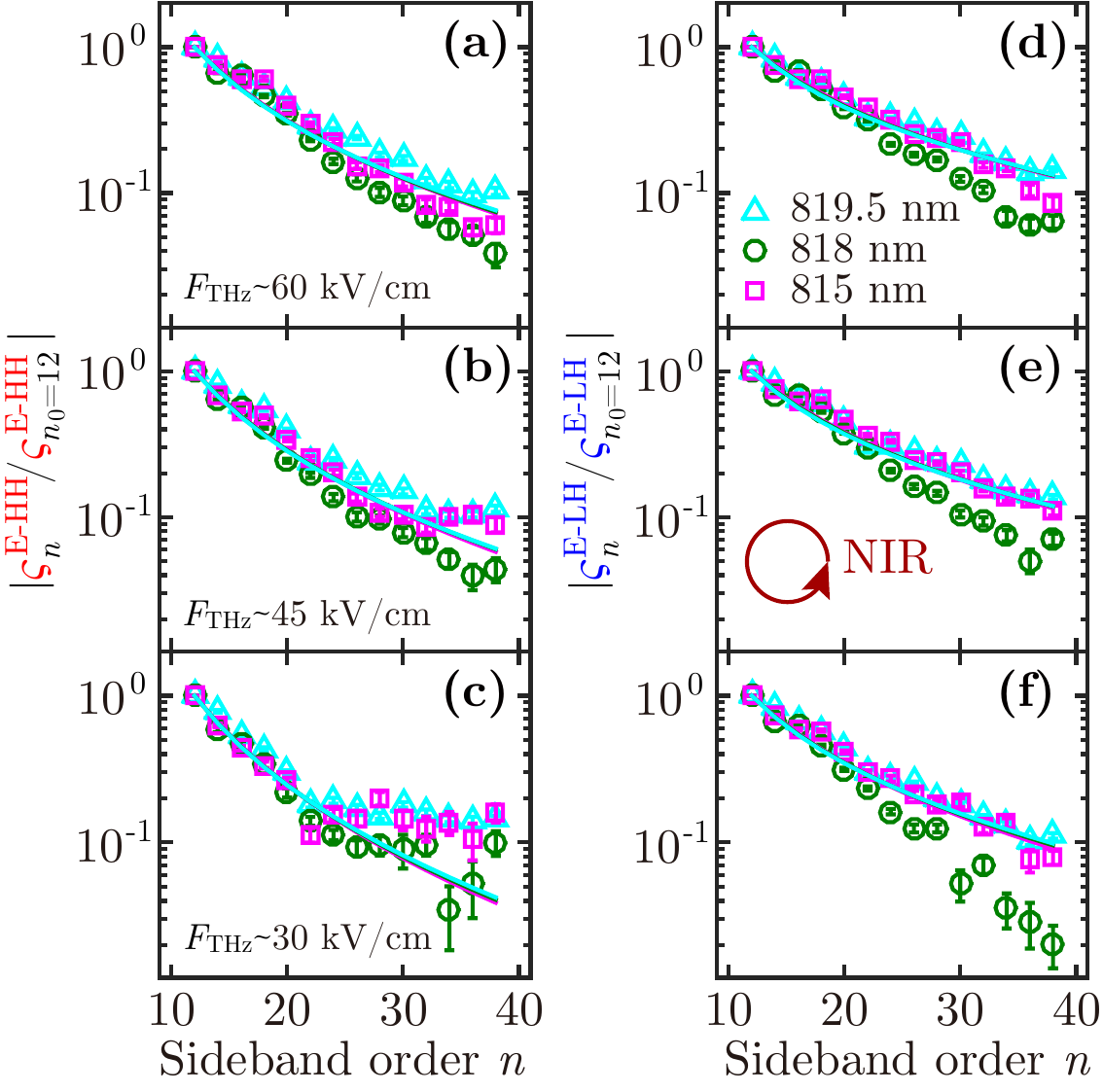}
	\caption{Absolute value of the propagator $\varsigma^{\nu}_n$ ($\nu={\rm E-HH,E-LH}$) relative to its value at the lowest detected sideband order $n_0=12$. The data were obtained by using a left-handed circular polarization (helicity -1) for the NIR laser. The first, second, and third rows show the data collected at three different THz-field strength levels: around 60\,kV/cm, 45\,kV/cm, and 30\,kV/cm, respectively (see Fig.~\ref{FIG:thz_field_strength} in Appendix~\ref{APP:thz_field_strength} for the exact THz-field strengths). In each panel, cyan triangles, dark green circles, and magenta squares represent the data obtained at three different NIR-laser wavelengths: 819.5\,nm, 818\,nm, and 815\,nm, respectively. For each set of laser parameters, two solid lines of the same color indicate the one-standard-deviation range resulting from uncertainties in the THz field strengths and the Hamiltonian parameters. The theoretical curves in each panel largely overlap. }
	\label{FIG:decay_propagator}
\end{figure}
For the absolute values of the propagators, $|{\varsigma}^{\rm E-HH}_n|$ and $|{\varsigma}^{\rm E-LH}_n|$, only their temperature dependences have been discussed previously~\cite{costello2023breaking}. To have a more systematic comparison between theory and experiment in the Hamiltonian reconstruction, we incorporate the information contained in the absolute values, $|{\varsigma}^{\rm E-HH}_n|$ and $|{\varsigma}^{\rm E-LH}_n|$. As mentioned in Sec.~\ref{SEC:eh_propagators} and discussed in Appendix~\ref{APP:propagation_sidebands}, the experimentally measured electron-hole propagator $\varsigma^{\nu}_n$ ($\nu={\rm E-HH,E-LH}$) differs from the theoretical electron-hole propagator ${\mathbb Q}^{\nu}_n$ by a proportionality factor ${\mathcal T}_n$, which accounts for the effects regarding propagation of the NIR-laser and sideband fields within the layered structure of the sample. The factor ${\mathcal T}_n$ depends on the thicknesses of the materials in the sample and on the refractive indices of the materials at the frequencies of the NIR laser and the $n$th-order sideband. These refractive indices may also be modified by the intense THz fields present during the HSG process, particularly at frequencies corresponding to optical transitions near the band edge of the bulk GaAs~\cite{liu2002adiabatic,tong2008excitonic}. An in-depth discussion of the THz-field-modulated dielectric functions of the materials in the sample, along with the determination of the factor ${\mathcal T}_n$, is beyond the scope of this paper. Figure~\ref{FIG:decay_propagator} shows the absolute value of the propagator $\varsigma^{\nu}_n$ with respect to its value at the lowest detected sideband order $n_0=12$ for nine of the 36 polarimetry experiments performed by using a left-handed circular polarization (helicity -1) for the NIR laser (see Fig.~\ref{FIG:propagator_decay_all} in Appendix~\ref{APP:sup_data} for data from all 36 polarimetry experiments and calculation of the error bars). Within the range of sideband orders considered in this paper, $12\le n\le 38$, which corresponds to approximately 48\,meV in the continuum states, we observe substantial variations in the absolute values, $|{\varsigma}^{\rm E-HH}_n|$ and $|{\varsigma}^{\rm E-LH}_n|$. Based on the theoretical analysis of the THz-field-modulated absorption spectra~\cite{liu2002adiabatic,tong2008excitonic} and the measured refractive index of GaAs~\cite{sell1974concentration}, we assume that the variation of ${\mathcal T}_n$ remains insignificant within the investigated sideband-order range, and equate the ratio $|\varsigma^{\nu}_n/\varsigma^{\nu}_{n_0}|$ to $|{\mathbb Q}^{\nu}_n/{\mathbb Q}^{\nu}_{n_0}|$ for fixed laser parameters. The ratio $|\varsigma^{\nu}_n/\varsigma^{\nu}_{n_0}|$ can also be qualitatively described by the analytic form in Eq.~\ref{EQ:propagator_analytic}, which gives
\begin{align}
|\frac{{\varsigma}^{\nu}_n}{{\varsigma}^{\nu}_{n_0}}|
=
\exp\{-\frac{\Gamma_{\nu}}{\hbar}(\tau^{\nu}_n-\tau^{\nu}_{n_0})\}.
\label{EQ:propagator_decay_analytic}
\end{align}
Since the acceleration time $\tau^{\nu}_n$ increases with increasing sideband order and decreasing THz-field strength, Eq.~\ref{EQ:propagator_decay_analytic} predicts that the absolute value of $\varsigma^{\nu}_n$ decays with increasing sideband order $n$, with a slower decay rate for stronger THz fields. This prediction is consistent with the data shown in Fig.~\ref{FIG:decay_propagator}. Note that, by taking ratios of propagators measured with identical NIR-laser parameters, propagation of experimental errors associated the NIR-laser intensity is avoided when the absolute value of the propagator $\varsigma^{\nu}_n$ is determined by using Eq.~(\ref{EQ:Stokes_to_Jonesmat_amp}).

In the next section, we will untilize the measured quantities ${{\varsigma}^{\rm E-HH}_n}/{{\varsigma}^{\rm E-LH}_n}$, $|{{\varsigma}^{\rm E-HH}_n}/{{\varsigma}^{\rm E-HH}_{n_0}}|$, and $|{{\varsigma}^{\rm E-LH}_n}/{{\varsigma}^{\rm E-LH}_{n_0}}|$, to reconstruct the effective electron-hole Hamiltonian for bulk GaAs based on a quantitative theory-experiment comparison by using Eq.~(\ref{EQ:propagator_algebraic_form}). Assuming that the reduced-mass parameter $\mu_{\rm ex}/m_0$ and Luttinger-parameter ratio $\gamma_3/\gamma_2$ are known from existing experiments, we will extract the dephasing constants $\Gamma_{\rm E-HH}$ and $\Gamma_{\rm E-LH}$, the NIR-laser detunings $\Delta^{\rm E-HH}_{\rm NIR}$ and $\Delta^{\rm E-LH}_{\rm NIR}$, and the combined Hamiltonian parameter $\gamma_2\mu_{\rm ex}/m_0$.

\section{Hamiltonian reconstruction}\label{SEC:hamiltonian_reconstruct}

The Hamiltonian reconstruction procedure begins with the definition of the following cost functions:
\begin{widetext}
\begin{align}
&R^{\rm E-HH}(\Gamma_{\rm E-HH},E_{\rm g,E-HH},\xi)
=\sqrt{
\frac{1}{N_{\rm data}'}
{\sum}'
\frac{
(|{
{\mathbb Q}^{\rm E-HH}_n
}
/{
{\mathbb Q}^{\rm E-HH}_{n_0}
}|_{\rm th}
-
|{
{\varsigma}^{\rm E-HH}_n
}
/{
{\varsigma}^{\rm E-HH}_{n_0}
}|_{\rm exp})^2
}
{\delta^2(|{{\varsigma}^{\rm E-HH}_n}/{{\varsigma}^{\rm E-HH}_{n_0}}|)}
},
\label{EQ:cost_f1}
\\
&R^{\rm E-LH}(\Gamma_{\rm E-LH},E_{\rm g,E-LH},\xi)
=\sqrt{
\frac{1}{N_{\rm data}'}
{\sum}'
\frac{
(|{
{\mathbb Q}^{\rm E-LH}_n
}
/{
{\mathbb Q}^{\rm E-LH}_{n_0}
}|_{\rm th}
-
|{
{\varsigma}^{\rm E-LH}_n
}
/{
{\varsigma}^{\rm E-LH}_{n_0}
}|_{\rm exp})^2
}
{\delta^2(|{{\varsigma}^{\rm E-LH}_n}/{{\varsigma}^{\rm E-LH}_{n_0}}|)}
},
\label{EQ:cost_f2}
\\
&R^{\rm phase}(\Gamma_{\rm E-HH},E_{\rm g,E-HH},\Gamma_{\rm E-LH},E_{\rm g,E-LH},\xi)\notag\\
=
&\sqrt{
\frac{1}{N_{\rm data}}
\sum
\{
\frac{
(\cos[{\rm Arg}(\frac{
{\mathbb Q}^{\rm E-HH}_n
}
{
{\mathbb Q}^{\rm E-LH}_n
})]_{\rm th}
-
\cos[{\rm Arg}(\frac{
{\varsigma}^{\rm E-HH}_n
}
{
{\varsigma}^{\rm E-LH}_n
})]_{\rm exp})^2
}
{\delta^2
(\cos[{\rm Arg}(\frac{
{\varsigma}^{\rm E-HH}_n}
{
{\varsigma}^{\rm E-LH}_n})]
)}
+
\frac{
(\sin[{\rm Arg}(\frac{
{\mathbb Q}^{\rm E-HH}_n
}
{
{\mathbb Q}^{\rm E-LH}_n
})]_{\rm th}
-
\sin[{\rm Arg}(\frac{
{\varsigma}^{\rm E-HH}_n
}
{
{\varsigma}^{\rm E-LH}_n
})]_{\rm exp})^2
}
{\delta^2(\sin[{\rm Arg}(\frac{
{\varsigma}^{\rm E-HH}_n}
{
{\varsigma}^{\rm E-LH}_n})]
)}
\}
},
\label{EQ:cost_f3}
\\
&R^{\rm abs}(\Gamma_{\rm E-HH},E_{\rm g,E-HH},\Gamma_{\rm E-LH},E_{\rm g,E-LH},\xi)
=\sqrt{
\frac{1}{N_{\rm data}}
\sum
\frac{
(|{
{\mathbb Q}^{\rm E-HH}_n
}
/{
{\mathbb Q}^{\rm E-LH}_n
}|_{\rm th}
-
|{
{\varsigma}^{\rm E-HH}_n
}
/{
{\varsigma}^{\rm E-LH}_n
}|_{\rm exp})^2
}
{\delta^2(|{{\varsigma}^{\rm E-HH}_n}/{{\varsigma}^{\rm E-LH}_n}|)}
},
\label{EQ:cost_f4}
\end{align}
\end{widetext}
which quantify the theory-experiment deviations for the quantities $|{{\varsigma}^{\rm E-HH}_n}/{{\varsigma}^{\rm E-HH}_{n_0}}|$, $|{{\varsigma}^{\rm E-LH}_n}/{{\varsigma}^{\rm E-LH}_{n_0}}|$, ${\rm Arg}({{\varsigma}^{\rm E-HH}_n}/{{\varsigma}^{\rm E-LH}_n})$, and $|{{\varsigma}^{\rm E-HH}_n}/{{\varsigma}^{\rm E-LH}_n}|$, respectively. Here, the quantities labeled with the subscript ``th" are calculated by using the theoretical model of the electron-hole propagator given in Eq.~(\ref{EQ:propagator_algebraic_form}). The quantities labeled with the subscript ``exp" are the experimentally measured values. The symbol $\delta$ denotes one standard deviation associated with each experimental data point. Each sum in Eqs.~(\ref{EQ:cost_f3}) and~(\ref{EQ:cost_f4}) contains
$N_{\rm data}=36\times14$ terms corresponding to the 36 polarimetry experiments and 14 sidebands detected in each experiment. The sums with prime symbols in Eqs.~(\ref{EQ:cost_f1}) and~(\ref{EQ:cost_f2}) exclude the terms with $n=n_0$, and each contains $N'_{\rm data}=36\times13$ terms. According to Eq.~(\ref{EQ:propagator_algebraic_form}), if the parameter $\mu_{ex}$ and the Luttinger-parameter ratio $\gamma_3/\gamma_2$ are known from existing experiments, these cost functions contain five parameters including the dephasing constants $\Gamma_{\rm E-HH}$ and $\Gamma_{\rm E-LH}$, the bandgaps $E_{\rm g, E-HH}$ and $E_{\rm g, E-LH}$, and the parameter $\xi=\gamma_2\mu_{\rm ex}/m_0$. The absolute value of the ratio ${{\mathbb Q}^{\nu}_n}/{{\mathbb Q}^{\nu}_{n_0}}$ ($\nu=\rm E-HH,E-LH$) is determined by the dephasing constant $\Gamma_{\nu}$, the bandgap $E_{{\rm g}, \nu}$, and the reduced mass $\mu^{\nu}_{xx}$, which contains the parameter $\xi$, whereas the ratio ${{\mathbb Q}^{\rm E-HH}_n}/{{\mathbb Q}^{\rm E-LH}_n}$ is determined by all five parameters. By imposing the constraint $E_{\rm g,E-LH}-E_{\rm g,E-HH}=\Delta_{\rm ex}$ based on the absorbance spectrum, the number of independent parameters is reduced to four. 

\begin{figure}
	\includegraphics[width=0.47\textwidth]{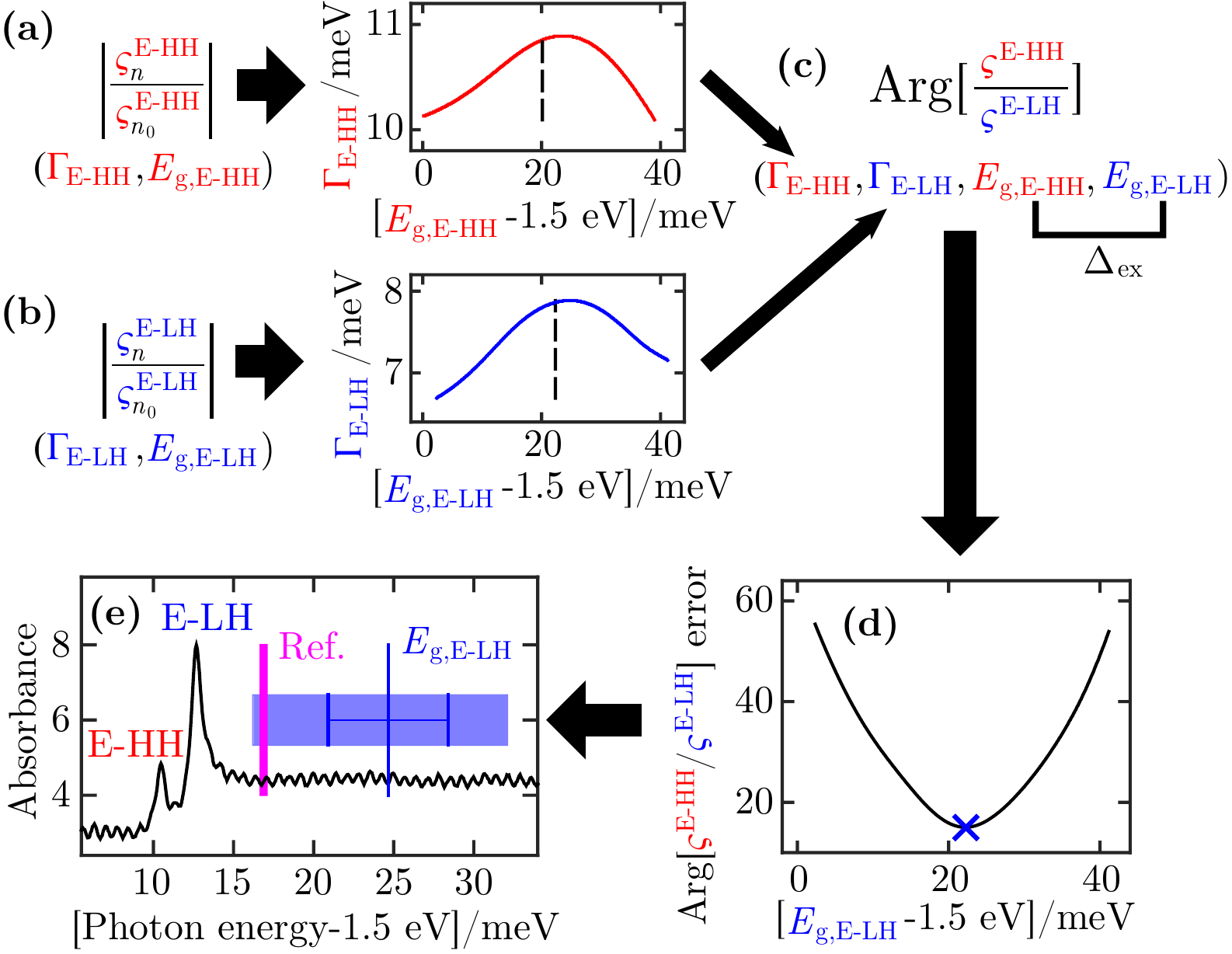}
	\caption{Extracting the dephasing constants and bandgaps. (a,b) First, the dephasing constant $\Gamma_{\nu}$ ($\nu={\rm E-HH, E-LH}$) is determined as a function of the bandgap $E_{{\rm g},\nu}$ by minimizing the cost function $R^{\nu}$ ([Eq.~(\ref{EQ:cost_f1}) and~(\ref{EQ:cost_f2})]), which quantifies the theory-experiment deviation for the absolute value of the propagator ratio $\varsigma^{\nu}_{n}/\varsigma^{\nu}_{n_0}$, which is a function of $\Gamma_{\nu}$ and $E_{{\rm g},\nu}$ with the assumption that the other Hamiltonian parameters are known from the literature. The quantity $|\varsigma^{\nu}_{n}/\varsigma^{\nu}_{n_0}|$ describes the propagator decay as a function of the sideband order $n$, with $n_0=12$ being the lowest sideband order detected in the experiment. (c) Second, by using the extracted relations between the dephasing constants and the bandgaps from (a) and (b), the relative phase between the E-HH and E-LH propagators, ${\rm Arg}[\varsigma^{\rm E-HH}/\varsigma^{\rm E-LH}]$, which is a function of the dephasing constants and bandgaps for both electron-hole species, becomes a function of a single variable---the bandgap $E_{\rm g,E-LH}$ when the constraint $E_{\rm g,E-LH}-E_{\rm g,E-HH}=\Delta_{\rm ex}$ is imposed based on the absorbance measurement [Fig.~\ref{FIG:exciton_splitting} (a)]. (d) Third, by minimizing the cost function $R^{\rm phase}$ [Eq.~(\ref{EQ:cost_f3})], which quantifies the theory-experiment deviation for ${\rm Arg}[\varsigma^{\rm E-HH}/\varsigma^{\rm E-LH}]$, an optimal value of $E_{\rm g,E-LH}$ is found (blue cross). The corresponding dephasing constants are indicated by dashed lines in (a) and (b). (e) Fourth, one standard deviation around the mean (blue error bar) and a 95\% confidence interval (blue shaded area) are obtained through a Monte Carlo simulation of the error propagation from the uncertainties in the reduced-mass parameters and in the THz-field strengths to the theoretically calculated electron-hole propagators. The extracted bandgap $E_{\rm g,E-LH}$ is shown on the absorbance spectrum together with a reference value $1.5169\pm0.0002$\,eV (magenta vertical bar) derived from the 1s-2s exciton energy difference measured at 2\,K in absorbance experiments~\cite{sell1972resolved}.}
	\label{FIG:extract_bandgap_dephasing}
\end{figure}
To extract the four parameters, we note that the absolute value of the ratio ${{\mathbb Q}^{\rm E-HH}_n}/{{\mathbb Q}^{\rm E-LH}_{n}}$ contains a factor $\sqrt{{ {\tilde\mu}^{\rm E-HH}_{yy}\mu^{\rm E-HH}_{zz} }}({ \mu^{\rm E-HH}_{xx}})^{3/8}/\sqrt{{ {\tilde\mu}^{\rm E-LH}_{yy}\mu^{\rm E-LH}_{zz} }}({ \mu^{\rm E-LH}_{xx}})^{3/8}$, which depends only on the parameter $\xi$ and the ratio $\gamma_3/\gamma_2$ [see Eqs.~(\ref{EQ:mu_xx}),~(\ref{EQ:mu_yy}),~(\ref{EQ:mu_zz}), and~(\ref{EQ:mu_xy}) ]. Since the measured values of $|{{\varsigma}^{\rm E-HH}_n}/{{\varsigma}^{\rm E-LH}_n}|$ are mostly on the order of unity [Figs.~\ref{FIG:abs_propagator_ratio} and~\ref{FIG:abs_ratio_all}], the cost function $R^{\rm abs}$ should be sensitive to the parameter $\xi$ once the other parameters are fixed. Therefore, we proceed by first assuming that the parameter $\xi$ is known, and attempt to extract the two dephasing constants and the two bandgaps from the remaining three cost functions. In this step, to set the parameters $\mu_{\rm ex}/m_0=(m_0/m_c+\gamma_1)^{-1}$, $\gamma_3/\gamma_2$, and $\xi=\gamma_2\mu_{\rm ex}/m_0$, we adopt the parameters from the literature with relatively small uncertainties: $m_c=(0.067\pm0.005)m_0$~\cite{ahmed1992far}, $\gamma_1=6.98\pm 0.45$, $\gamma_2=2.2\pm 0.1$, and $\gamma_3=2.9\pm 0.2$~\cite{skolnick1976investigation}. With these parameters fixed, the cost function $R^{\nu}$ ($\nu=\rm E-HH,E-LH$) constrains the relationship between the dephasing constant $\Gamma_{\nu}$ and the bandgap $E_{{\rm g}, \nu}$, which are the only free parameters in $R^{\nu}$. For each value of $E_{{\rm g}, \nu}$, minimization of $R^{\nu}$ yields an optimal value of the dephasing constant $\Gamma_{\nu}$. In this way, the dephasing constant $\Gamma_{\nu}$ can be effectively expressed as a function of $E_{{\rm g}, \nu}$. Figures~\ref{FIG:extract_bandgap_dephasing} (a) and (b) show the optimal values of the dephasing constants $\Gamma_{\rm E-HH}$ and $\Gamma_{\rm E-LH}$, which are calculated as functions of the bandgaps $E_{\rm g, E-HH}$ and $E_{\rm g, E-LH}$, respectively, by using the mean values of the THz-field strengths, the parameter $m_0$, and the three Luttinger parameters (see Fig.~\ref{FIG:dephasing_vs_gap} in Appendix~\ref{APP:sup_data_reconstruction} for the values of the cost function $R^{\nu}$). The optimal dephasing constants exhibit only weak dependence on the bandgaps, consistent with expectations from the analytic model discussed in Sec.~\ref{SEC:information}. These extracted values of the dephasing constants are also close to the mean dephasing constant $\bar\Gamma\approx 9$\,meV reported in Ref.~\cite{o2024bloch}. Under these constraints relating the bandgaps and dephasing constants, along with the constraint $E_{\rm g,E-LH}-E_{\rm g,E-HH}=\Delta_{\rm ex}$, the ratio ${{\mathbb Q}^{\rm E-HH}_n}/{{\mathbb Q}^{\rm E-LH}_n}$ and therefore the cost function $R^{\rm phase}$ becomes a function of a single variable $E_{\rm g,E-LH}$ [Fig.~\ref{FIG:extract_bandgap_dephasing} (c)]. As shown in Fig.~\ref{FIG:extract_bandgap_dephasing} (d), by using the mean values of the THz-field strengths, the parameter $m_0$, and the three Luttinger parameters, the calculated cost function $R^{\rm phase}$ exhibits a clear minimum at the optimal bandgap value $E_{\rm g,E-LH}=1.522$\,eV (blue cross), which, in turn, determines the dephasing constants through the bandgap-dephasing constraints [vertical dashed lines in Figs.~\ref{FIG:extract_bandgap_dephasing} (a) and (b)]. To estimate the confidence intervals for the extracted bandgap $E_{\rm g,E-LH}$, we perform a Monte Carlo simulation of the error propagation from the uncertainties in the parameters $\mu_{\rm ex}/m_0$, $\gamma_3/\gamma_2$, and $\xi$, as well as the uncertainties in the THz-field strengths, to the theoretically calculated electron-hole propagators (see Appendix~\ref{APP:sup_data_reconstruction} for the distributions of the parameters and more details about the Monte Carlo simulation). The resulting mean value, one standard deviation around the mean (error bar), and a 95\% confidence interval for the extracted bandgap $E_{\rm g,E-LH}$, [1.516,1.532]\,eV, are shown in Fig.~\ref{FIG:extract_bandgap_dephasing} (e). The 1s-exciton binding energy in bulk GaAs has been determined to be $4.2\pm0.2$\,meV from low-temperature absorbance measurements~\cite{sell1972resolved}. By using this binding-energy value, the bandgap $E_{\rm g,E-LH}$ is expected to be $1.5169\pm0.0002$\,eV (magenta vertical bar), which lies within the 95\% confidence interval of our extracted bandgap $E_{\rm g,E-LH}$.

\begin{figure}
	\includegraphics[width=0.47\textwidth]{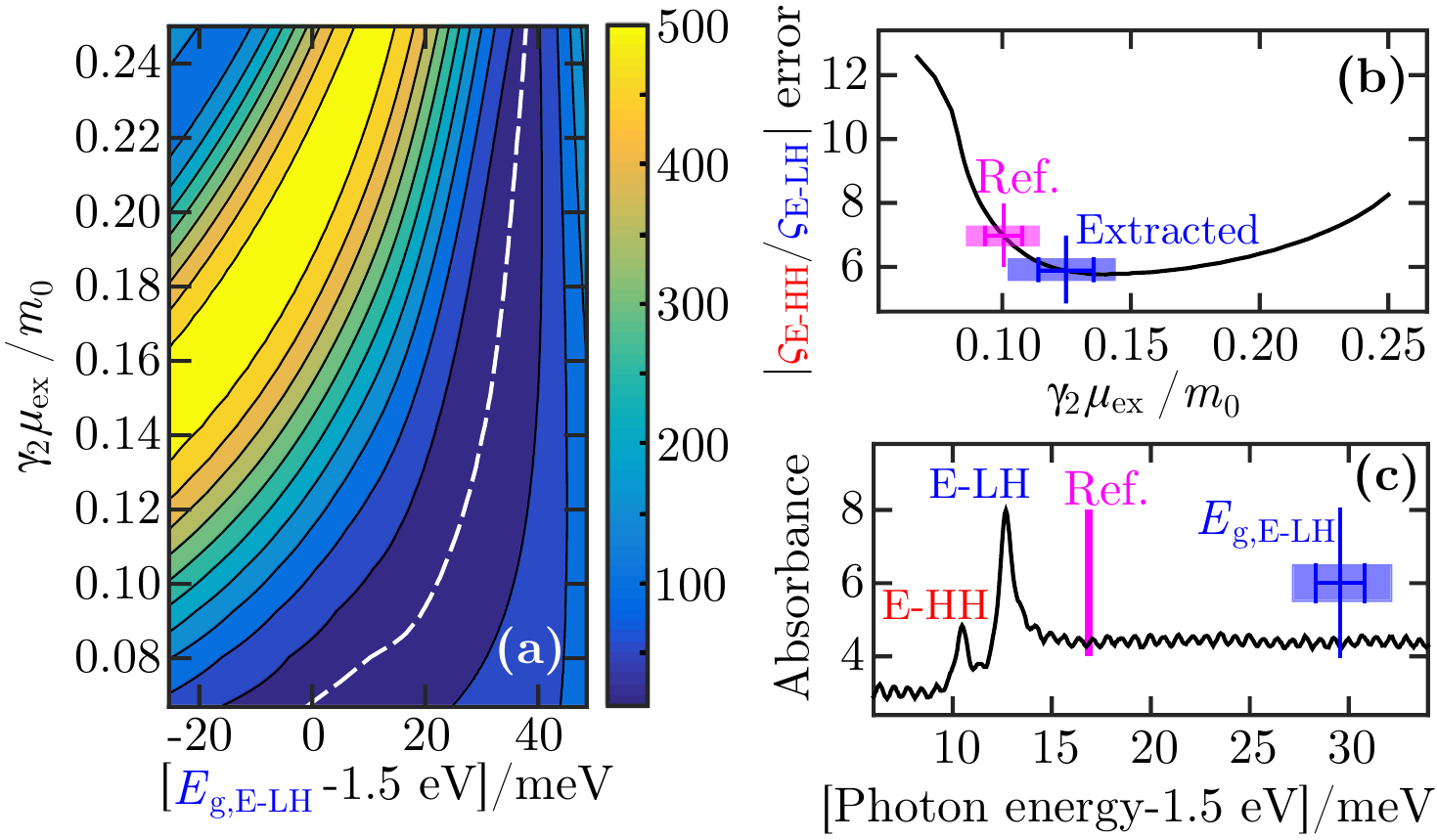}
	\caption{Extracting the dephasing constants, bandgaps, and the parameter $\xi=\gamma_2\mu_{\rm ex}/m_0$. (a) The values of the cost function $R^{\rm phase}$ [Eq.~(\ref{EQ:cost_f3})]. For each value of $\xi$, the optimal dephasing constant $\Gamma_{\nu}$ ($\nu=\rm E-HH,E-LH$) is expressed as a function of the bandgap $E_{{\rm g},\nu}$ by minimizing the cost function $R^{\nu}$ ([Eq.~(\ref{EQ:cost_f1}) and~(\ref{EQ:cost_f2})]). The white dashed line indicate the optimal bandgap $E_{\rm g,E-LH}$ as a function of $\xi$. (b) Extracted parameter $\xi$. With $\Gamma_{\nu}$ expressed as a function of $E_{{\rm g},\nu}$ and $E_{\rm g,E-LH}$ expressed as a function of $\xi$, the cost function $R^{\rm abs}$ [Eq.~(\ref{EQ:cost_f4})], which quantifies the theory-experiment deviation for $|\varsigma^{\rm E-HH}_{n}/\varsigma^{\rm E-LH}_{n}|$, becomes a function of a single variable $\xi$. The black curve shows $R^{\rm abs}$, as a function of $\xi$, calculated by using the parameters $\mu_{\rm ex}/m_0=(1/0.067+6.98)^{-1}$ and $\gamma_3/\gamma_2=2.9/2.2$. The mean optimal value of $\xi$ (blue vertical line), one standard deviation around the mean (blue error bar), and a 95\% confidence interval (blue shaded area) are obtained via a Monte Carlo simulation of the error propagation from the uncertainties in the parameters, $\mu_{\rm ex}/m_0$ and $\gamma_3/\gamma_2$, and in the THz-field strengths to the theoretically calculated electron-hole propagators. The magenta vertical line, error bar, and shaded area represent the mean value, one standard deviation around the mean, and a 95\% confidence interval estimated from the literature values of the Hamiltonian parameters. (c) Extracted bandgap $E_{\rm g,E-LH}$. The extracted $E_{\rm g,E-LH}$ is shown on the absorbance spectrum together with a reference value $1.5169\pm0.0002$\,eV (magenta vertical bar) derived from the 1s-2s exciton energy difference measured at 2\,K in absorbance experiments~\cite{sell1972resolved}. The blue vertical line, error bar, and shaded area represent the mean value of $E_{\rm g,E-LH}$, one standard deviation around the mean, and a 95\% confidence interval obtained from the same Monte Carlo simulation used for extracting the value of $\xi$ in (b).}
	\label{FIG:extract_reduced_mass_ratio}
\end{figure}
For a fixed value of the parameter $\xi$, we follow the procedure described above to express the optimal dephasing constant $\Gamma_{\nu}$ as a function of the bandgap $E_{{\rm g},\nu}$ ($\nu=\rm E-HH,E-LH$), and to determine the optimal bandgap $E_{\rm g,E-LH}$, which then becomes a function of $\xi$. As shown in Fig.~\ref{FIG:extract_reduced_mass_ratio} (a), for each value of $\xi$, by using the mean values of the THz-field strengths together with the parameters $\mu_{\rm ex}/m_0=(1/0.067+6.98)^{-1}$ and $\gamma_3/\gamma_2=2.9/2.2$, the cost function $R^{\rm phase}$ exhibits a minimum associated with an optimal bandgap $E_{\rm g,E-LH}$. With $\Gamma_{\nu}$ expressed as a function of $E_{{\rm g},\nu}$ and $E_{\rm g,E-LH}$ expressed as a function of $\xi$, the cost function $R^{\rm abs}$ becomes a function of a single variable, $\xi$ [black curve in Fig.~\ref{FIG:extract_reduced_mass_ratio} (b)], whose minimum corresponds to the optimal value of $\xi$. To estimate the confidence intervals for the extracted parameters, we perform a Monte Carlo simulation of the error propagation from the uncertainties in the parameters $\mu_{\rm ex}/m_0$ and $\gamma_3/\gamma_2$, as well as the uncertainties in the THz-field strengths to the theoretically calculated electron-hole propagators (see Appendix~\ref{APP:sup_data_reconstruction} for the distributions of the parameters and more details about the Monte Carlo simulation). As shown in Fig.~\ref{FIG:extract_reduced_mass_ratio} (b), the resulting value of $\xi$ is $0.125\pm0.011$, with a 95\% confidence interval of [0.102,0.144]. For comparison, with the parameters $m_c$, $\gamma_1$, and $\gamma_2$ randomly drawn 10,000 times from normal distributions whose means and standard deviations correspond to the reported values---$m_c=(0.067\pm0.005)m_0$~\cite{ahmed1992far}, $\gamma_1=6.98\pm 0.45$, $\gamma_2=2.2\pm 0.1$~\cite{skolnick1976investigation}---we obtain a reference value $\xi=0.101\pm0.007$ with a 95\% confidence interval [0.086,0.115] [see Fig.~\ref{FIG:monte_carlo_one} (a), (b), and (c) in Appendix~\ref{APP:sup_data_reconstruction} for the distributions of the parameters], which overlaps with the 95\% confidence interval for the extracted value of $\xi$. The 95\% confidence interval associated with the extracted bandgap value $E_{\rm g,E-LH}=1.530\pm0.001$\,eV, [1.527,1.532]\,eV, now lies about 10\,meV above the reference value $1.5169\pm0.0002$\,eV derived from low-temperature absorbance measurements~\cite{sell1972resolved} [Figures.~\ref{FIG:extract_reduced_mass_ratio} (c)]. The extracted dephasing constants are $\Gamma_{\rm E-HH}=10.4\pm0.2$\,meV and $\Gamma_{\rm E-LH}=7.6\pm0.2$\,meV, with narrow 95\% confidence intervals of [10.0,10.8]\,meV and  [7.2,7.9]\,meV, respectively. In previous work on distinguishing the dephasing rates of the two electron-hole species in the same GaAs sample~\cite{costello2023breaking}, only the temperature dependences of the dephasing constants associated with finite phonon occupations were extracted. According to this earlier work, the temperature-dependent components of the dephasing constants $\Gamma_{\rm E-HH}$ and $\Gamma_{\rm E-LH}$ are expected to be much less than 1\,meV at 30\,K. Therefore, the values of the dephasing constants extracted in the present study suggest that any dominant dephasing mechanism of the electron-hole coherence in our HSG experiments, if related to electron-phonon interactions, should not involve thermal phonons.

To have a direct theory-experiment comparison, we use 10,000 sets of the extracted dephasing constants, bandgaps, and the parameter $\xi$ from the Monte Carlo simulation, along with the corresponding randomly sampled values of the THz-field strengths and the parameters $\mu_{\rm ex}/m_0$ and $\gamma_3/\gamma_2$, to calculate the theoretical values of $|{{\varsigma}^{\rm E-HH}_n}/{{\varsigma}^{\rm E-HH}_{n_0}}|$, $|{{\varsigma}^{\rm E-LH}_n}/{{\varsigma}^{\rm E-LH}_{n_0}}|$, and ${{\varsigma}^{\rm E-HH}_n}/{{\varsigma}^{\rm E-LH}_n}$. In each panel of Figs.~\ref{FIG:abs_propagator_ratio},~\ref{FIG:arg_propagator},~\ref{FIG:decay_propagator},~\ref{FIG:abs_ratio_all},~\ref{FIG:arg_ratio_all}, and~\ref{FIG:propagator_decay_all}, two solid lines of the same color are used to indicate the range corresponding to one standard deviation in the theoretical results. The close agreement between theory and experiment validates our Hamiltonian reconstruction procedure.

The deviations of the experimental data from the theoretical curves are more prominent for sideband orders above 28, particularly at the lowest level of THz-field strength. These deviations do not arise from breakdown of the approximations that are used to derive the analytic expression of the electron-hole propagator [Eq.~(\ref{EQ:propagator_algebraic_form})] from its form of Feynman path integrals [Eq.~(\ref{EQ:propagator_isotropic})], as can be verified directly through numerical integration. Instead, they are most likely attributable to the reduced signal-to-noise ratio associated with the smaller intensities of the higher-order sidebands measured at weaker THz fields. Compared to the experimental data obtained at the two shorter NIR-laser wavelengths, the data collected at the longest NIR-laser wavelength (819.5\,nm) exhibit the smallest deviation from the theoretical curves because of the stronger HSG signals resulting from longer integration times. The theory-experiment agreement may be improved by extending the effective electron-hole Hamiltonian to include additional terms, such as those describing electron-hole and electron-phonon interactions. Since the number of real quantities measured in our experiment ($|{{\varsigma}^{\rm E-HH}_n}/{{\varsigma}^{\rm E-HH}_{n_0}}|$, $|{{\varsigma}^{\rm E-LH}_n}/{{\varsigma}^{\rm E-LH}_{n_0}}|$, ${\rm Arg}({{\varsigma}^{\rm E-HH}_n}/{{\varsigma}^{\rm E-LH}_n})$, and $|{{\varsigma}^{\rm E-HH}_n}/{{\varsigma}^{\rm E-LH}_n}|$ for more than ten sidebands) is much larger than the number of the extracted parameters, the same data set can, in principle, be used to determine additional parameters for a more refined effective Hamiltonian and a more complete dephasing model.

In the Hamiltonian reconstruction demonstrated above, to reduce the uncertainties in the extracted Hamiltonian parameters, the only two input parameters, $\mu_{\rm ex}/m_0$ and $\gamma_3/\gamma_2$, have been fixed based on previous cyclotron resonance experiments. By identifying the bandgap difference $E_{\rm g, E-LH}-E_{\rm g, E-HH}$ as the exciton-peak splitting $\Delta_{\rm ex}$ measured in our absorbance experiment, the four parameters $E_{\rm g, E-LH}$, $\Gamma_{\rm E-HH}$, $\Gamma_{\rm E-LH}$, and $\xi$ are determined unambiguously from our polarimetry measurements in the sense that the cost functions $R^{\nu}$, $R^{\rm abs}$, and $R^{\rm phase}$ exhibit well-defined minima. With the static dielectric constant $\varepsilon(0)$ determined from low-frequency capacitance measurements~\cite{samara1983temperature}, the Hamiltonian can also be reconstructed based solely on absorbance and HSG experiments. The parameter $\mu_{\rm ex}/m_0$ can be calculated by using the 1s-exciton binding energy $E_{\rm 1s}$ determined from low-temperature absorbance measurements~\cite{sell1972resolved} as $\mu_{\rm ex}/m_0=E_{\rm 1s}\varepsilon^2(0)/R_{\infty}$~\cite{willatzen2009kp}, where $R_{\infty}$ is the Rydberg constant, while the ratio $\gamma_3/\gamma_2$ can be extracted from HSG experiments~\cite{costello2021reconstruction}. By using $E_{\rm 1s}=4.2\pm0.2$\,meV~\cite{sell1972resolved} and $\gamma_3/\gamma_2=1.47\pm0.48$, a similar Monte Carlo simulation results in similar values for the dephasing constants $\Gamma_{\rm E-HH}$ and $\Gamma_{\rm E-LH}$, and slightly smaller values for the bandgap $E_{\rm g,E-LH}$ and the parameter $\xi$. The 95\% confidence intervals associated with $\Gamma_{\rm E-HH}$ and $\Gamma_{\rm E-LH}$ remain narrow, while the 95\% confidence intervals associated with $E_{\rm g,E-LH}$ and $\xi$ become wider because of the larger uncertainty in the ratio $\gamma_3/\gamma_2$ (see Appendix~\ref{APP:sup_data_reconstruction} for more details about the Monte Carlo simulation). The ratio $\gamma_3/\gamma_2$ can also be determined from the polarimetry data presented in this paper by using Eq.~(\ref{EQ:Tmat_offdiagonal})~\cite{costello2021reconstruction}. However, because the polarization of the THz field in our polarimetry experiments was nearly perpendicular to the [110] axis such that $\sin(2\theta)\approx 0$ in Eq.~(\ref{EQ:Tmat_offdiagonal}), the ratio $\gamma_3/\gamma_2$ obtained in this way is subject to large uncertainty. In future HSG experiments in GaAs or similar materials, the uncertainty in the extracted ratio $\gamma_3/\gamma_2$ may be reduced by suppressing the substrate-induced strain and by driving electron-hole pairs along directions away from high-symmetry axes.

\section{Effects of quantum fluctuations on HSG spectra}\label{SEC:discussion_approx}

In the Hamiltonian reconstruction, quantum fluctuations have been incorporated in the theoretical model to achieve quantitative agreement between theory and experiment. We have also shown in Sec.~\ref{SEC:information} that the dependence of the measured quantities ${{\varsigma}^{\rm E-HH}_n}/{{\varsigma}^{\rm E-LH}_n}$ and $|{{\varsigma}^{\nu}_n}/{{\varsigma}^{\nu}_{n_0=12}}|$ ($\nu=\rm E-HH,E-LH$) on the THz-field strength and sideband order can be qualitatively described by the analytic model given in Eq.~(\ref{EQ:propagator_analytic}), which does not include the effects of quantum fluctuations. To have a quantitative understanding of the effects of quantum fluctuations on the HSG spectra, we use the extracted material parameters to analyze the contributions of each term in our propagator model [Eq.~(\ref{EQ:propagator_algebraic_form})] to the quantities ${{\varsigma}^{\rm E-HH}_n}/{{\varsigma}^{\rm E-LH}_n}$ and $|{{\varsigma}^{\nu}_n}/{{\varsigma}^{\nu}_{n_0=12}}|$ (see Appendix~\ref{APP:app_propagator} for details). From this analysis, we find that, the propagator model employed in the Hamiltonian reconstruction can be approximately replaced by
\begin{align}
{\mathbb Q}^{\nu}_n
\propto
&
\frac
{\sqrt{{ {\tilde\mu}^{\nu}_{yy}\mu^{\nu}_{zz} }}({ \mu^{\nu}_{xx}})^{3/8}}
{\sqrt{|q^{\nu}_{0}(n,i{\tilde\Gamma}_{\nu}+{\tilde\Delta}^{\nu}_{\rm NIR})|}}
\notag\\
&
\times
\exp[i(\frac{8}{15}n+i{\tilde\Gamma}^{\prime}_{\nu,n}+{\tilde\Delta}^{\nu,\prime}_{\rm NIR,n})(\frac{18n}{{\tilde U}^{\nu}_{\rm p}})^{1/4}],
\label{EQ:propagator_rewritten}
\end{align}
where the function $q^{\nu}_{0}$ is defined by Eq.~(\ref{EQ:definition_q0}), and the $n$-dependent auxiliary variables ${\tilde\Gamma}^{\prime}_{\nu,n}$ and ${\tilde\Delta}^{\nu,{\prime}}_{\rm NIR,n}$ are defined by the dephasing constant ${\tilde\Gamma}_{\nu}$ and the NIR-laser detuning ${\tilde\Delta}^{\nu}_{\rm NIR}$ in units of the THz photon energy as:
\begin{align}
{\tilde\Gamma}^{\prime}_{\nu,n}
&\equiv
{\tilde\Gamma}_{\nu}
-
\frac{1}{3}\frac{{\rm Im}[(i{\tilde\Gamma}_{\nu}+{\tilde\Delta}^{\nu}_{\rm NIR})^{3/2}]}{\sqrt{n}},
\label{EQ:auxiliary_dephasing}
\\
{\tilde\Delta}^{\nu,\prime}_{\rm NIR,n}
&\equiv
{\tilde\Delta}^{\nu}_{\rm NIR}
-
\frac{1}{3}\frac{{\rm Re}[(i{\tilde\Gamma}_{\nu}+{\tilde\Delta}^{\nu}_{\rm NIR})^{3/2}]}{\sqrt{n}}.
\label{EQ:auxiliary_detuning}
\end{align}
Compared to the analytic model given in Eq.~(\ref{EQ:propagator_analytic}), Eq.~(\ref{EQ:propagator_rewritten}) can be viewed as arising from the physical picture of classical electron-hole recollisions, supplemented with the $n$-dependent variables ${\tilde\Gamma}^{\prime}_{\nu,n}$ and ${\tilde\Delta}^{\nu,{\prime}}_{\rm NIR,n}$ to phenomenologically describe the dephasing and NIR-laser detuning, and with an additional factor proportional to $\sqrt{{ {\tilde\mu}^{\nu}_{yy}\mu^{\nu}_{zz} }}({ \mu^{\nu}_{xx}})^{3/8}/\sqrt{|q^{\nu}_{0}|}$ to account for the quantum fluctuations.

The factor $\sqrt{{ {\tilde\mu}^{\nu}_{yy}\mu^{\nu}_{zz} }}({ \mu^{\nu}_{xx}})^{3/8}/\sqrt{|q^{\nu}_{0}|}$ results from evaluating the propagator ${\mathbb Q}^{\nu}_n$ in the form of Feynman path integrals [Eq.~(\ref{EQ:propagator_isotropic})] by summing over the $k$-space trajectories in the vicinity of a shortest electron-hole recollision 
associated with the saddle point $({\bf P}^{\nu}_n, t^{\prime\nu}_n, t^{\nu}_n)$ of the action $S^{\nu}_n({\bf P},t',t)$, which governs the acceleration of an electron-hole pair with reduced mass $\mu^{\nu}_{xx}$ in the THz field [Eq.~(\ref{EQ:propagator_action})], as discussed in Sec.~\ref{SEC:theory}. Each $k$-space trajectory is given by ${\bf k}(t'')={\bf P}+(e/\hbar){\bf A}_{\rm THz}(t'')$ with $t'\le t''\le t$, characterized by the canonical momentum $\hbar\bf P$, the electron-hole creation time $t'$, and the electron-hole annihilation time $t$. The summation over $\bf P$ yields the electron-hole wavefunction at the origin, i.e., the spatial overlap between the electron and hole wavefunctions, which determines the oscillator strength for sideband emission. For an electron-hole pair created at the origin at time $t'$, the initial electron-hole wavefunction contains all quasi-momentum components with equal weight. As the THz field drives the electron-hole pair from $t'$ to $t$, each of the quasi-momentum components accumulates a dynamic phase given by the time integral of the corresponding quadratic kinetic energy. This results in broadening of the electron-hole wavefunction and thus reduction of the oscillator strength by a factor proportional to $(|t-t'|/\mu^{\nu}_{xx})^{3/2}$, which is larger for longer acceleration time $|t-t'|$. When $\bf P$, $t'$ and $t$ are all allowed to vary, Gaussian quantum fluctuations around the saddle point $({\bf P}^{\nu}_n, t^{\prime\nu}_n, t^{\nu}_n)$ lead to reduction of the oscillator strength by a factor proportional to the square root of the Hessian determinant ${\rm det}(\partial^2S^{\nu}_n)$ of the action $S^{\nu}_n$~\cite{wu2023explicit}. Under the linear-in-time approximation for the THz field, ${\rm det}(\partial^2S^{\nu}_n)$ is proportional to $\sqrt{|q^{\nu}_{0}|}/({ \mu^{\nu}_{xx}})^{11/8}$, and scales with the complex wavevectors ${\bf k}^{\nu}_n(t^{\prime\nu}_n)$ and ${\bf k}^{\nu}_n(t^{\nu}_n)$, which satisfy the generalized conditions of energy conservation at the complex electron-hole creation time $t^{\prime\nu}_n$ and recollision time $t^{\nu}_n$ [Eqs.~(\ref{EQ:saddle_point2}) and~(\ref{EQ:saddle_point3})], and the acceleration time $|t^{\nu}_n-t^{\prime\nu}_n|$, as $|{\bf k}^{\nu}_n(t^{\prime\nu}_n)||{\bf k}^{\nu}_n(t^{\nu}_n)||t^{\nu}_n-t^{\prime\nu}_n|^{5/2}$. The additional factor ${\sqrt{ {\tilde\mu}^{\nu}_{yy}\mu^{\nu}_{zz} } }/{\mu^{\nu}_{xx}}$ arises from the reduced-mass anisotropy.

According to the discussion above on the scaling of the function $q^{\nu}_{0}$, in addition to dephasing, quantum fluctuations introduce a second mechanism of propagator decay through $q^{\nu}_{0}$. When the sideband energy $\hbar\omega_{\rm NIR}+n\hbar\omega_{\rm THz}$ lies above the bandgap $E^{\nu}_g$, the magnitude of the complex wavevector ${\bf k}^{\nu}_n(t^{\nu}_n)$ at electron-hole recollision increases with increasing sideband order. In this case, the magnitude of $q^{\nu}_{0}$ is larger for a higher sideband order because of the longer acceleration time $|t^{\nu}_n-t^{\prime\nu}_n|$, corresponding to a stronger reduction of the oscillator strength. Using the extracted dephasing constants and bandgaps as inputs, we find that, for both electron-hole species, the factor $1/\sqrt{|q^{\nu}_{0}|}$ decreases by roughly a factor of four as the sideband order increases from 12 to 38. For most experimental parameters, this decay factor is larger than that induced by dephasing. As an exception, the E-HH propagator decay induced by dephasing is slightly more significant for the lowest level of the THz-field strength and the shortest NIR-laser wavelength used in our experiments (see Figs.~\ref{FIG:expfac_approx} and~\ref{FIG:q0fac_approx} in Appendix~\ref{APP:app_propagator} for the numerical details).

According to Eq.~(\ref{EQ:propagator_rewritten}), quantum fluctuations also determine the absolute value of ${\mathbb Q}^{\rm E-HH}_n/{\mathbb Q}^{\rm E-LH}_n$ through the factor $\sqrt{{ {\tilde\mu}^{\nu}_{yy}\mu^{\nu}_{zz} }}({ \mu^{\nu}_{xx}})^{3/8}$. By using the mean value of the extracted parameter $\xi=0.125$ and the ratio $\gamma_3/\gamma_2=2.9/2.2$, this factor contributes a multiplicative factor of approximately 2.2 to $|{\mathbb Q}^{\rm E-HH}_n/{\mathbb Q}^{\rm E-LH}_n|$.
 
The need to include quantum fluctuations in the quantitative theoretical modeling can be directly inferred from the experimental data on the electron-hole propagators. Indeed, without including the effects of quantum fluctuations, the phase of the propagator ratio ${{\varsigma}^{\rm E-HH}_n}/{{\varsigma}^{\rm E-LH}_n}$, as well as the propagator decay described by the quantity $|{{\varsigma}^{\nu}_n}/{{\varsigma}^{\nu}_{n_0=12}}|$ can still be reasonably quantified by the analytic model given in Eq.~(\ref{EQ:propagator_analytic}). However, an inconsistency arises when the absolute value of the propagator ratio ${{\varsigma}^{\rm E-HH}_n}/{{\varsigma}^{\rm E-LH}_n}$ is analyzed in parallel with the propagator decay. This model assumes a simple exponential form $|{{\varsigma}^{\nu}_n}|\propto\exp(-\Gamma_{\nu}\tau^{\nu}_n/\hbar)$ with the acceleration time $\tau^{\nu}_n\propto n^{1/4}$, implying a relation between the quantities $|{{\varsigma}^{\rm E-HH}_n}/{{\varsigma}^{\rm E-LH}_n}|$ and $|{{\varsigma}^{\nu}_n}/{{\varsigma}^{\nu}_{n_0=12}}|$:
\begin{align}
\frac{|{{\varsigma}^{\rm E-HH}_n}/{{\varsigma}^{\rm E-HH}_{n_0}}|}{|{{\varsigma}^{\rm E-LH}_n}/{{\varsigma}^{\rm E-LH}_{n_0}}|}
=
|\frac{{\varsigma}^{\rm E-HH}_n}{{\varsigma}^{\rm E-LH}_n}|^{1-(n_0/n)^{1/4}},
\label{EQ:r_cl}
\end{align}
which is inconsistent with the experimental observation (see Fig.~\ref{FIG:r_cl} in Appendix~\ref{APP:app_propagator} for inconsistency between Eq.~(\ref{EQ:r_cl}) and the experimental data). Therefore, the analytic model given in Eq.~(\ref{EQ:propagator_analytic}), which is based on classical electron-hole recollisions, is insufficient for quantitatively describing the polarimetry data. Moreover, since most qualitative features of the electron-hole propagators are captured by this model, as discussed in Sec.~\ref{SEC:information}, it is natural to incorporate quantum fluctuations around the classical recollision pathways to achieve quantitative agreement between theory and experiment.

\section{extension of the effective Hamiltonian}\label{SEC:discussion}
The reconstructed three-band electron-hole Hamiltonian is not perfect in explaining the data from our HSG experiments, particularly at the lowest THz-field strengths. The main discrepancy between the theory and experiment is that the extracted bandgap is approximately 10\,meV larger than the value inferred from previous absorbance measurements~\cite{sell1972resolved}. 

An additional and seemingly unrelated surprise is that we have never seen evidence in HSG spectra of a threshold near the 36.6-meV longitudinal optical (LO) phonon energy~\cite{irmer1996temperature} for LO-phonon emission, which is ubiquitous in experiments on GaAs. As we will see in this section, both the bandgap blue shift and the absence of the expected threshold can be attributed to modulation of the Fr\"ohlich interaction by strong THz fields.

This larger bandgap should not arise from artifacts such as sample heating in the presence of the laser fields, because the bandgap of bulk GaAs shrinks with increasing temperature~\cite{uesugi2000temperature}. Note also that the bandgap we extract in this paper is the free electron-hole bandgap but not the gap for optical absorption, which can be modulated by strong THz fields in the dynamic Franz-Keldysh effects~\cite{liu2002adiabatic,tong2008excitonic}.

Since the effective-Hamiltonian parameters have been optimized based on the HSG experiments, it is necessary to extend the effective Hamiltonian in order to achieve better theory-experiment agreement. In the following, we discuss two possible extensions of the effective Hamiltonian by introducing two interaction effects, electron-hole Coulomb attraction and Fr\"ohlich interaction.

\subsection{Effects of Coulomb interaction}\label{SEC:discussion_coulomb}

One possibility to obtain a smaller bandgap from the Hamiltonian reconstruction based on HSG is to add electron-hole Coulomb attraction. In the presence of strong Coulomb interaction, it has been shown that the electron-hole acceleration times in the recollision processes in HSG can be shorter than those in the absence of electron-hole attraction~\cite{freudenstein2022attosecond}. Consider the analytic model of the electron-hole propagator given in Eq.~(\ref{EQ:propagator_analytic}). If the acceleration time $\tau^{\nu}_n$ is effectively reduced, then the extracted bandgap $E_{\rm g, E-LH}$ will be smaller in order to maintain the same value of the propagator ratio ${\varsigma}^{\rm E-HH}_n/{\varsigma}^{\rm E-LH}_n$, according to Eq.~(\ref{EQ:propagator_ratio_analytic}). Incorporating Coulomb interaction into the Hamiltonian reconstruction will be a topic of future work. 

\subsection{Energy renormalization and absence of thresholds for LO-phonon emission under THz-field-modulated Fr\"ohlich interaction}\label{SEC:discussion_frolich}

Another possibility is that the bandgap of GaAs in HSG may be enlarged through Fr\"ohlich interaction~\cite{frohlich1950xx} that is modulated by the strong THz field. It has long been known that an electron moving in a polar crystal can be dressed with optical phonons to form a polaron, a quasiparticle exhibiting an energy dispersion different from that of a bare electron~\cite{lee1952motion}. In the presence of a sufficiently strong electric field, acceleration of electrons in a scattering event can become non-negligible~\cite{haug2008quantum}, and therefore the polaronic effects may be modified. In a preliminary study of electron-phonon scattering in an intense THz field, by considering the quantum kinetics of electron-hole coherences and correlations between electron-hole pairs and phonons, it was shown in a quasi-one-dimensional two-band model that the dephasing of electron-hole pairs due to Fr\"ohlich interaction between electrons and LO phonons can be strongly modulated by a THz field with a strength on the order of 10 kV/cm~\cite{renbao2002electron}. In this study, it was also shown that modulation of electron-phonon interactions by an intense THz field can lead to renormalization of the electron-hole energy.

To investigate how the electron-hole energy of GaAs in HSG can be renormalized through electron-phonon interactions under the strong THz field, we consider a generic electron-phonon system interacting with classical electromagnetic fields. Following Ref.~\cite{renbao2002electron}, we derive a closed set of quantum-kinetic equations truncated at the lowest-order correlations between electron-hole pairs and phonons (see Appendix~\ref{APP:frolich_thz} for the equations). Based on this general quantum-kinetic analysis, we examine HSG from a generic electron-phonon system with a single conduction band and a single valence band, which are associated with negligible Berry curvature and a constant dipole moment. Under the Markovian and second-order Born approximations, we find that, at low temperatures, where phonon occupations are negligible, phonon emission under a linearly polarized THz field ${\bf E}_{\rm THz}(t)=\hat{x}F_{\rm THz}\cos(\omega_{\rm THz}t)$ can lead to modification of the dynamics of the interband polarization through the following quantity:
\begin{align}
&
{\mathcal Q}_{\bf P}
\approx
\sum_{n,{\bf q},j}
i
[
\frac{
|J_n(
\frac{eF_{\rm THz}q_x}{m_{c}\omega_{\rm THz}^2}
)|^2
|G^{\prime{\bf q},j}_{cc}|^2}
{E_{c,{\bf P}}
-
E_{c,{\bf P}-{\bf q}}
-\hbar\Omega_{{\bf q},j}
+n\hbar\omega_{\rm THz}
+i\Gamma_{\rm e-ph}}
\notag\\
&
+
\frac{
|J_n(
\frac{eF_{\rm THz}q_x}{m_{v}\omega_{\rm THz}^2}
)|^2
|G^{\prime{\bf q},j}_{vv}|^2}
{E_{v,{\bf P}-{\bf q}}
-
E_{v,{\bf P}}
-\hbar\Omega_{{\bf q},j}
+n\hbar\omega_{\rm THz}+i\Gamma_{\rm e-ph}}
],
\label{EQ:constant_QlowT}
\end{align}
whose real and imaginary parts introduce dephasing and electron-hole energy renormalization, respectively. Here, ``c" and ``v" label the conduction and valence bands associated with effective masses $m_c$ and $m_v$, which define the band energies $E_{c,{\bf P}}=\hbar^2P^2/(2m_c)$ and $E_{v,{\bf P}}=-\hbar^2P^2/(2m_v)$, respectively. The coupling constant $G^{\prime{\bf q},j}_{cc}$ ($G^{\prime{\bf q},j}_{vv}$) is associated with the electron-phonon scattering processes, where an electron (hole) with quasi-momentum $\hbar{\bf P}$ ($-\hbar{\bf P}$) in the conduction (valence) band is scattered to the state with quasi-momentum $\hbar({\bf P-q})$ ($-\hbar({\bf P-q})$) in same band, accompanied by emission of a phonon with quasi-momentum $\hbar{\bf q}$ ($-\hbar{\bf q}$) in the $j$-th phonon branch with phonon dispersion $\hbar\Omega_{{\bf q},j}$ (see Appendix~\ref{APP:frolich_thz} for the derivation). The constant $\Gamma_{\rm e-ph}$ describes dephasing of the so-called phonon-assisted density matrix. In the limit of zero THz field and small $\Gamma_{\rm e-ph}$, the real part of Eq.~(\ref{EQ:constant_QlowT}) reduces to the result given by Fermi's golden rule. The THz field effectively renormalizes the electron-phonon coupling strength through the Bessel functions and opens up new electron-phonon scattering channels that are assisted by THz photons. 

\begin{figure}
	\includegraphics[width=0.47\textwidth]{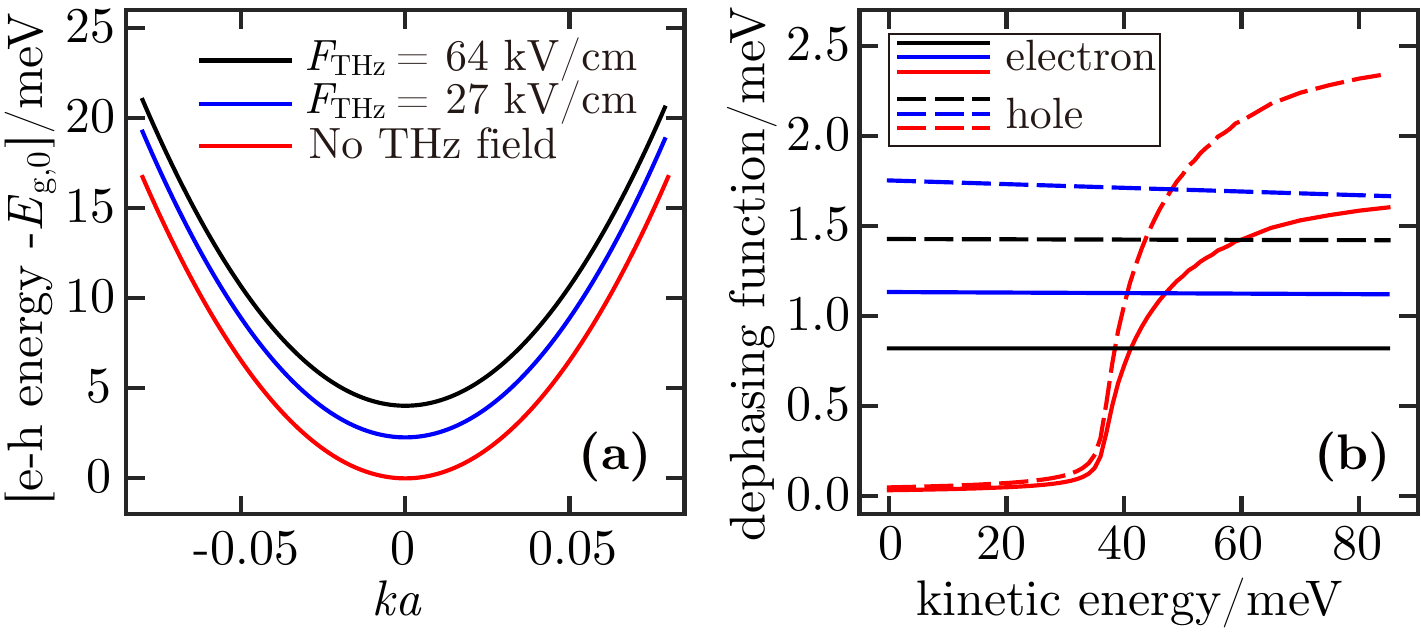}
	\caption{Renormalization of electron-hole energy and dephasing by THz-field-modulated Fr\"ohlich interaction in a parabolic two-band model. (a) Electron-hole energy referenced to the bandgap $E_{\rm g,0}$ at zero THz-field strength. Here, the dimensionless wavevector $ka$ is used with $a=5.65$\,\AA\, being the lattice constant of GaAs~\cite{soma1982thermal,driscoll1975precision}. (b) Dephasing function for the electrons (solid lines) and holes (dashed lines). The red, blue, and black curves represent the results calculated with $F_{\rm THz}=0$, $27$, and $64$\,kV/cm, respectively. }
	\label{FIG:frolich}
\end{figure}
To estimate the electron-hole energy shifts in bulk GaAs due to the THz-field-modulated electron-phonon interactions, we consider the long-range Fr\"ohlich interaction between LO phonons and free electron-hole pairs in a parabolic two-band model. In this model, we take the effective masses of the conduction and valence bands to be respectively $m_c=0.067m_0$~\cite{ahmed1992far} and $m_v=m_0/\gamma_1=m_0/6.98$~\cite{skolnick1976investigation}, which yield the electron-hole reduced mass $\mu_{\rm ex}$ determining the exciton binding energy of bulk GaAs. Because of the negligible carrier occupations in HSG, we use the unscreened coupling constants
\begin{align}
|G^{\prime{\bf q},j}_{cc}|^2
=
|G^{\prime{\bf q},j}_{vv}|^2
=
\frac{e^2}{q^2}
\frac
{\hbar\omega_{\rm LO}}
{2\varepsilon_0V}
[
\frac{1}{\varepsilon(\infty)}
-
\frac{1}{\varepsilon(0)}
],
\end{align}
where $\hbar\omega_{\rm LO}=36.56$\,meV is the LO phonon energy~\cite{irmer1996temperature}, $V$ is the volume of the system, $\varepsilon(\infty)=10.8$ and $\varepsilon(0)=12.7$~\cite{samara1983temperature} are the high- and zero-frequency dielectric constants, respectively. Here, the high-frequency dielectric constant is obtained from the Lyddane-Sachs-Teller relation $\varepsilon(0)/\varepsilon(\infty)=\omega^2_{\rm LO}/\omega^2_{\rm TO}$~\cite{moore1996infrared}, where $\hbar\omega_{\rm TO}=33.72$\,meV is the energy of a transverse optical phonon~\cite{irmer1996temperature}. The quantity ${\mathcal Q}_{\bf P}$ generally depends on the direction of the quasi-momentum ${\bf P}$. Since the electron-hole recollision pathways passing ${\bf k}={\bf 0}$ dominantly describe the HSG in bulk GaAs, we evaluate ${\mathcal Q}_{\bf P}$ for electron-hole quasi-momenta parallel to the THz electric field. The imaginary part of ${\mathcal Q}_{\bf P}$, referenced to its values at zero THz-field strength, determines the electron-hole energy shifts. As shown in Fig.~\ref{FIG:frolich} (a), along the $k$-space trajectory parallel to the THz electric field, the electron-hole energy is effectively shifted upward by about 2 to 4\,meV for the experimental THz-field strength ranging from 27 to 64\,kV/cm, with only minor changes in the curvature of the electron-hole energy dispersion. The results are not sensitive to the choice of the dephasing constant $\Gamma_{\rm e-ph}$ for the phonon-assisted density matrix as long as it is sufficiently small. In the calculation, $\Gamma_{\rm e-ph}=1$\,meV is used.

Examining the quantity ${\mathcal Q}_{\bf P}$ also provides insights into why the dephasing rates of the electron-hole pairs in our Hamiltonian reconstruction can be treated as constant in the presence of the Fr\"ohlich interaction. Figure~\ref{FIG:frolich} (b) shows the individual contributions to the dephasing function ${\rm Re}({\mathcal Q}_{\bf P})$ from electron-LO phonon scattering (solid curves) and hole-LO phonon scattering (dashed curves) under a THz field with three different strengths, $F_{\rm THz}=0$, $27$, and $64$\,kV/cm. In the absence of strong THz fields, an electron or a hole must acquire sufficiently large kinetic energy to effectively emit LO phonons. As expected, the electron and hole dephasing rates both exhibit a threshold at the LO-phonon energy when there is no THz field. However, under the strong THz fields used in this work, the dephasing function becomes nearly independent of the carrier kinetic energy, effectively rendering the dephasing rate constant. The total dephasing rate of the electron-hole pairs, given by ${\rm Re}({\mathcal Q}_{\bf P})$, for $F_{\rm THz}=27$ and $64$\,kV/cm is between 2 to 3\,meV. This indicates that dephasing of the electron-hole coherences in the GaAs sample does not occur solely through Fr\"ohlich interaction, as the extracted dephasing constants are above 7\,meV. As discussed in Sec.~\ref{SEC:hamiltonian_reconstruct}, based on previous HSG experiments conducted at varying temperatures~\cite{costello2023breaking}, the contribution of thermal phonons to the dephasing of the electron-hole coherences in our GaAs sample at 30\,K is expected to be negligible. Note also that, compared to the LH band, the HH band has a higher density of states at the same energy, providing more channels for intraband elastic scattering. Since the extracted dephasing constant $\Gamma_{\rm E-HH}$ is larger than $\Gamma_{\rm E-LH}$, it is likely that elastic scattering through defects is the dominant dephasing mechanism of the electron-hole coherences in our HSG experiments.

The analysis above suggests that, as a simple extension of the effective three-band electron-hole Hamiltonian, one may incorporate a dependence of the bandgap on the THz-field strength. A more precise characterization of the electron-hole energy renormalization and dephasing due to the THz-field-modulated Fr\"ohlich interaction in GaAs can be achieved by analyzing the quantum kinetics of correlations between the LO phonons and the two species of electron-hole pairs, the E-HH and E-LH pairs. In the three-band model, the quantum kinetics of the electron-phonon system is more involved, because both intraband and interband electron-phonon scattering processes can occur for the holes~\cite{scholz1995hole}. A full quantum-kinetic analysis within the three-band electron-hole model is left for future work.

\section{Conclusion}\label{SEC:conclusion}

In conclusion, we have reconstructed an effective three-band electron-hole Hamiltonian in bulk GaAs based on HSG induced by quasi-continuous NIR and THz lasers. We performed polarimetry of high-order sidebands emitted from a bulk GaAs epilayer while varying the wavelength and polarization of the exciting NIR laser, as well as the THz-field strength, to systematically explore the information encoded in the sidebands. Based on previous understanding of HSG in bulk GaAs in terms of Bloch-wave interferometry~\cite{o2024bloch}, information about the effective electron-hole Hamiltonian is compactly wrapped into two electron-hole propagators, which govern the acceleration of two species of electron-hole pairs under the strong THz fields. An analytic model of the interband polarization in HSG is derived to strengthen the theoretical foundation of the Bloch-wave interferometry and to express the electron-hole propagators as functions of the effective-Hamiltonian parameters. By using sufficiently strong THz fields with a sufficiently low frequency, each electron-hole propagator is associated with a shortest electron-hole recollision pathway. Assuming that the effective-Hamiltonian parameters $\mu_{\rm ex}/m_0$ and $\gamma_3/\gamma_2$, which determine the 1s-exciton binding energy and the hole Bloch wavefunctions, respectively, are known from existing experiments, we show that the two dephasing constants associated with the two electron-hole species, the bandgap of GaAs, and the effective-Hamiltonian parameter $\xi=\gamma_2\mu_{\rm ex}/m_0$, which defines the electron-hole reduced masses, can all be unambiguously determined from the measured electron-hole propagators. Since the parameter $\mu_{\rm ex}/m_0$ can be determined from low-temperature absorbance spectra~\cite{sell1972resolved}, and the parameter $\gamma_3/\gamma_2$ can be extracted based on HSG~\cite{costello2021reconstruction}, we have thus demonstrated that reconstruction of the three-band electron-hole Hamiltonian in bulk GaAs can be achieved by combining HSG experiments with absorbance spectroscopy. The reconstruction procedure is validated by quantitative agreement between the measured and calculated electron-hole propagators.

Confidence intervals for the extracted parameters are determined via Monte Carlo simulations. The extracted bandgap of GaAs exceeds the value inferred from absorbance measurements~\cite{sell1972resolved} by about 10\,meV. The extracted bandgap can be smaller, if electron-hole Coulomb attraction, which shortens the electron-hole acceleration times in the recollision processes, is taken into account. We also show that the Fr\"ohlich interaction between the electron-hole pairs and the longitudinal optical phonons may be modulated by the strong THz fields, leading to a larger bandgap and more $k$-independent dephasing rates of the electron-hole coherences. Incorporating electron-hole Coulomb attraction and THz-field-modulated Fr\"ohlich interaction into the Hamiltonian reconstruction is left for future work.

\section*{Acknowledgments}

We acknowledge N. Agladze for assistance in maintaining and operating the UCSB Millimeter-Wave Free Electron Laser. This work was supported by NSF-DMR 2004995 and 2333941. 

Q.W. conceptualized the study, developed the theory, performed all calculations, analyzed the data,  and wrote the original draft. S.D.O. conducted the experiments and compiled the raw data with assistance from J. B. C.. S.D.O. and J.B.C. fabricated the sample. L.N.P. and K.W.W. grew the GaAs crystal. M.S.S. acquired funding and supervised the project.

The data that support the findings of this article are openly available~\cite{qile2025electron}.

\appendix

\section{THz-field strengths}\label{APP:thz_field_strength}

\begin{figure}
	\includegraphics[width=0.47\textwidth]{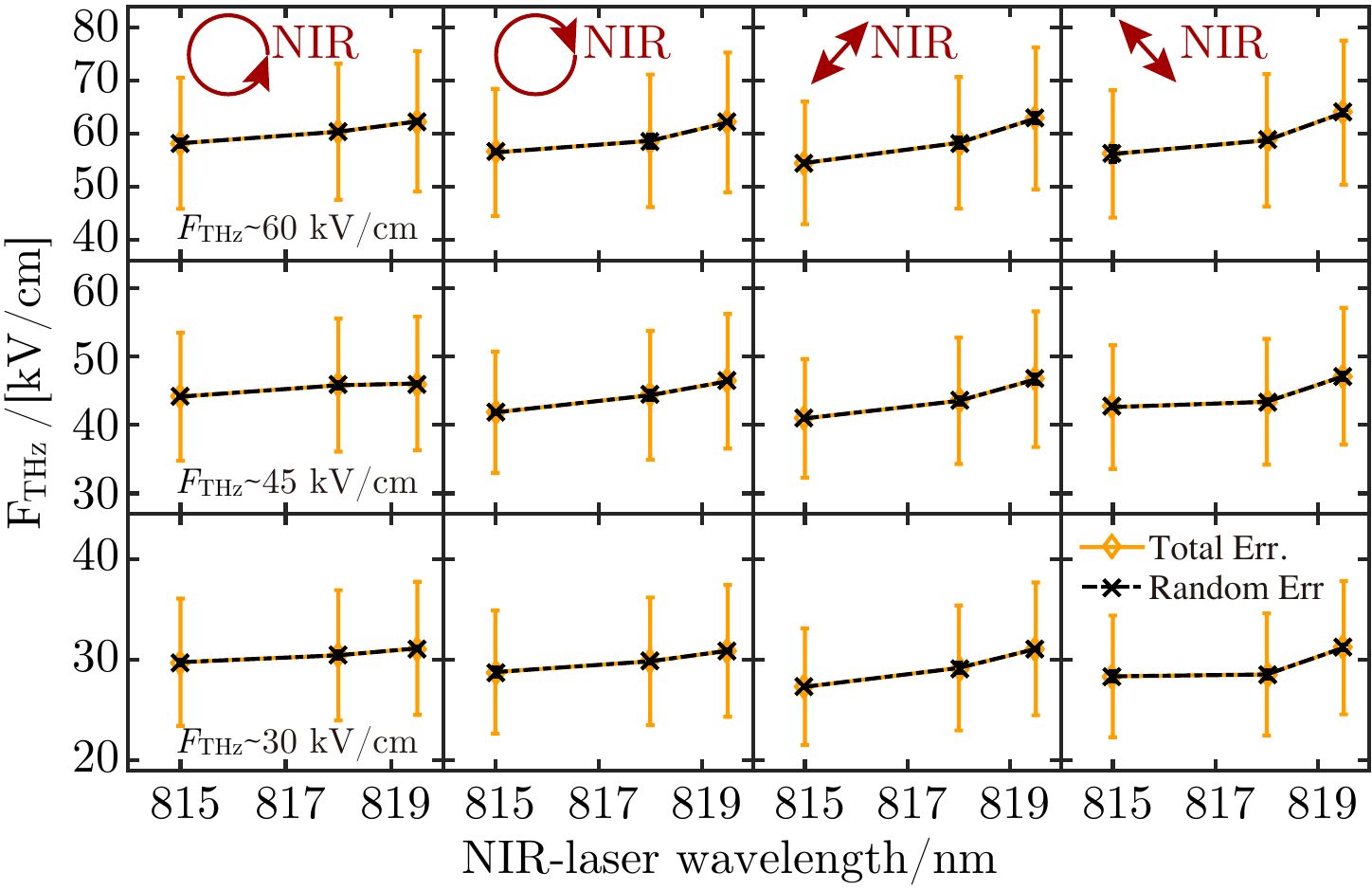}
	\caption{The THz-field strengths in the 36 repeated polarimetry experiments. Each column shows the THz-field strengths for a specific NIR-laser polarization (from left to right: left-handed circular polarization with helicity -1, right-handed circular polarization with helicity +1, linear polarization at $45^\circ$ to the $x$ axis, and linear polarization at $-45^\circ$ to the $x$ axis). The error bar associated with each data point represents one standard deviation of the THz-field strength from the mean value. The data shown in orange incorporate systematic errors arising from the uncertainties in the spatial beam profile of the THz field, as well as random errors. For the data shown in black, only random errors are included.}
	\label{FIG:thz_field_strength}
\end{figure}
The strength of the THz field was constantly monitored by splitting the THz beam generated from the FEL into two beam paths with a beam splitter. Along one beam path, 10\% of the THz output power was directed to a pyroelectric detector, which measured the output power of each FEL pulse. A Thomas Keating (TK) absolute power/energy meter was used to calibrate the pyroelectric detector. The THz beam containing the remaining 90\% of the output power was reflected by a flat mirror, focused by a gold-coated off-axis parabolic mirror with a 12.5 cm focal length, and directed by an ITO slide into the cryogenic chamber, where the GaAs sample resided. The THz beam incident on the GaAs epilayer had a spot diameter of approximately 1.2\,mm. The NIR laser beam was focused to a spot with a diameter of about 0.5 mm at the center of the THz-beam spot. The THz-field strength in the beam spot of the NIR laser is calculated by assuming that the gold-coated off-axis parabolic mirror and flat mirror are both $100\%$ reflective, the ITO slide is $70\%$ reflective, and the cryostat window is $95\%$ transmissive. A vector network analyzer was used to measure the reflection coefficient of the sample near the frequency of the THz field, to characterize the enhancement  of the THz field at the GaAs epilayer resulting from the constructive interference between the incident THz field and the THz field reflected from the ITO film in the sample. Besides the random errors, a systematic uncertainty in the THz-field strength, arising from the uncertainties in the spatial beam profile, is estimated by using two limiting beam profiles: a Gaussian and a flat-top shape. As illustrated in Fig.~\ref{FIG:thz_field_strength}, the systematic errors dominate over the random errors.

The first row of Fig.~\ref{FIG:thz_field_strength} shows the THz-field strengths for 12 of the 36 repeated polarimetry experiments without attenuation of the THz field. The four columns show respectively the THz-field strengths for four NIR-laser polarizations: a left-handed circular polarization with helicity -1, a right-handed circular polarization with helicity +1, a linear polarization at $45^\circ$ to the $x$ axis, and a linear polarization at $-45^\circ$ to the $x$ axis. The bottom two rows of Fig.~\ref{FIG:thz_field_strength} show the smaller THz-field strengths for the remaining 24 polarimetry experiments, in which two wire-grid polarizers were used to attenuate the THz field. The angle between the two wire grids was tuned to $30^\circ$ and $60^\circ$, respectively, to obtain the THz-field strengths shown in the second and third row of Fig.~\ref{FIG:thz_field_strength}, while the polarization of the THz field in the GaAs epilayer was maintained. The TK power/energy meter was used to calibrate the wire-grid rotation angles.

\section{Calculation of the dynamical Jones matrices}\label{APP:cal_Jones_mat}

To calculate the dynamical Jones matrices by using Eqs.~(\ref{EQ:Stokes_to_Jonesmat_amp}),~(\ref{EQ:Stokes_to_Jonesmat_phase}),~(\ref{EQ:Tmat_diagonal}), and~(\ref{EQ:Tmat_offdiagonal}), we write down the following normalized Jones vectors for the four different NIR-laser polarizations:
\begin{align}
\begin{pmatrix}
{\tilde E}_{+,{\rm NIR}}\\
{\tilde E}_{-,{\rm NIR}}
\end{pmatrix}
=
&
\begin{pmatrix}
0\\
1
\end{pmatrix}
,
\begin{pmatrix}
1\\
0
\end{pmatrix}
,
\notag\\
&
\frac{1}{\sqrt{2}}
\begin{pmatrix}
1\\
-ie^{-i2\varphi}
\end{pmatrix}
,
\frac{1}{\sqrt{2}}
\begin{pmatrix}
1\\
ie^{-i2\varphi}
\end{pmatrix}
        \label{EQ:NIR_jones4},
\end{align}
which correspond to a left-handed circular polarization with helicity -1, a right-handed circular polarization with helicity +1, a linear polarization at $45^\circ$ to the $x$ axis, and a linear polarization at $-45^\circ$ to the $x$ axis, respectively (see small cartoons in Fig.~\ref{FIG:thz_field_strength}). For each Jones vector in Eq.~(\ref{EQ:NIR_jones4}), we first obtain the ratios $T_{-+,n}/T_{++,n}$ and $T_{+-,n}/T_{++,n}$ by solving Eqs.~(\ref{EQ:Stokes_to_Jonesmat_phase}),~(\ref{EQ:Tmat_diagonal}), and~(\ref{EQ:Tmat_offdiagonal}), and then use Eq.~(\ref{EQ:Stokes_to_Jonesmat_amp}) to determine the absolute value of $T_{++,n}$. In this way, each dynamical Jones matrix is determined up to an overall phase factor.

\section{Propagation of sideband fields}\label{APP:propagation_sidebands}

\begin{figure}
	\includegraphics[width=0.47\textwidth]{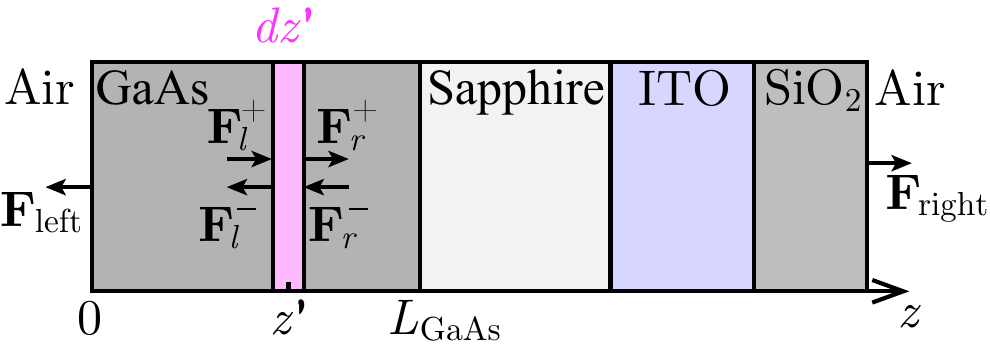}
	\caption{Propagation of sideband fields. The interband polarization in a thin GaAs layer of thickness $dz$ (pink region) results in sideband radiation propagating in two opposite directions along the $z$ axis.}
	\label{FIG:transfer_matrix}
\end{figure}
The relation between the interband polarization and the detected sideband electric fields can be established by solving the Maxwell's equations with a polarization source. We assume here that the sample is homogeneous in the planes perpendicular to the $z$-axis (Fig.~\ref{FIG:transfer_matrix}). As the NIR laser propagates through the sample, it decays and acquires a $z$-dependent phase, resulting in a $z$ dependence in the NIR-laser electric field. Given the Jones vector of the NIR laser in the air, ${\bf E}_{\rm NIR}$, the NIR-laser electric field in the GaAs epilayer can be generally represented by the complex form $g_{\rm NIR}(z){\bf E}_{\rm NIR}\exp(-i\omega_{\rm NIR}t)$ with a $z$-dependent factor containing two counter-propagating components:
\begin{align}
g_{\rm NIR}(z)=g_{+}e^{in^{\rm NIR}_{\rm GaAs}q_0z}+ g_{-}e^{-in^{\rm NIR}_{\rm GaAs}q_0z},
        \label{EQ:NIR_zdependent}
\end{align}
where $n^{\rm NIR}_{\rm GaAs}$ is the refractive index of GaAs at the NIR-laser frequency and $q_0=\omega_{\rm NIR}/c$ is the wavevector of the NIR laser in free space. The coefficients $g_{\pm}$ depend on the dielectric functions and thicknesses of the materials in the sample and can be calculated by considering the propagation of the NIR-laser field in the sample as a stratified medium~\cite{born2013principles}. By using the Fourier component ${\mathbb P}_n$ given in Eq.~(\ref{EQ:Jonesmat_theory_P}), where a $z$-independent NIR-laser field is assumed, the polarization source associated with the $n$th-order sideband in the GaAs epilayer can be written as
\begin{align}
{\mathcal P}_n(z,t)
= 
 g_{\rm NIR}(z){\mathbb P}_ne^{-i(\omega_{\rm NIR}+n\omega_{\rm THz})t}.
        \label{EQ:polarization_source}
\end{align}
Note that the coupling between sideband fields emitted from different locations in the GaAs epilayer is negligible, since the sidebands are much weaker than the NIR-laser field. Under this consideration, the total sideband electric field can be calculated as a superposition of the field components arising from the interband polarization at different locations in the GaAs epilayer. By restricting the polarization source ${\mathcal P}_n(z,t)$ within a thin layer of thickness $dz'$ at $z'$ (Fig.~\ref{FIG:transfer_matrix}), the sideband electric field ${\mathbb E}_n(z,t)={\bf F}_n(z)\exp[-i(\omega_{\rm NIR}+n\omega_{\rm THz})t]$ satisfies the following wave equation:
\begin{align}
\frac{\partial^2{\bf F}_n}{\partial z^2}+\varepsilon_n(z) q^2_n{\bf F}_n
=
-\varepsilon_0q^2_n g_{\rm NIR}(z'){\mathbb P}_n\Theta(z),
        \label{EQ:wave_equation}
\end{align}
where $\Theta(z)$ is 1 for $z\in[z'-dz/2,z'+dz/2]$ and zero everywhere else, $\varepsilon_n(z)$ is the dielectric function at the sideband frequency $f_{{\rm SB},n}$, and $q_n=(\omega_{\rm NIR}+n\omega_{\rm THz})/c$ is the wavevector of the $n$th-order sideband in free space. Away from $z'$, in each material of the sample, the sideband electric field is a superposition of two counter-propagating components along the $z$ axis. The continuity of the sideband electric field and its derivative with respect to $z$ requires that the sideband radiation near the source at $z'$ is connected to the radiation propagating out of the sample through via
\begin{align}
&
\begin{pmatrix}
{\bf F}_{\rm right}\\
0
\end{pmatrix}
=
{\mathbb M}_{\rm sub}M^{{n}_{\rm GaAs}}_{L_{\rm GaAs}-z'}
\begin{pmatrix}
{\bf F}^{+}_{r}\\
{\bf F}^{-}_{r}
\end{pmatrix},
\label{EQ:field_right_prop}
\\
&
\begin{pmatrix}
{\bf F}^{+}_{l}\\
{\bf F}^{-}_{l}
\end{pmatrix}
=
M^{{n}_{\rm GaAs}}_{z'}
M_{\rm int}(n_{\rm Air},{n}_{\rm GaAs})
\begin{pmatrix}
0\\
{\bf F}_{\rm left}
\end{pmatrix},
\label{EQ:field_left_prop}
\end{align}
where ${\bf F}^{\pm}_{r(l)}$ are the two components of ${\bf F}_n(z)$ on the right (left) surface of the polarization source, ${\bf F}_{\rm left (right)}$ is the component of ${\bf F}_n(z)$ in the sideband field that just leaves the sample from the left (right), the matrix
\begin{align}
M^{n_0}_{L}
=
\begin{pmatrix}
e^{in_0 q_nL} & 0\\
0 & e^{-in_0 q_nL}
\end{pmatrix}
\label{EQ:transfer_mat_bulk}
\end{align}
describes the propagation of the sideband field in a material with refractive index $n_0$ over a distance $L$, the matrix
\begin{align}
M_{\rm int}(n_1,n_2)
=
\begin{pmatrix}
\frac{1}{2}(1+\frac{n_1}{n_2}) & \frac{1}{2}(1-\frac{n_1}{n_2})\\
\frac{1}{2}(1-\frac{n_1}{n_2}) & \frac{1}{2}(1+\frac{n_1}{n_2})
\end{pmatrix}
\label{EQ:transfer_mat_interface}
\end{align}
connects the sideband electric field at the interface between two materials with refractive indices $n_1$ and $n_2$, respectively, and the matrix
\begin{align}
{\mathbb M}_{\rm sub}
&
=
M_{\rm int}(n_{\rm GaAs},n_{\rm Sap})
M^{n_{\rm Sap}}_{L_{\rm Sap}}
M_{\rm int}(n_{\rm Sap},n_{\rm ITO})
M^{n_{\rm ITO}}_{L_{\rm ITO}}
\notag\\
&
\times
M_{\rm int}(n_{\rm ITO},n_{{\rm SiO}_2})
M^{n_{{\rm SiO}_2}}_{L_{{\rm SiO}_2}}
M_{\rm int}(n_{{\rm SiO}_2},n_{\rm Air})
\label{EQ:transfer_mat_substrate}
\end{align}
describes the propagation of the sideband electric field from the left-hand side of the sapphire substrate to the right-hand side of the ${\rm SiO}_2$ layer. Here, ${n}_{\rm Air}$, ${n}_{\rm GaAs}$, ${n}_{\rm Sap}$, and $n_{{\rm SiO}_2}$ are the refractive indices at the sideband frequency $f_{{\rm SB},n}$ for the air, GaAs, sapphire substrate, and ${\rm SiO}_2$, respectively, and $L_{\rm GaAs}$, $L_{\rm Sap}$, $L_{\rm ITO}$, and $L_{{\rm SiO}_2}$ are the thicknesses of the corresponding layers in the sample. Integrating Eq.~(\ref{EQ:wave_equation}) across the source layer yields a relation between the components ${\bf F}^{\pm}_{r}$ and ${\bf F}^{\pm}_{l}$:
\begin{align}
&
\frac{\partial{\bf F}_n}{\partial z}|_{z=z_0+dz/2}-\frac{\partial{\bf F}_n}{\partial z}|_{z=z_0-dz/2}\notag\\
=
&
i{n}_{\rm GaAs}q_n[({\bf F}^{+}_{r}-{\bf F}^{-}_{r})-({\bf F}^{+}_{l}-{\bf F}^{-}_{l})]
\notag\\
=
&
-\varepsilon_0q^2_n g_{\rm NIR}(z'){\mathbb P}_ndz',
        \label{EQ:source_layer}
\end{align}
which, together with the continuity condition ${\bf F}^{+}_{r}+{\bf F}^{-}_{r}={\bf F}^{+}_{l}+{\bf F}^{-}_{l}$, gives the following equation relating ${\bf F}^{\pm}_{l}$ and ${\bf F}^{\pm}_{r}$:
\begin{align}
\begin{pmatrix}
{\bf F}^{+}_{r}\\
{\bf F}^{-}_{r}
\end{pmatrix}
=
\begin{pmatrix}
{\bf F}^{+}_{l}\\
{\bf F}^{-}_{l}
\end{pmatrix}
+
\frac{i\varepsilon_0q_n}{2{n}_{\rm GaAs}} g_{\rm NIR}(z'){\mathbb P}_ndz'
\begin{pmatrix}
1\\
-1
\end{pmatrix}.
        \label{EQ:connection_lr}
\end{align}
Using Eqs.~(\ref{EQ:field_right_prop}),~(\ref{EQ:field_left_prop}), and~(\ref{EQ:connection_lr}) to eliminate ${\bf F}^{\pm}_{l}$ and ${\bf F}^{\pm}_{r}$, we obtain the following equation connecting ${\bf F}_{\rm left}$, ${\bf F}_{\rm right}$, and ${\mathbb P}_n$:
\begin{align}
\begin{pmatrix}
{\bf F}_{\rm right}\\
0
\end{pmatrix}
=
&
{\mathbb M}_{\rm tot}
\begin{pmatrix}
0\\
{\bf F}_{\rm left}
\end{pmatrix}
+
i\frac{\varepsilon_0q_n}{2{n}_{\rm GaAs}} g_{\rm NIR}(z'){\mathbb P}_ndz'\notag\\
&
\times {\mathbb M}_{\rm sub}
\begin{pmatrix}
e^{i{n}_{\rm GaAs}q_n(L_{\rm GaAs}-z')}\\
-e^{-i{n}_{\rm GaAs}q_n(L_{\rm GaAs}-z')}
\end{pmatrix},
        \label{EQ:F_left_right_pn}
\end{align}
which gives
\begin{align}
{\bf F}_{\rm right}
&
=
[
(
{\mathbb M}_{\rm sub,11}
-
{\mathbb M}_{\rm sub,21}
\frac{{\mathbb M}_{\rm tot,12}}{{\mathbb M}_{\rm tot,22}}
)e^{i{n}_{\rm GaAs}q_n(L_{\rm GaAs}-z')}
\notag\\
&-
(
{\mathbb M}_{\rm sub,12}
-
{\mathbb M}_{\rm sub,22}
\frac{{\mathbb M}_{\rm tot,12}}{{\mathbb M}_{\rm tot,22}}
)e^{-i{n}_{\rm GaAs}q_n(L_{\rm GaAs}-z')}
]\notag\\
&\times i\frac{\varepsilon_0q_n}{2{n}_{\rm GaAs}} g_{\rm NIR}(z'){\mathbb P}_ndz',
        \label{EQ:F_right_pn}
\end{align}
where the matrix ${\mathbb M}_{\rm tot}$ is defined as 
\begin{align}
{\mathbb M}_{\rm tot}
=
{\mathbb M}_{\rm sub}M^{{n}_{\rm GaAs}}_{L_{\rm GaAs}}M_{\rm int}(n_{\rm Air},{n}_{\rm GaAs}).
\label{EQ:transfer_mat_total}
\end{align}
The electric field of the $n$th-order sideband on the right-hand side of the sample can be written as
${\mathbb E}_n(z,t)={\bf F}_{\rm right}\exp\{iq_n[n_{\rm Air}(z-L_{\rm tot})-ct]\}$, where $L_{\rm tot}$ is the total thickness of the sample. Therefore, the Jones vector of the $n$th-order sideband, ${\bf E}_{n}$, is proportional to the Fourier component ${\mathbb P}_n$, with a proportionality factor ${\mathcal T}_n$ given by
\begin{align}
{\mathcal T}_n
&
=
 ie^{-in_{\rm Air}q_nL_{\rm tot}}\frac{\varepsilon_0q_n}{2{n}_{\rm GaAs}} 
[
(
{\mathbb M}_{\rm sub,11}
-
{\mathbb M}_{\rm sub,21}
\frac{{\mathbb M}_{\rm tot,12}}{{\mathbb M}_{\rm tot,22}}
)
\notag\\
&
\times
\int_0^{L_{\rm GaAs}}dz'g_{\rm NIR}(z')e^{i{n}_{\rm GaAs}q_n(L_{\rm GaAs}-z')}
\notag\\
&
-
(
{\mathbb M}_{\rm sub,12}
-
{\mathbb M}_{\rm sub,22}
\frac{{\mathbb M}_{\rm tot,12}}{{\mathbb M}_{\rm tot,22}}
)
\notag\\
&
\times
\int_0^{L_{\rm GaAs}}dz'g_{\rm NIR}(z')e^{-i{n}_{\rm GaAs}q_n(L_{\rm GaAs}-z')}].
        \label{EQ:proportionality_factor}
\end{align}

\section{Band structure of GaAs under biaxial strains}\label{APP:biaxial_strain}

We consider here a biaxial strain in bulk GaAs described by the following strain tensor:
\begin{align}
\epsilon
=
\begin{pmatrix}
\epsilon_{XX} & 0 & 0\\
0 & \epsilon_{XX} & 0\\
0 & 0 & \epsilon_{ZZ}
\end{pmatrix},
        \label{EQ:strain_tensor}
\end{align}
where the coordinate system is defined by the $X$, $Y$, and $Z$ axes. To ensure that the normal stress along the [001] crystal direction vanishes, we impose the condition
\begin{align}
2c_{12}\epsilon_{XX}+c_{11}\epsilon_{ZZ}=0,
        \label{EQ:strain_constraint}
\end{align}
where $c_{12}=566$\,GPa and $c_{11}=1221$\,GPa~\cite{vurgaftman2001band} are two components of the stiffness tensor of GaAs. Under this biaxial strain, the conduction band is shifted by $H_{c,\epsilon}=a_c {\rm tr}(\epsilon)$~\cite{willatzen2009kp}, where $a_c=-7.17$\,eV~\cite{vurgaftman2001band} is a deformation potential for the conduction band. For the highest two valence bands, the Luttinger Hamiltonian is modified to lowest order in $k$ by adding the diagonal term~\cite{willatzen2009kp}
\begin{align}
H_{v,\epsilon}
=
a_v {\rm tr}(\epsilon){\bf 1}_4
+
b_v\sum_{j=X,Y,Z}\epsilon_{jj}(J_j^2-\frac{1}{3}J^2),
        \label{EQ:strain_hamiltonian}
\end{align}
where $a_v=-1.16$\,eV and $b_v=-2.0$\,eV~\cite{vurgaftman2001band} are two deformation potentials for the valence bands. The valence-band-edge states $|\pm3/2\rangle$ are shifted by $b_v(\epsilon_{ZZ}-\epsilon_{XX})$, while the valence-band-edge states $|\pm1/2\rangle$ are shifted by $-b_v(\epsilon_{ZZ}-\epsilon_{XX})$. Because the dipole moment associated with the states $|\pm3/2\rangle$ is greater than that associated with the states $|\pm1/2\rangle$ by a factor of $\sqrt{3}$,
to match the exciton-peak splitting in the absorbance spectrum [Fig.~\ref{FIG:exciton_splitting} (a)], we set 
\begin{align}
-2b_v(\epsilon_{ZZ}-\epsilon_{XX})=\Delta_{\rm ex}.
        \label{EQ:strain_splitting}
\end{align}
Equations~(\ref{EQ:strain_constraint}) and~(\ref{EQ:strain_splitting}) are solved to obtain the strain tensor, which determines the strain Hamiltonians $H_{c,\epsilon}$ and $H_{v,\epsilon}$. The energies of the valence bands are obtained by diagonalizing the Hamiltonian $H_{v}+H_{v,\epsilon}$. In the calculation, we use the parameter $m_c=0.067m_0$~\cite{ahmed1992far}, and the Luttinger parameters $\gamma_1=6.98$, $\gamma_2=2.2$, and $\gamma_3=2.9$~\cite{skolnick1976investigation}.

\section{Supplementary data of the propagator ratios}\label{APP:sup_data}

\begin{figure*}
	\includegraphics[width=0.98\textwidth]{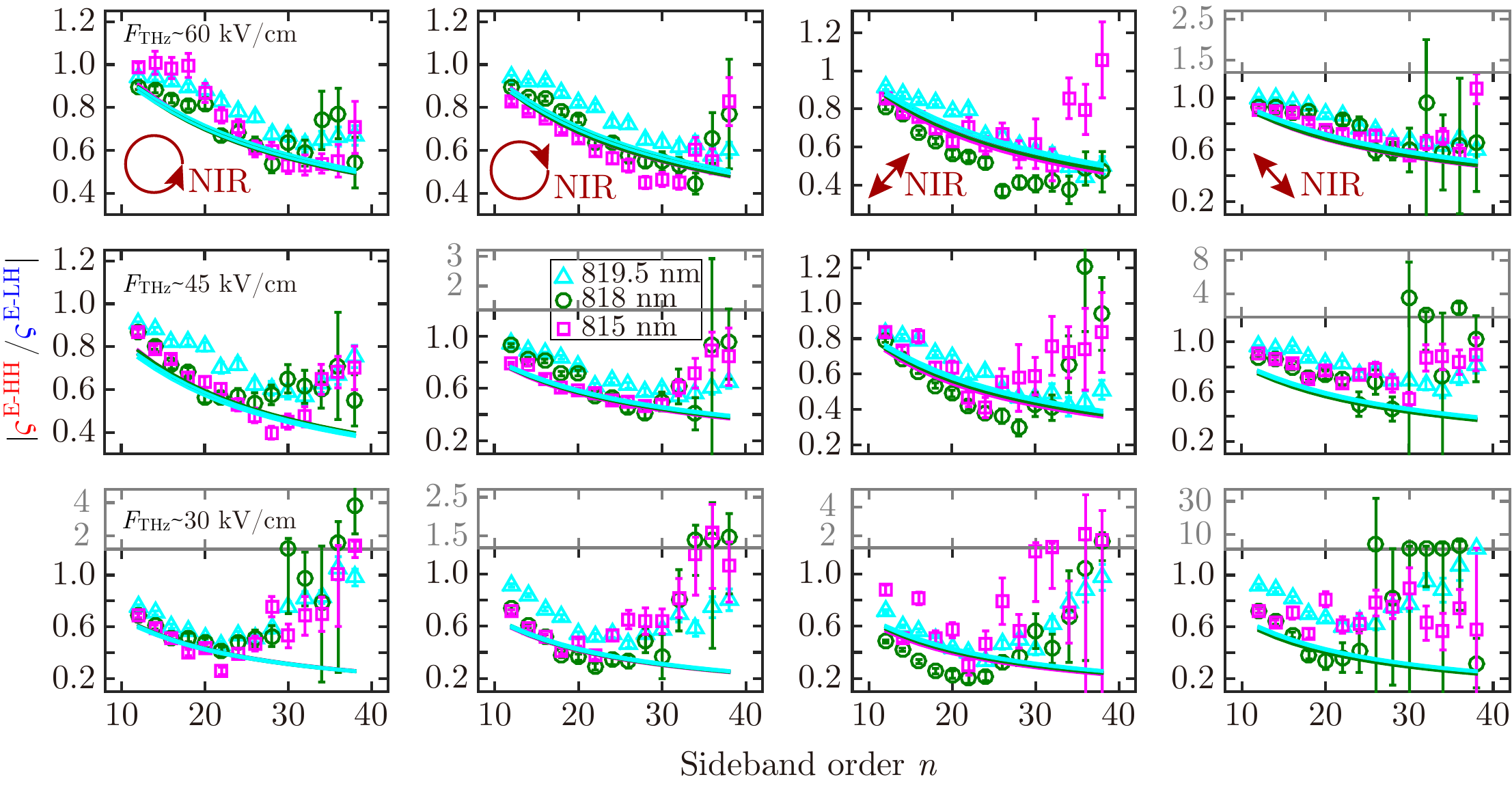}
	\caption{Absolute values of the propagator ratio $\varsigma^{\rm E-HH}/\varsigma^{\rm E-LH}$. Each column shows the data for a specific NIR-laser polarization (from left to right: left-handed circular polarization with helicity -1, right-handed circular polarization with helicity +1, linear polarization at $45^\circ$ to the $x$ axis, and linear polarization at $-45^\circ$ to the $x$ axis). The first, second, and third rows show the data collected at three different THz-field strength levels: around 60\,kV/cm, 45\,kV/cm, and 30\,kV/cm, respectively (see Fig.~\ref{FIG:thz_field_strength} in Appendix~\ref{APP:thz_field_strength} for the exact THz-field strengths). In each panel, cyan triangles, dark green circles, and magenta squares represent the data obtained at three different NIR-laser wavelengths: 819.5\,nm, 818\,nm, and 815\,nm, respectively. The cyan, dark green, and magenta solid lines represent the corresponding theoretical results. For each set of laser parameters, two solid lines of the same color indicate the one-standard-deviation range resulting from uncertainties in the THz field strengths and the Hamiltonian parameters. The theoretical curves in each panel largely overlap. In the grey boxes, larger $y$ scales are used.}
	\label{FIG:abs_ratio_all}
\end{figure*}
\begin{figure*}
	\includegraphics[width=0.98\textwidth]{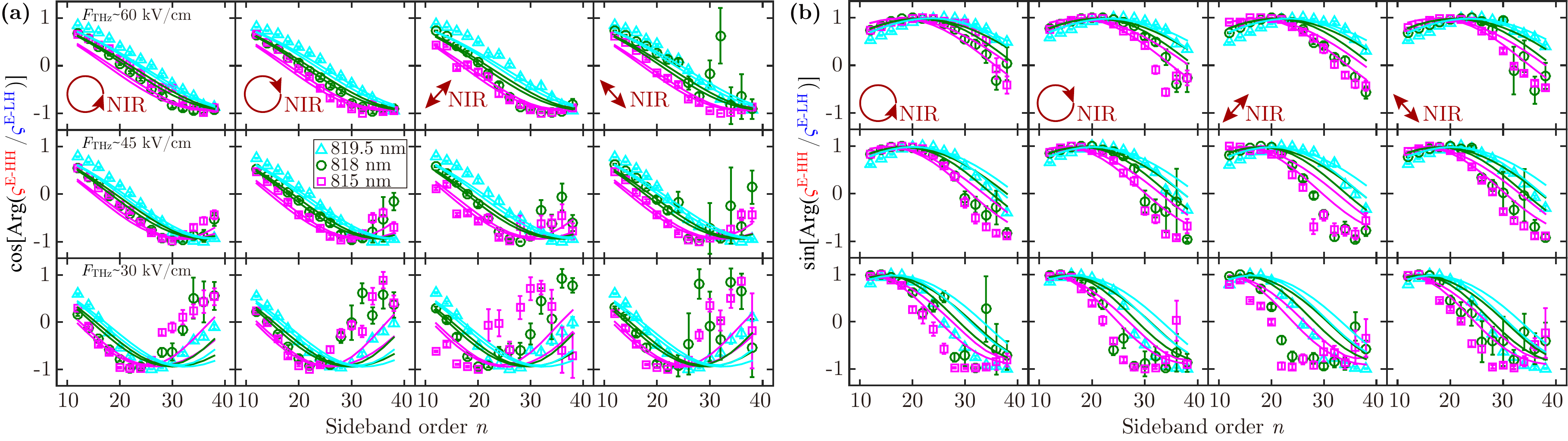}
	\caption{Phases of the propagator ratio $\varsigma^{\rm E-HH}/\varsigma^{\rm E-LH}$ expressed in terms of cosine and sine functions. Each column shows the data for a specific NIR-laser polarization (from left to right in (a) or (b): left-handed circular polarization with helicity -1, right-handed circular polarization with helicity +1, linear polarization at $45^\circ$ to the $x$ axis, and linear polarization at $-45^\circ$ to the $x$ axis). The first, second, and third rows show the data collected at three different THz-field strength levels: around 60\,kV/cm, 45\,kV/cm, and 30\,kV/cm, respectively (see Fig.~\ref{FIG:thz_field_strength} in Appendix~\ref{APP:thz_field_strength} for the exact THz-field strengths). In each panel, cyan triangles, dark green circles, and magenta squares represent the data obtained at three different NIR-laser wavelengths: 819.5\,nm, 818\,nm, and 815\,nm, respectively. The cyan, dark green, and magenta solid lines represent the corresponding theoretical results. For each set of laser parameters, two solid lines of the same color indicate the one-standard-deviation range resulting from uncertainties in the THz field strengths and the Hamiltonian parameters.}
	\label{FIG:arg_ratio_all}
\end{figure*}
\begin{figure*}
	\includegraphics[width=0.98\textwidth]{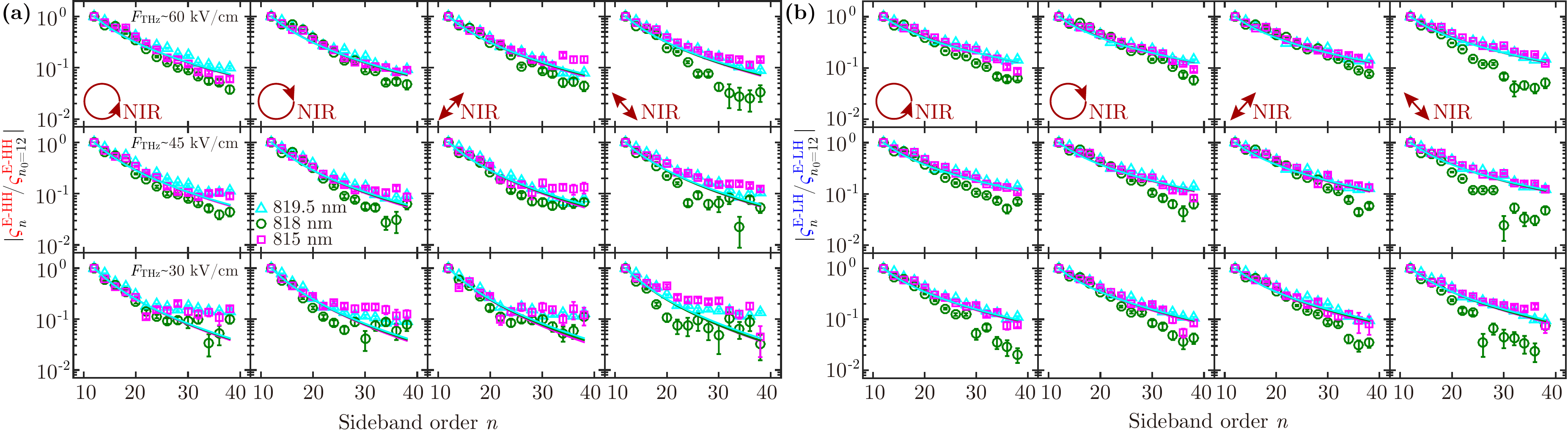}
	\caption{Absolute values of the propagators $\varsigma^{\rm E-HH}$ and $\varsigma^{\rm E-LH}$ relative to their values at the lowest detected sideband order $n_0=12$. Each column shows the data for a specific NIR-laser polarization (from left to right in (a) or (b): left-handed circular polarization with helicity -1, right-handed circular polarization with helicity +1, linear polarization $45^\circ$ to the $x$ axis, and linear polarization $-45^\circ$ to the $x$ axis). The first, second, and third rows show the data collected at three different THz-field strength levels: around 60\,kV/cm, 45\,kV/cm, and 30\,kV/cm, respectively (see Fig.~\ref{FIG:thz_field_strength} in Appendix~\ref{APP:thz_field_strength} for the exact THz-field strengths). In each panel, cyan triangles, dark green circles, and magenta squares represent the data corresponding to three different NIR-laser wavelengths: 819.5\,nm, 818\,nm, and 815\,nm, respectively. For each set of laser parameters, two solid lines of the same color indicate the one-standard-deviation range resulting from uncertainties in the THz field strengths and Hamiltonian parameters. The theoretical curves in each panel largely overlap. }
	\label{FIG:propagator_decay_all}
\end{figure*}
Figures~\ref{FIG:abs_ratio_all},~\ref{FIG:arg_ratio_all}, and~\ref{FIG:propagator_decay_all} show, respectively, the absolute values of the ratio $\varsigma^{\rm E-HH}/\varsigma^{\rm E-LH}$, the phases of the ratio $\varsigma^{\rm E-HH}/\varsigma^{\rm E-LH}$ expressed in terms of sine and cosine functions, and the absolute values of $\varsigma^{\rm E-HH}_n/\varsigma^{\rm E-HH}_{n_0}$ and $\varsigma^{\rm E-LH}_n/\varsigma^{\rm E-LH}_{n_0}$ for all 36 polarimetry experiments. The mean values of these quantities and their standard deviations (error bars) are determined via a Monte Carlo simulation of the propagation of the uncertainties in the sideband intensity spectra. First, each data point in a sideband-intensity spectrum is randomly sampled 1,000 times from a normal distribution, whose mean and standard deviation are determined by four CCD cans, yielding the mean value and standard deviation of the area of each sideband peak $I(n,\theta_{\rm QWP})$ in Eq.~(\ref{EQ:intensiry_stokes}). Second, the value of $I(n,\theta_{\rm QWP})$ is randomly sampled 1,000 times from a normal distribution defined by these mean and standard-deviation values. Third, 1,000 sets of the Stokes parameters are computed by using Eq.~(\ref{EQ:fourier_stokes}). Fourth, Eqs.~(\ref{EQ:Stokes_to_Jonesmat_amp}),~(\ref{EQ:Stokes_to_Jonesmat_phase}),~(\ref{EQ:Tmat_diagonal}), and~(\ref{EQ:Tmat_offdiagonal}) are used to calculate the absolute value of $T_{++,n}$ and the ratio $T_{-+,n}/T_{++,n}$. In this step, the Luttinger-parameter ratio $\gamma_3/\gamma_2$ is generated by randomly drawing the parameters $\gamma_2$ and $\gamma_3$ from normal distributions whose mean and standard deviations correspond to the reported values: $\gamma_2=2.2\pm0.1$ and $\gamma_3=2.9\pm0.2$~\cite{skolnick1976investigation}. Fifth, the unit vector $\hat{n}$ is calculated by using the randomly sampled values of $\gamma_3/\gamma_2$. Last, the propagator ratios $\varsigma^{\rm E-HH}/\varsigma^{\rm E-LH}$, $\varsigma^{\rm E-HH}_n/\varsigma^{\rm E-HH}_{n_0}$, and $\varsigma^{\rm E-LH}_n/\varsigma^{\rm E-LH}_{n_0}$ are calculated 1,000 times by using Eqs.~(\ref{EQ:ehh_propagator_tppmp}) and~(\ref{EQ:elh_propagator_tppmp}).

\section{Supplementary data in the Hamiltonian reconstruction}\label{APP:sup_data_reconstruction}

\begin{figure}
	\includegraphics[width=0.47\textwidth]{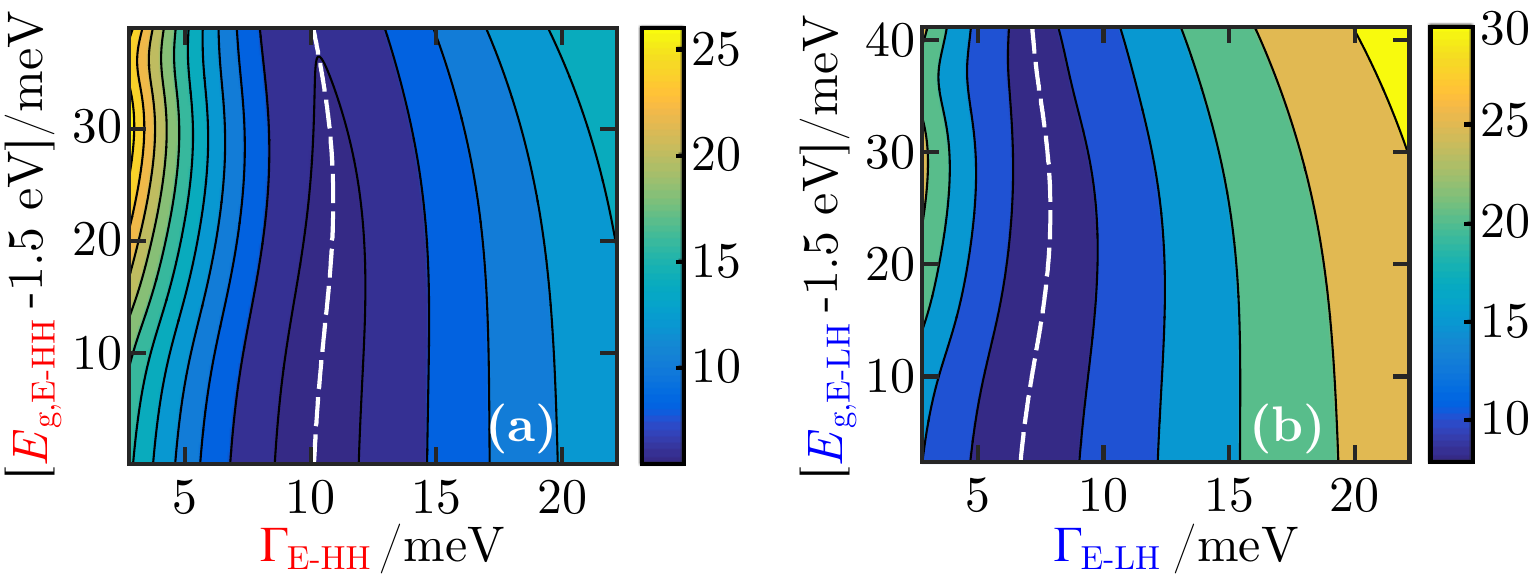}
	\caption{Values of the cost functions (a) $R^{\rm E-HH}$ and (b) $R^{\rm E-LH}$ defined in Eqs.~(\ref{EQ:cost_f1}) and~(\ref{EQ:cost_f2}), respectively. In the calculations, we use the parameter $m_c=0.067m_0$~\cite{ahmed1992far}, and the Luttinger parameters $\gamma_1=6.98$, $\gamma_2=2.2$, and $\gamma_3=2.9$~\cite{skolnick1976investigation}. The white dashed lines indicate the optimal values of the dephasing constants, $\Gamma_{\rm E-HH}$ and $\Gamma_{\rm E-LH}$, as functions of the bandgaps, $E_{\rm g,E-HH}$ and $E_{\rm g,E-LH}$, respectively.}.
	\label{FIG:dephasing_vs_gap}
\end{figure}
Constraints on the relationship between the dephasing constant $\Gamma_{\nu}$ ($\nu=\rm E-HH,E-LH$) and the bandgap $E_{\rm g,\nu}$ are obtained by minimizing the cost function $R^{\nu}$.
Figure~\ref{FIG:dephasing_vs_gap} shows the values of the cost functions $R^{\rm E-HH}$ and $R^{\rm E-LH}$ defined by Eqs.~(\ref{EQ:cost_f1}) and~(\ref{EQ:cost_f2}), respectively. In the calculations, we use the parameter $m_c=0.067m_0$~\cite{ahmed1992far} and the Luttinger parameters $\gamma_1=6.98$, $\gamma_2=2.2$, and $\gamma_3=2.9$~\cite{skolnick1976investigation}. For each value of the bandgap $E_{\rm g,\nu}$, there is an optimal value of the dephasing constant $\Gamma_{\nu}$ corresponding to the minimum of the cost function $R^{\nu}$ (white dashed lines).

To determine the confidence intervals for the extracted dephasing constants, bandgaps, and the parameter $\xi=\gamma_2\mu_{\rm ex}/m_0$, we perform Monte Carlo simulations of the error propagation from the uncertainties in the Hamiltonian parameters and THz-field strengths to the theoretically calculated electron-hole propagators. 

\begin{figure}
	\includegraphics[width=0.47\textwidth]{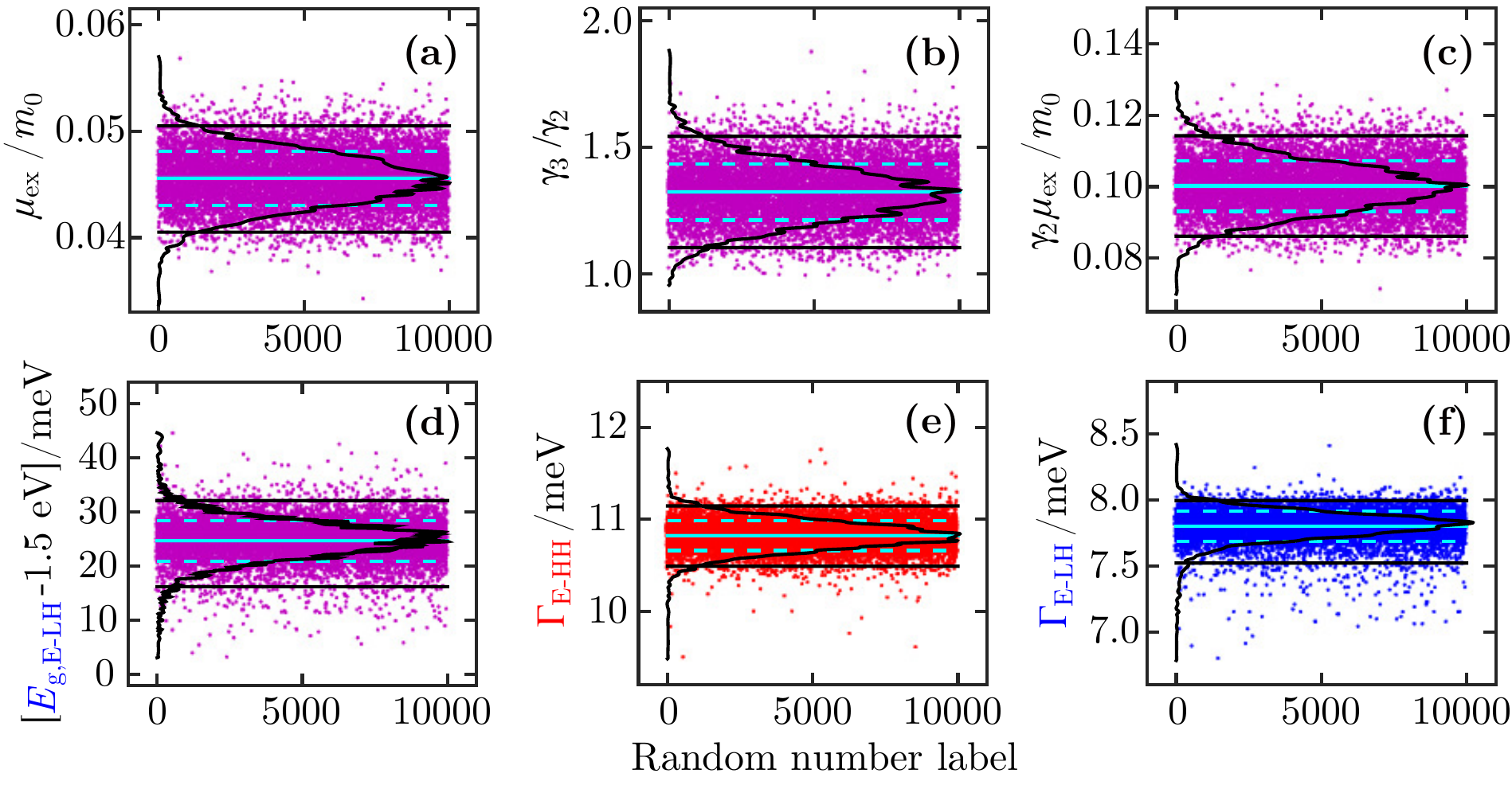}
	\caption{Determination of the confidence intervals for the extracted dephasing constants and bandgaps via Monte Carlo simulation. (a) Randomly sampled values of $\mu_{\rm ex}/m_0$. (b) Randomly sampled values of $\gamma_3/\gamma_2$. (c) Randomly sampled values of $\xi=\gamma_2\mu_{\rm ex}/m_0$. (d) Extracted values of $E_{\rm g,E-LH}$. (e) Extracted values of $\Gamma_{\rm E-HH}$. (f) Extracted values of $\Gamma_{\rm E-LH}$. In each panel, the black curve shows the variable distribution; the cyan solid line represents the mean value; the two cyan dashed lines show one standard deviation around the mean; and the two black solid lines mark a 95\% confidence interval.}
	\label{FIG:monte_carlo_one}
\end{figure}
When only the two dephasing constants and the two bandgaps are extracted with the parameter $\xi$ assumed to be known, each THz-field-strength value is randomly drawn 10,000 times from a normal distribution whose mean and standard deviation correspond to the measured values shown in Fig.~\ref{FIG:thz_field_strength} in Appendix~\ref{APP:thz_field_strength}. In parallel, 10,000 sets of the parameters $\mu_{\rm ex}/m_0$, $\gamma_3/\gamma_2$, and $\xi$ are generated by randomly sampling the parameters $m_c$, $\gamma_1$, $\gamma_2$, and $\gamma_3$ 10,000 times from normal distributions whose means and standard deviations correspond to the reported values---$m_c=(0.067\pm0.005)m_0$~\cite{ahmed1992far}, $\gamma_1=6.98\pm 0.45$, $\gamma_2=2.2\pm 0.1$, and $\gamma_3=2.9\pm 0.2$~\cite{skolnick1976investigation}. The resulting distributions of the parameters $\mu_{\rm ex}/m_0$, $\gamma_3/\gamma_2$, and $\xi$ are shown in Figs.~\ref{FIG:monte_carlo_one} (a), (b), and (c). The reference value of $\xi$ displayed in Fig.~\ref{FIG:extract_reduced_mass_ratio} (b) corresponds to the distribution shown in Fig.~\ref{FIG:monte_carlo_one} (c). For each set of THz-field-strength values and parameters $\mu_{\rm ex}/m_0$, $\gamma_3/\gamma_2$, and $\xi$, constraints on the relationship between the dephasing constant $\Gamma_{\nu}$ ($\nu=\rm E-HH,E-LH$) and the bandgap $E_{\rm g,\nu}$ are obtained by minimizing the cost function $R^{\nu}$. The optimal value of the bandgap $E_{\rm g,E-LH}$ is then obtained by minimizing the cost function $R^{\rm phase}$. For each optimal bandgap $E_{\rm g,E-LH}$, the bandgap $E_{\rm g,E-HH}$ is determined from the constraint $E_{\rm g,E-LH}-E_{\rm g,E-HH}=\Delta_{\rm ex}$, and the dephasing constant $\Gamma_{\nu}$ is determined from its functional relationship with $E_{\rm g,\nu}$. The extracted bandgap $E_{\rm g,E-LH}$ and the two dephasing constants are shown in Figs.~\ref{FIG:monte_carlo_one} (d), (e), and (f), respectively.

\begin{figure}
	\includegraphics[width=0.47\textwidth]{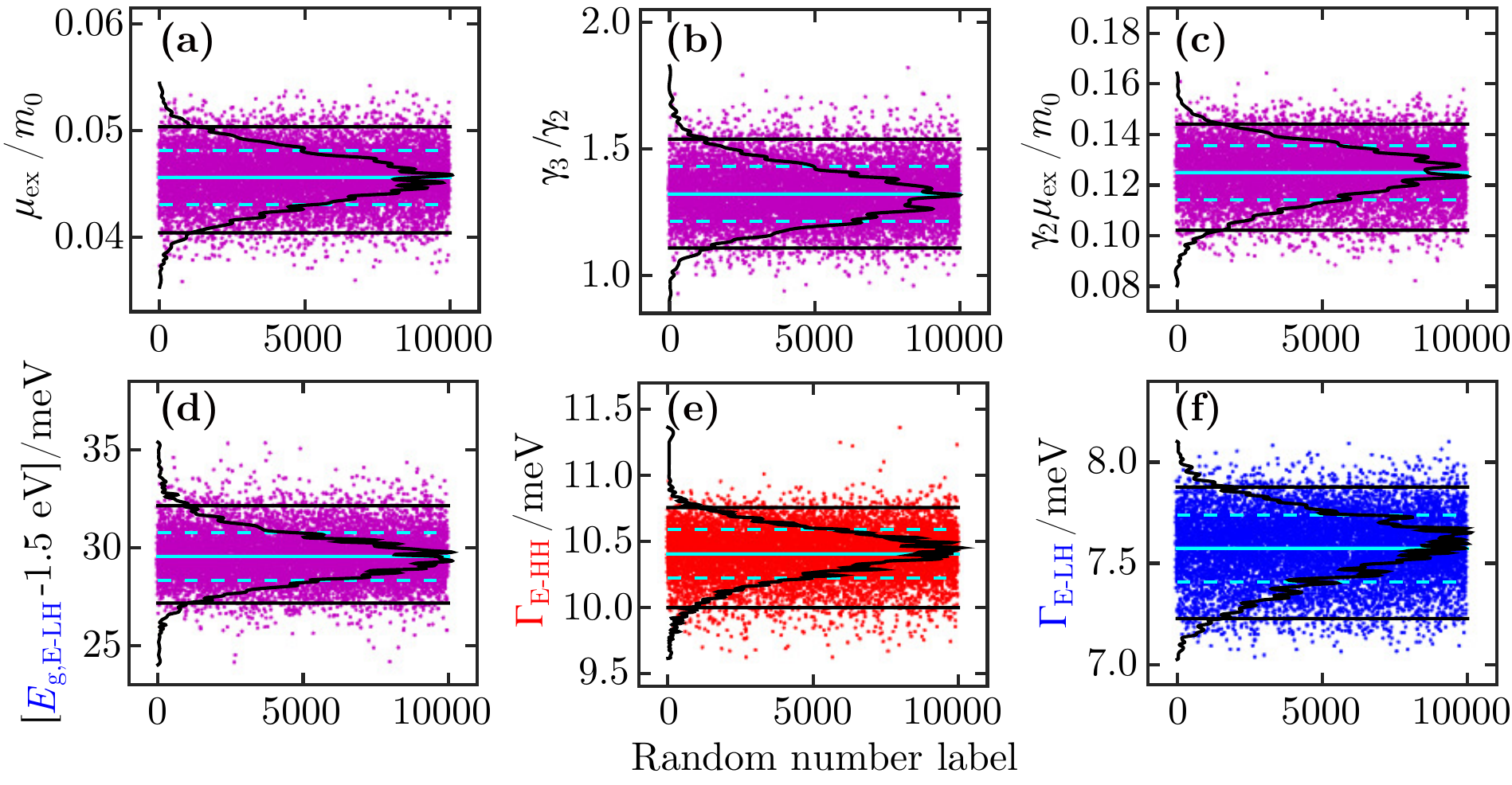}
	\caption{Determination of the confidence intervals for the extracted dephasing constants, bandgaps, and the parameter $\xi=\gamma_2\mu_{\rm ex}/m_0$ via Monte Carlo simulation, with the parameters $\mu_{\rm ex}/m_0$ and $\gamma_3/\gamma_2$ determined from the cyclotron resonance experiments in Refs.~\cite{ahmed1992far} and ~\cite{skolnick1976investigation}. (a) Randomly sampled values of $\mu_{\rm ex}/m_0$. (b) Randomly sampled values of $\gamma_3/\gamma_2$. (c) Extracted values of $\xi$. (d) Extracted values of $E_{\rm g,E-LH}$. (e) Extracted values of $\Gamma_{\rm E-HH}$. (f) Extracted values of $\Gamma_{\rm E-LH}$. In each panel, the black curve shows the variable distribution; the cyan solid line represents the mean value; the two cyan dashed lines show one standard deviation around the mean; and the two black solid lines mark a 95\% confidence interval.}
	\label{FIG:monte_carlo_two}
\end{figure}
When the two dephasing constants, the two bandgaps, and the parameter $\xi$ are extracted simultaneously, each THz-field-strength value is again randomly drawn 10,000 times from a normal distribution whose mean and standard deviation correspond to the measured values shown in Fig.~\ref{FIG:thz_field_strength} in Appendix~\ref{APP:thz_field_strength}. In parallel, 10,000 sets of the parameters $\mu_{\rm ex}/m_0$ and $\gamma_3/\gamma_2$ are generated by randomly sampling the parameters $m_c$, $\gamma_1$, $\gamma_2$, and $\gamma_3$ 10,000 times from normal distributions whose means and standard deviations correspond to the reported values---$m_c=(0.067\pm0.005)m_0$~\cite{ahmed1992far}, $\gamma_1=6.98\pm 0.45$, $\gamma_2=2.2\pm 0.1$, and $\gamma_3=2.9\pm 0.2$~\cite{skolnick1976investigation}. The resulting distributions of the parameters $\mu_{\rm ex}/m_0$ and $\gamma_3/\gamma_2$ are shown in Figs.~\ref{FIG:monte_carlo_two} (a) and (b). For each set of THz-field-strength values and parameters $\mu_{\rm ex}/m_0$ and $\gamma_3/\gamma_2$, with $\xi$ fixed at a certain value, constraints on the relationship between the dephasing constant $\Gamma_{\nu}$ ($\nu=\rm E-HH,E-LH$) and the bandgap $E_{\rm g,\nu}$ are obtained by minimizing the cost function $R^{\nu}$. The optimal value of the bandgap $E_{\rm g,E-LH}$ is then obtained as a function of $\xi$ by minimizing the cost function $R^{\rm phase}$. With $\Gamma_{\nu}$ expressed as a function of $E_{{\rm g},\nu}$, and $E_{\rm g,E-LH}$ expressed as a function of $\xi$, the cost function $R^{\rm abs}$ becomes a function of a single variable, $\xi$, with a minimum corresponding to the optimal value of $\xi$. For each optimal value of $\xi$, the bandgap $E_{\rm g,E-LH}$ is determined through its functional dependence on $\xi$ from the $R^{\rm phase}$ minimization. The bandgap $E_{\rm g,E-HH}$ is then determined from the constraint $E_{\rm g,E-LH}-E_{\rm g,E-HH}=\Delta_{\rm ex}$, and the dephasing constant $\Gamma_{\nu}$ is determined from its functional relationship with $E_{\rm g,\nu}$. The extracted value of $\xi$, the bandgap $E_{\rm g,E-LH}$, and the two dephasing constants are shown in Figs.~\ref{FIG:monte_carlo_two} (c), (d), (e), and (f), respectively.

\begin{figure}
	\includegraphics[width=0.47\textwidth]{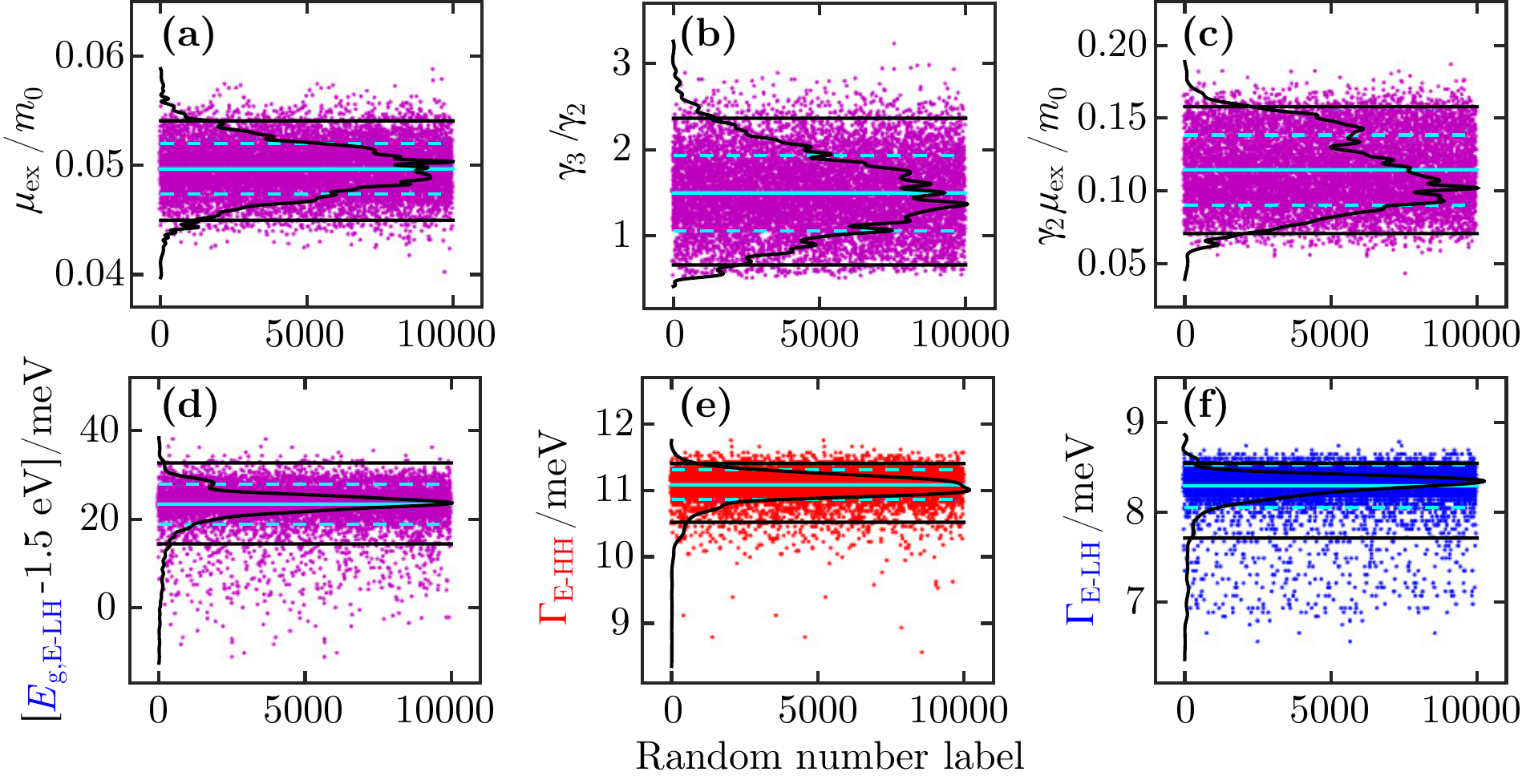}
	\caption{Determination of the confidence intervals for the extracted dephasing constants, bandgaps, and the parameter $\xi=\gamma_2\mu_{\rm ex}/m_0$ via Monte Carlo simulation, with the parameters $\mu_{\rm ex}/m_0$ determined from the absorbance measurements in Ref.~\cite{sell1972resolved} and the ratio $\gamma_3/\gamma_2$ extracted from the HSG experiments in Ref.~\cite{costello2021reconstruction}. (a) Randomly sampled values of $\mu_{\rm ex}/m_0$. (b) Randomly sampled values of $\gamma_3/\gamma_2$. (c) Extracted values of $\xi$. (d) Extracted values of $E_{\rm g,E-LH}$. (e) Extracted values of $\Gamma_{\rm E-HH}$. (f) Extracted values of $\Gamma_{\rm E-LH}$. In each panel, the black curve shows the variable distribution; the cyan solid line represents the mean value; the two cyan dashed lines show one standard deviation around the mean; and the two black solid lines mark a 95\% confidence interval.}
	\label{FIG:monte_carlo_three}
\end{figure}
We also perform a similar Monte Carlo simulation to extract the two dephasing constants, the two bandgaps, and the parameter $\xi$ simultaneously, using the parameters $\mu_{\rm ex}/m_0$ and $\gamma_3/\gamma_2$ determined from previous absorbance and HSG experiments, respectively. In this simulation, the ratio $\gamma_3/\gamma_2$ is randomly sampled from a normal distribution whose mean and standard deviation correspond to the reported value, $\gamma_3/\gamma_2=1.47\pm0.48$~\cite{costello2021reconstruction}. The parameter $\mu_{\rm ex}/m_0$ is calculated as $\mu_{\rm ex}/m_0=E_{\rm 1s}\varepsilon^2(0)/R_{\infty}$~\cite{willatzen2009kp}, where $E_{\rm 1s}$ is the 1s-exciton binding energy randomly sampled from a normal distribution whose mean and standard deviation correspond to the value determined from low-temperature absorbance measurements, $4.2\pm0.2$\,meV~\cite{sell1972resolved}, $R_{\infty}$ is the Rydberg constant, and $\varepsilon(0)=12.69$ is the corresponding static dielectric constant~\cite{samara1983temperature}.
Without modifying the other steps in the Monte Carlo simulation described in the previous paragraph, we obtain the values of $\xi$, the bandgap $E_{\rm g,E-LH}$, and the two dephasing constants, as shown in Figs.~\ref{FIG:monte_carlo_three} (c), (d), (e), and (f), respectively. Now the extracted value of $\xi$ is $0.114\pm0.024$, with a 95\% confidence interval of [0.071,0.158], which covers the reference value $\xi=0.101\pm0.007$. The mean value of the extracted bandgap $E_{\rm g,E-LH}$ is 1.523\,eV, which lies about 6\,meV above the reference value $1.5169\pm0.0002$\,eV derived from low-temperature absorbance measurements~\cite{sell1972resolved}, and the associated 95\% confidence interval, [1.514,1.533]\,eV, encloses the reference value. The extracted dephasing constants are $\Gamma_{\rm E-HH}=11.1\pm0.2$\,meV and $\Gamma_{\rm E-LH}=8.3\pm0.2$\,meV, with 95\% confidence intervals of [10.5,11.4]\,meV and  [7.7,8.5]\,meV, respectively. The randomly sampled $\mu_{\rm ex}/m_0$ and $\gamma_3/\gamma_2$ are shown in Figs.~\ref{FIG:monte_carlo_three} (a) and (b). Because of the larger uncertainty in the ratio $\gamma_3/\gamma_2$, we calculate the propagator ratios $\varsigma^{\rm E-HH}/\varsigma^{\rm E-LH}$, $\varsigma^{\rm E-HH}_n/\varsigma^{\rm E-HH}_{n_0}$, and $\varsigma^{\rm E-LH}_n/\varsigma^{\rm E-LH}_{n_0}$ used in this Monte Carlo simulation based on the same procedure described in Appendix~\ref{APP:sup_data} but with the number of random samples increased from 1,000 to 10,000.

\section{Approximate model of electron-hole propagators}\label{APP:app_propagator}

\begin{figure}
	\includegraphics[width=0.47\textwidth]{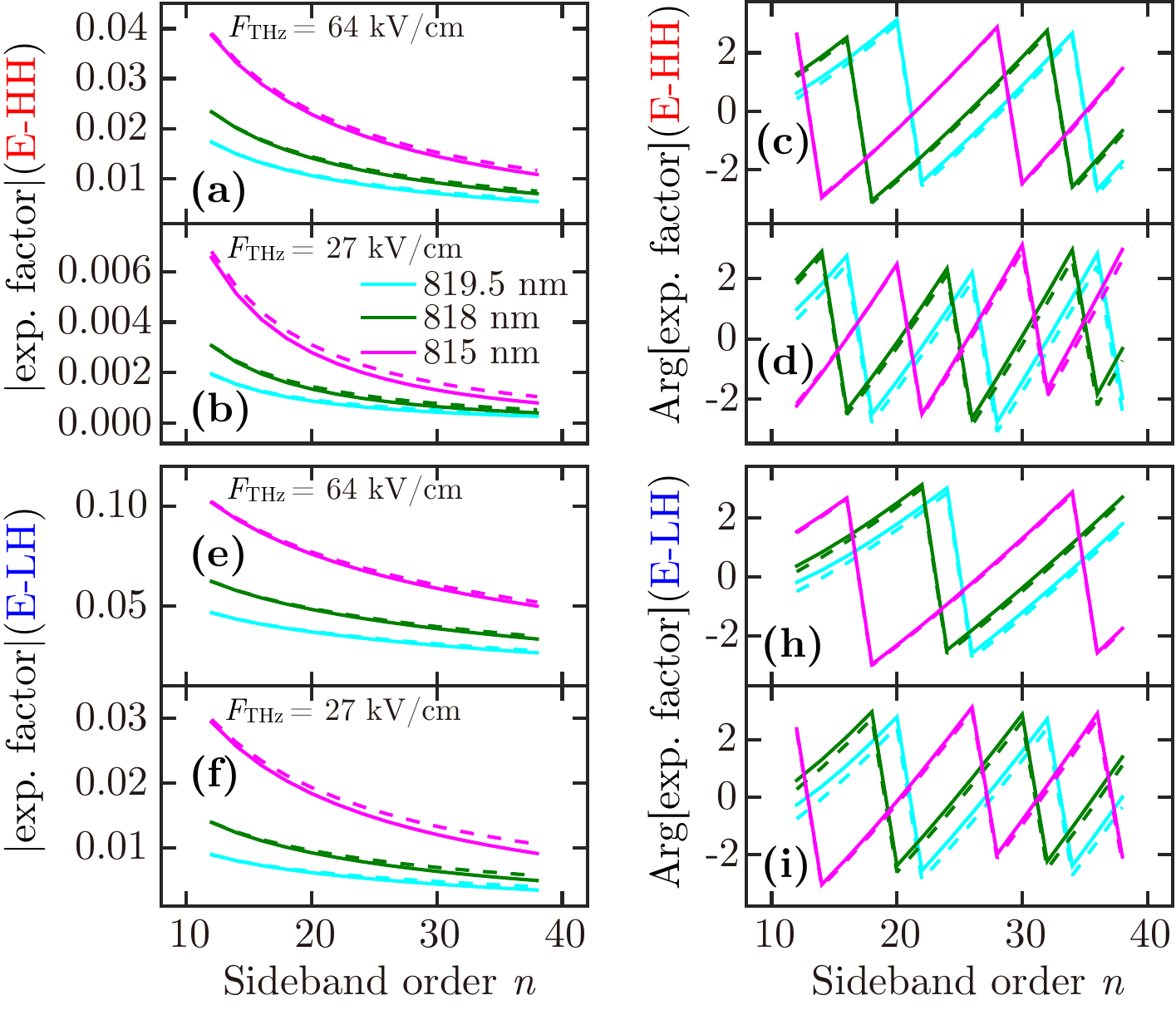}
	\caption{Comparison of the exponential factor $\exp\{i[q^{\nu}_{1/4}({\tilde U}^{\nu}_{\rm p})^{-1/4}+q^{\nu}_{3/4}({\tilde U}^{\nu}_{\rm p})^{-3/4}]\}$ ($\nu=\rm E-HH,E-LH$) in Eq.~(\ref{EQ:propagator_algebraic_form}) (solid lines) and the exponential factor $\exp[iq^{\nu,(2)}_{1/4}({\tilde U}^{\nu}_{\rm p})^{-1/4}]$ in Eq.~(\ref{EQ:propagator_app}) (dashed lines). Left (right) column: absolute values (phases) of the exponential factors. In each panel, cyan, dark green, and magenta curves represent the results corresponding to three different NIR-laser wavelengths: 819.5\,nm, 818\,nm, and 815\,nm, respectively. The functions $q^{\nu}_{1/4}$ and $q^{\nu}_{3/4}$ are calculated by using Eqs.~(\ref{EQ:definition_q1o4}) and ~(\ref{EQ:definition_q3o4}), respectively, with the NIR-laser detunings for the two electron-hole species distinguished as $\Delta_{\rm NIR}\rightarrow\Delta^{\nu}_{\rm NIR}$. The function $q^{\nu,(2)}_{1/4}$ is calculated by using Eq.~(\ref{EQ:app_q1o4}). The first and third (second and fourth) rows show the results by using a THz-field strength of $64$ (27)\,kV/cm, which corresponds to the strongest (weakest) THz field used in the polarimetry experiments. The parameters $\mu_{\rm ex}/m_0$, $\gamma_3/\gamma_2$, $\xi=\gamma_2\mu_{\rm ex}/m_0$, $E_{\rm g,E-LH}$, $\Gamma_{\rm E-HH}$, and $\Gamma_{\rm E-LH}$ are set to the mean values obtained from the Monte Carlo simulation for simultaneously extracting the two dephasing constants, the two bandgaps, and the parameter $\xi$ (Fig.~\ref{FIG:monte_carlo_two}).}
	\label{FIG:expfac_approx}
\end{figure}
\begin{figure}
	\includegraphics[width=0.47\textwidth]{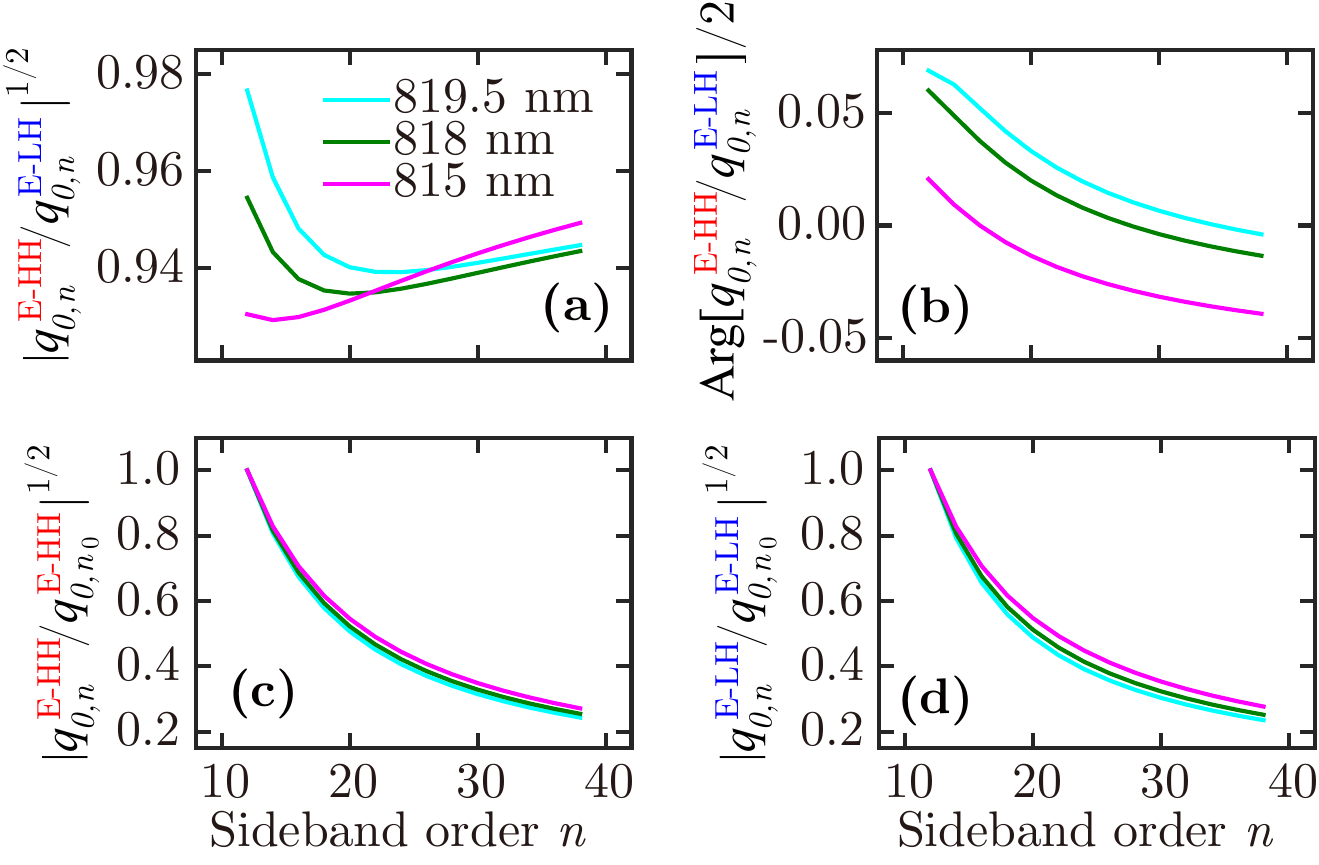}
	\caption{Numerical values of the ratios $q^{\rm E-HH}_{0,n}/q^{\rm E-LH}_{0,n}$ and $q^{\nu}_{0,n}/q^{\nu}_{0,n_0=12}$ ($\nu=\rm E-HH,E-LH$). (a) The value of $(q^{\rm E-HH}_{0,n}/q^{\rm E-LH}_{0,n})^{1/2}$. (b) The value of ${\rm Arg}[{q^{\rm E-HH}_{0,n}/q^{\rm E-LH}_{0,n}}]/2$. (c) The value of $|q^{\rm E-HH}_{0,n}/q^{\rm E-HH}_{0,n_0}|^{1/2}$. (d) The value of $|q^{\rm E-LH}_{0,n}/q^{\rm E-LH}_{0,n_0}|^{1/2}$. The function $q^{\nu}_{0}$ is calculated by using Eqs.~(\ref{EQ:definition_q0}), with the NIR-laser detunings for the two electron-hole species distinguished as $\Delta_{\rm NIR}\rightarrow\Delta^{\nu}_{\rm NIR}$. In each panel, cyan, dark green, and magenta curves represent the results corresponding to three different NIR-laser wavelengths: 819.5\,nm, 818\,nm, and 815\,nm, respectively. The parameters $E_{\rm g,E-LH}$, $\Gamma_{\rm E-HH}$, and $\Gamma_{\rm E-LH}$ are set to the mean values obtained from the Monte Carlo simulation for simultaneously extracting the two dephasing constants, the two bandgaps, and the parameter $\xi$ (Fig.~\ref{FIG:monte_carlo_two}).}
	\label{FIG:q0fac_approx}
\end{figure}
As discussed in Sec.~\ref{SEC:information}, the analytic model of electron-hole propagators given in Eq.~(\ref{EQ:propagator_analytic}) qualitatively agrees with the experimental data for the electron-hole propagator ratios (Figs.~\ref{FIG:abs_propagator_ratio},~\ref{FIG:arg_propagator},~\ref{FIG:decay_propagator},~\ref{FIG:abs_ratio_all},~\ref{FIG:arg_ratio_all}, and~\ref{FIG:propagator_decay_all}). To understand the connection between this analytic model and the propagator model used in the Hamiltonian reconstruction [Eq.~(\ref{EQ:propagator_algebraic_form})], we analyze the functions $q^{\nu}_{0}$, $q^{\nu}_{1/4}$ and $q^{\nu}_{3/4}$  ($\nu=\rm E-HH,E-LH$) in Eq.~(\ref{EQ:propagator_algebraic_form}) by using the parameters $\mu_{\rm ex}/m_0$, $\gamma_3/\gamma_2$, $\xi=\gamma_2\mu_{\rm ex}/m_0$, $E_{\rm g,E-LH}$, $\Gamma_{\rm E-HH}$, and $\Gamma_{\rm E-LH}$ set to the mean values obtained from the Monte Carlo simulation for the Hamiltonian reconstruction (Fig.~\ref{FIG:monte_carlo_two}). As mentioned in Sec.~\ref{SEC:theory}, the analytic model given in Eq.~(\ref{EQ:propagator_analytic}), apart from a constant factor, is just the exponential function $\exp[iq^{\nu,(1)}_{1/4}({\tilde U}^{\nu}_{\rm p})^{-1/4}]$, where $q^{\nu,(1)}_{1/4}=(18n)^{1/4}[(8/15)n+i{\tilde\Gamma}_{\nu}+{\tilde\Delta}_{\rm NIR}]$ is the Taylor expansion of $q^{\nu}_{1/4}$ [Eq.~(\ref{EQ:taylor_q1o4})] retained up to the first-order term in $i{\tilde\Gamma}_{\nu}+{\tilde\Delta}_{\rm NIR}$. In fact, as shown in Fig.~\ref{FIG:expfac_approx}, for the THz-field strengths used in the polarimetry experiments, the exponential factor $\exp\{i[q^{\nu}_{1/4}({\tilde U}^{\nu}_{\rm p})^{-1/4}+q^{\nu}_{3/4}({\tilde U}^{\nu}_{\rm p})^{-3/4}]\}$ in Eq.~(\ref{EQ:propagator_algebraic_form}) can be closely approximated by the simpler exponential factor $\exp[iq^{\nu,(2)}_{1/4}({\tilde U}^{\nu}_{\rm p})^{-1/4}]$, with
\begin{align}
q^{\nu,(2)}_{1/4}(n,& i{\tilde\Gamma}_{\nu}+{\tilde\Delta}^{\nu}_{\rm NIR})
=
(18n)^{1/4}
[
\frac{8}{15}n
+
(i{\tilde\Gamma}_{\nu}+{\tilde\Delta}^{\nu}_{\rm NIR})
\notag\\
&
\times
(1
-
\frac{1}{3}\sqrt{\frac{i{\tilde\Gamma}_{\nu}+{\tilde\Delta}^{\nu}_{\rm NIR}}{n}}
)
],
\label{EQ:app_q1o4}
\end{align}
which is the Taylor expansion of $q^{\nu}_{1/4}$ [Eq.~(\ref{EQ:taylor_q1o4})] retained up to the third-order term in $\sqrt{i{\tilde\Gamma}_{\nu}+{\tilde\Delta}^{\nu}_{\rm NIR}}$. Here, we distinguish the NIR-laser detunings for the two electron-hole species ($\Delta_{\rm NIR}\rightarrow\Delta^{\nu}_{\rm NIR}$). Moreover, using the extracted dephasing constants and bandgaps, we find that the difference in the factor ${\exp[-i\arg(q^{\nu}_{0})/2]}/{\sqrt{|q^{\nu}_{0}|}}$ for the two electron-hole species is negligible, as shown in Figs.~\ref{FIG:q0fac_approx} (a) and (b). The propagator ratio ${\mathbb Q}^{\rm E-HH}_n/{\mathbb Q}^{\rm E-LH}_n$, which determines the sideband polarizations
(see Sec.~\ref{SEC:information}), can thus be approximated as:
\begin{align}
&\frac{{\mathbb Q}^{\rm E-HH}_n}{{\mathbb Q}^{\rm E-LH}_n}
\approx 
\sqrt{
\frac{ {\tilde\mu}^{\rm E-HH}_{yy}\mu^{\rm E-HH}_{zz} }
{ {\tilde\mu}^{\rm E-LH}_{yy}\mu^{\rm E-LH}_{zz} }
}
(\frac{ \mu^{\rm E-HH}_{xx}}{ \mu^{\rm E-LH}_{xx}})^{3/8}
\notag\\
&
\times
\exp
\{
i[
q^{\rm E-HH,(2)}_{1/4}(n,i{\tilde\Gamma}_{\nu}+{\tilde\Delta}^{\nu}_{\rm NIR})({\tilde U}^{\rm E-HH}_{\rm p})^{-1/4}
\notag\\
&
-
q^{\rm E-LH,(2)}_{1/4}(n,i{\tilde\Gamma}_{\nu}+{\tilde\Delta}^{\nu}_{\rm NIR})({\tilde U}^{\rm E-LH}_{\rm p})^{-1/4}
]
\},
\label{EQ:propagator_app}
\end{align}
where $\sqrt{{ {\tilde\mu}^{\rm E-HH}_{yy}\mu^{\rm E-HH}_{zz} }/{ {\tilde\mu}^{\rm E-LH}_{yy}\mu^{\rm E-LH}_{zz} }}({ \mu^{\rm E-HH}_{xx}}/{ \mu^{\rm E-LH}_{xx}})^{3/8}$ is a factor independent of the sideband order $n$ and the THz-field strength $F_{\rm THz}$. Moreover, within the range of sideband orders considered in this paper, $12 \le n\le 38$, the extracted mean value of $i{\tilde\Gamma}_{\nu}+{\tilde\Delta}^{\nu}_{\rm NIR}$ has a modulus less than the sideband orders. 
Therefore, Eqs.~(\ref{EQ:app_q1o4}) and~(\ref{EQ:propagator_app}) imply that the analytic model given in Eq.~(\ref{EQ:propagator_analytic}) can be used to qualitatively describe the dependence of the sideband polarizations on the THz-field strength and the sideband order. For the ratio ${\mathbb Q}^{\nu}_n/{\mathbb Q}^{\nu}_{n_0}$, which describes the propagator decay with increasing sideband order, the function $q^{\nu}_0$ can not be ignored. The absolute value of $q^{\nu}_0$ significantly contributes to the propagator decay as a function of sideband order, as shown in Figs.~\ref{FIG:q0fac_approx} (c) and (d). As an approximation, the absolute value of the propagator ratio ${\mathbb Q}^{\nu}_n/{\mathbb Q}^{\nu}_{n_0}$ can be written as:
\begin{align}
|\frac{{\mathbb Q}^{\nu}_n}{{\mathbb Q}^{\nu}_{n_0}}|
\approx 
&
\sqrt{
\frac{|q^{\nu}_{0}(n_0,i{\tilde\Gamma}_{\nu}+{\tilde\Delta}^{\nu}_{\rm NIR})|}
{|q^{\nu}_{0}(n,i{\tilde\Gamma}_{\nu}+{\tilde\Delta}^{\nu}_{\rm NIR})|}
}
\notag\\
&
\times
\exp
\{
i[
q^{\nu,(2)}_{1/4}(n,i{\tilde\Gamma}_{\nu}+{\tilde\Delta}^{\nu}_{\rm NIR})({\tilde U}^{\nu}_{\rm p})^{-1/4}
\notag\\
&
-
q^{\nu,(2)}_{1/4}(n_0,i{\tilde\Gamma}_{\nu}+{\tilde\Delta}^{\nu}_{\rm NIR})({\tilde U}^{\nu}_{\rm p})^{-1/4}
]
\}.
\label{EQ:propagator_decay_app}
\end{align}
The dependence of the propagator decay on the THz-field strength is captured by the factor $\exp[iq^{\nu,(2)}_{1/4}({\tilde U}^{\nu}_{\rm p})^{-1/4}]$, and therefore by the analytic model given in Eq.~(\ref{EQ:propagator_analytic}). By introducing real auxiliary variables ${\tilde\Gamma}^{\prime}_{\nu,n}$ and ${\tilde\Delta}^{\nu,{\prime}}_{\rm NIR,n}$ satisfying
\begin{align}
i{\tilde\Gamma}^{\prime}_{\nu,n}+{\tilde\Delta}^{\nu,\prime}_{\rm NIR,n}
=
&
(i{\tilde\Gamma}_{\nu}+{\tilde\Delta}^{\nu}_{\rm NIR})
\notag\\
&
\times
(1
-
\frac{1}{3}\sqrt{\frac{i{\tilde\Gamma}_{\nu}+{\tilde\Delta}^{\nu}_{\rm NIR}}{n}}
)
],
\label{EQ:auxiliary_igamma_delta}
\end{align}
Eq.~(\ref{EQ:app_q1o4}) can be rewritten as
\begin{align}
&q^{\nu,(2)}_{1/4}(n, i{\tilde\Gamma}^{\prime}_{\nu,n}+{\tilde\Delta}^{\nu,{\prime}}_{\rm NIR,n})
\notag\\
=
&
(18n)^{1/4}
(
\frac{8}{15}n
+
i{\tilde\Gamma}^{\prime}_{\nu,n}+{\tilde\Delta}^{\nu,\prime}_{\rm NIR,n}).
\label{EQ:app_q1o4_variant}
\end{align}
Equations.~(\ref{EQ:propagator_app}) and~(\ref{EQ:propagator_decay_app}) can thus be viewed as derived from the analytic model given in Eq.~(\ref{EQ:propagator_analytic}), with ${\tilde\Gamma}_{\nu}$ and ${\tilde\Delta}^{\nu}_{\rm NIR}$ replaced by the $n$-dependent auxiliary variables ${\tilde\Gamma}^{\prime}_{\nu,n}$ and ${\tilde\Delta}^{\nu,{\prime}}_{\rm NIR,n}$, and with an additional factor proportional to $\sqrt{{ {\tilde\mu}^{\nu}_{yy}\mu^{\nu}_{zz} }}({ \mu^{\nu}_{xx}})^{3/8}/\sqrt{|q^{\nu}_{0}(n,i{\tilde\Gamma}_{\nu}+{\tilde\Delta}^{\nu}_{\rm NIR})|}$ to account for the quantum fluctuations.

\begin{figure*}
	\includegraphics[width=0.94\textwidth]{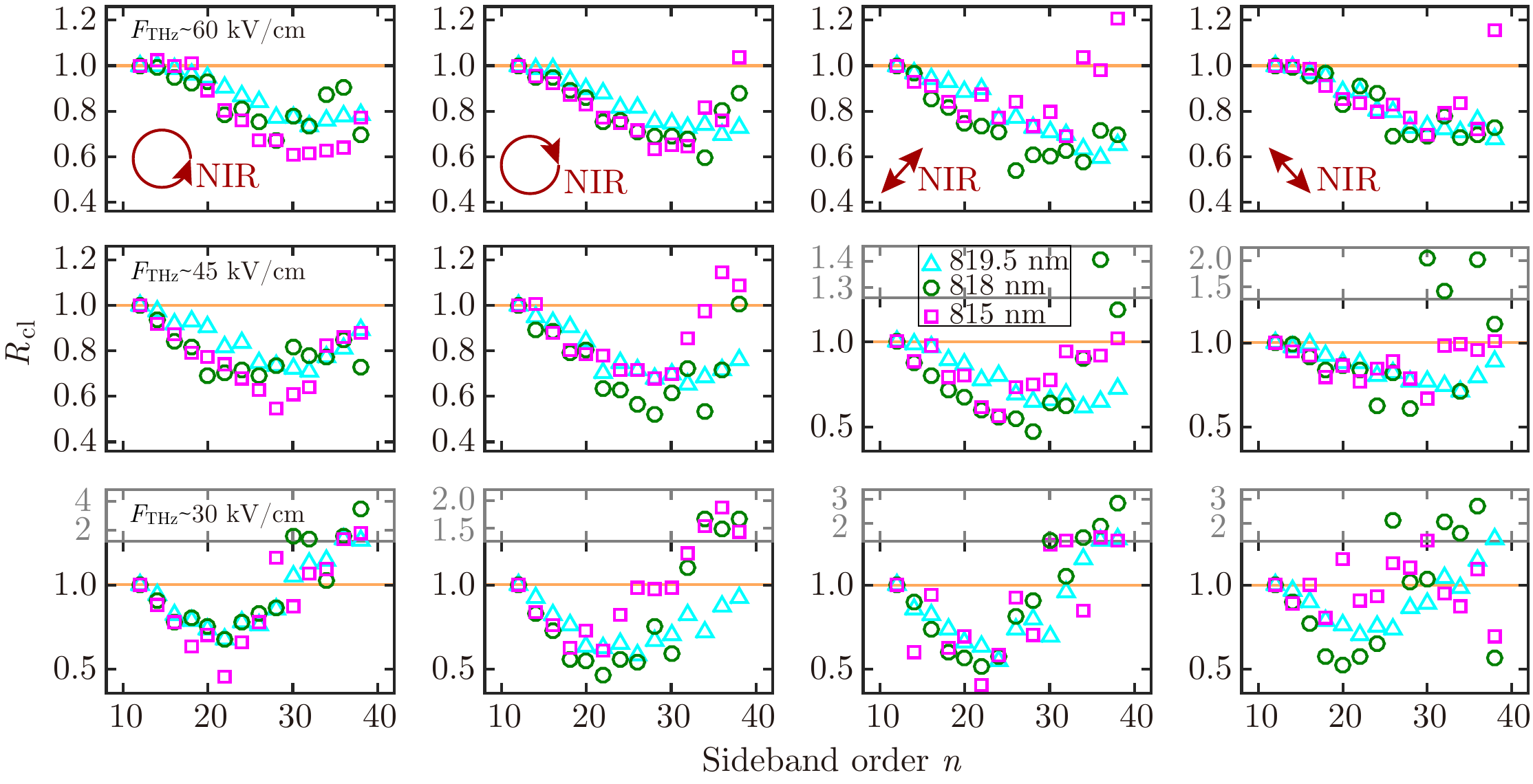}
	\caption{Comparison of the ratio $r_{\rm abs}\equiv|\varsigma^{\rm E-HH}/\varsigma^{\rm E-LH}|$ with the ratio of $r_{\rm E-HH}\equiv|\varsigma^{\rm E-HH}_{n}/\varsigma^{\rm E-HH}_{n_0=12}|$ to $r_{\rm E-LH}\equiv|\varsigma^{\rm E-LH}_{n}/\varsigma^{\rm E-LH}_{n_0=12}|$. The quantity $R_{\rm cl}\equiv (r_{\rm E-HH}/r_{\rm E-LH})/r_{\rm abs}^{1-(n_0/n)^{1/4}}$ is plotted. Each column shows the data for a specific NIR-laser polarization (from left to right in (a) or (b): left-handed circular polarization with helicity -1, right-handed circular polarization with helicity +1, linear polarization at $45^\circ$ to the $x$ axis, and linear polarization at $-45^\circ$ to the $x$ axis). The first, second, and third rows show the data collected at three different THz-field strength levels: around 60\,kV/cm, 45\,kV/cm, and 30\,kV/cm, respectively (see Fig.~\ref{FIG:thz_field_strength} in Appendix~\ref{APP:thz_field_strength} for the exact THz-field strengths). In each panel, cyan triangles, dark green circles, and magenta squares represent the data obtained at three different NIR-laser wavelengths: 819.5\,nm, 818\,nm, and 815\,nm, respectively. The orange lines highlight the value $R_{\rm cl}=1$.}
	\label{FIG:r_cl}
\end{figure*}
Without including the effects of quantum fluctuations, the analytic model given in Eq.~(\ref{EQ:propagator_analytic}) imposes a constraint relating the propagator decay described by the quantity $|{{\varsigma}^{\nu}_n}/{{\varsigma}^{\nu}_{n_0=12}}|$ to the absolute value of the propagator ratio ${{\varsigma}^{\rm E-HH}_n}/{{\varsigma}^{\rm E-LH}_n}$ [Eq.~(\ref{EQ:r_cl})]:
\begin{align}
\frac{|{{\varsigma}^{\rm E-HH}_n}/{{\varsigma}^{\rm E-HH}_{n_0}}|}{|{{\varsigma}^{\rm E-LH}_n}/{{\varsigma}^{\rm E-LH}_{n_0}}|}
=
|\frac{{\varsigma}^{\rm E-HH}_n}{{\varsigma}^{\rm E-LH}_n}|^{1-(n_0/n)^{1/4}},
\label{EQ:r_clapp}
\end{align}
which is inconsistent with the experimental data. Figure~\ref{FIG:r_cl} shows the experimentally determined ratio $R_{\rm cl}$ of the left-hand side to the right-hand side of Eq.~(\ref{EQ:r_clapp}). The analytic model given in Eq.~(\ref{EQ:propagator_analytic}) predicts $R_{\rm cl}=1$ for all sidebands. However, the values of $R_{\rm cl}$ calculated from the measured electron-hole propagators are not constant and can be as low as 0.5 for sideband orders around 20.

\section{Quantum kinetics of electron-phonon systems}\label{APP:frolich_thz}

We consider here HSG from free electron-hole pairs interacting with phonons. Under the dipole approximation, we write the effective Hamiltonian as $H=H_{\rm el}+H_{\rm ph}+H_{\rm el-ph}$, with the Hamiltonians of the electron, phonon, and electron-phonon interactions in the following generic forms:
\begin{align}
H_{\rm el}
=
\sum_{\lambda,\lambda',{\bf P}} 
&
h^{\lambda\lambda'}_{{\bf P}+\frac{e}{\hbar}{\bf A}(t)}
a^{\dag}_{\lambda,{\bf P}}a_{\lambda',{\bf P}},
\label{EQ:hamiltonian_el}
\\
H_{\rm ph}
=
\sum_{{\bf q},j}
&
\hbar\Omega_{{\bf q},j} (b_{{\bf q},j}^{\dag}b_{{\bf q},j}+\frac{1}{2}),
\label{EQ:hamiltonian_ph}
\\
H_{\rm el-ph}
=
\sum_{\lambda,\lambda',{\bf P}}
\sum_{{\bf q},j}
&
G^{{\bf q},j}_{\lambda\lambda',{\bf P}+\frac{e}{\hbar}{\bf A}(t)}
(b_{{\bf q},j}+b^{\dag}_{-{\bf q},j})
\notag\\
&
\times
a^{\dag}_{\lambda,{\bf P}}a_{\lambda',{\bf P}-{\bf q}},
\label{EQ:hamiltonian_elph}
\end{align} 
where ${\bf A}(t)$ is the vector potential of the total electromagnetic field, $a^{\dag}_{\lambda,\bf P}$ ($a_{\lambda,\bf P}$) creates (annihilates) an electron in the Bloch state $e^{i{\bf P}\cdot{\bf r}}u_{\lambda}({\bf r})$ defined by the canonical momentum $\hbar\bf P$ and the band-edge function $u_{\lambda}({\bf r})$ for the $\lambda$ band, $b^{\dag}_{{\bf q},j}$ ($b_{{\bf q},j}$) creates (annihilates) a phonon in the $j$-th branch with wavevector $\bf q$ and energy $\hbar\Omega_{{\bf q},j}$, $h^{\lambda\lambda'}_{{\bf P}+\frac{e}{\hbar}{\bf A}(t)}$ is a generic effective electron Hamiltonian, and $G^{{\bf q},j}_{\lambda\lambda',{\bf P}+\frac{e}{\hbar}{\bf A}(t)}$ is a generic electron-phonon coupling constant. Here, we ignore the Umklapp processes in the electron-phonon interaction.

Electromagnetic fields radiated from the electron-phonon system can be generally studied by calculating the electric current as the expectation value of the functional derivative of the Hamiltonian with respect to the vector potential ${\bf A}(t)$:
\begin{align}
{\bf J}
=
-\langle
\frac{\delta H}{\delta {\bf A}(t)}
\rangle,
\label{EQ:current_general}
\end{align} 
which is determined by the density matrix $\rho^{\lambda\lambda'}_{\bf P}\equiv\langle a^{\dag}_{\lambda',{\bf P}}a_{\lambda,{\bf P}}\rangle$. We will analyze the dynamics of the density matrix $\rho_{\bf P}$ by treating the operators $a^{\dag}_{\lambda',{\bf P}}$ and $a_{\lambda,{\bf P}}$ as time-dependent in the Heisenberg picture. For convenience, we denote ${\tilde a}_{\lambda,{{\bf k}(t)}}(t)\equiv a_{\lambda,{\bf P}}(t)$, with $\hbar{\bf k}(t)$ being the kinetic momentum $\hbar{\bf k}(t)=\hbar{\bf P}+e{\bf A}(t)$. Equivalently, one can define a polarization field ${\mathbb P}$ satisfying $\partial_t{\mathbb P}={\bf J}/V$:
\begin{align}
{\mathbb P}
=
\frac{-ie}{V}
\sum_{\lambda,\bf P}
\langle
a^{\dag}_{\lambda,{\bf P}}\partial_{\bf P}a_{\lambda,{\bf P}}
\rangle,
\label{EQ:polarization_general}
\end{align} 
where $V$ is the volume of the system. In HSG, only the interband polarization is relevant. The interband and intraband polarization components can be separated through a unitary transformation that diagonalizes the electron Hamiltonian $h_{{\bf k}}$: $U_{\bf k}h_{\bf k}U^{\dag}_{\bf k}=\Lambda_{\bf k}$, where $\Lambda^{\lambda\lambda'}_{\bf k}=\delta_{\lambda\lambda'}E_{\lambda,\bf k}$ is a diagonal matrix defined by the energy dispersion $E_{\lambda,\bf k}$ labeled by the band index $\lambda$. By using the unitary matrix $U_{\bf k}$, the operator ${\tilde a}_{\lambda,{{\bf k}}}$ is transformed into ${\tilde a}'_{\lambda,{{\bf k}}}\equiv\sum_{\lambda'}U^{\lambda\lambda'}_{\bf k}{\tilde a}_{\lambda',{{\bf k}}}$, which annihilates an electron in the Bloch state associated with the band energy $E_{\lambda,\bf k}$. Accordingly, the polarization ${\mathbb P}$ can be transformed into
\begin{align}
{\mathbb P}
=
\frac{-e}{V}
\sum_{\lambda,\bf k}
(
\langle
{\tilde a}^{\prime\dag}_{\lambda,{\bf k}}
i\partial_{\bf k}
{\tilde a}'_{\lambda,{\bf k}}
\rangle
+
\sum_{\lambda'}
{\bf R}^{\lambda\lambda'}_{\bf k}
\langle
{\tilde a}^{\prime\dag}_{\lambda,{\bf k}}
{\tilde a}'_{\lambda',{\bf k}}
\rangle
),
\label{EQ:polarization_decomposition}
\end{align} 
where ${\bf R}_{\bf k}\equiv iU_{\bf k}\partial_{\bf k}U^{\dag}_{\bf k}$ is a connection matrix. For a two-band semiconductor model, the connection matrix has the general form:
\begin{align}
{\bf R}_{\bf k}
=
\begin{pmatrix}
{\mathcal A}^{c}_{\bf k} & -{\bf d}^{cv}_{\bf k}/e\\
-{\bf d}^{cv *}_{\bf k}/e & {\mathcal A}^{v}_{\bf k}
\end{pmatrix},
\label{EQ:connection_matrix}
\end{align} 
which contains the Berry connection matrices ${\mathcal A}^{c}_{\bf k}$ and ${\mathcal A}^{v}_{\bf k}$ for the conduction band and valence band, respectively, and the dipole vector ${\bf d}^{cv}_{\bf k}$. We will ignore the intraband contribution and write the polarization as
\begin{align}
{\mathbb P}
=
\frac{1}{V}
\sum_{\lambda\ne\lambda',\bf k}
(-e
{\bf R}^{\lambda'\lambda}_{\bf k})
{\tilde\rho}^{\prime\lambda\lambda'}_{\bf k}.
\label{EQ:polarization_interband}
\end{align} 
which is determined by the transformed density matrix ${\tilde\rho}^{\prime\lambda\lambda'}_{\bf k}\equiv\langle{\tilde a}^{\prime\dag}_{\lambda',{\bf k}}{\tilde a}'_{\lambda,{\bf k}}\rangle$.

To calculate the interband polarization field ${\mathbb P}$, we consider the dynamics of the density matrix $\rho_{\bf P}$, which satisfies the equations of motion
\begin{align}
i\hbar\partial_t\rho^{\lambda\lambda'}_{\bf P}
=
\langle
[
a^{\dag}_{\lambda',{\bf P}}a_{\lambda,{\bf P}}
,
H_{\rm el}
+
H_{\rm el-ph}
]
\rangle.
\label{EQ:density_matrix}
\end{align} 
After some algebra, we obtain the following matrix equation:
\begin{align}
i\hbar\partial_t\rho_{\bf P}
&
=
[
h_{{\bf P}+\frac{e}{\hbar}{\bf A}(t)}
,\rho_{\bf P}]
\notag\\
&
+
\sum_{{\bf q},j}
G^{{\bf q},j}_{{\bf P}+\frac{e}{\hbar}{\bf A}(t)}
[
\Xi^{\dag}_{{\bf P},{\bf q},j}
+
\Xi_{{\bf P-q},-{\bf q},j}
]
\notag\\
&
-
\sum_{{\bf q},j}
[
\Xi_{{\bf P},{\bf q},j}
+
\Xi^{\dag}_{{\bf P-q},-{\bf q},j}
]
[G^{{\bf q},j}_{{\bf P}+\frac{e}{\hbar}{\bf A}(t)}]^{\dag},
\label{EQ:density_matrix_dynamics}
\end{align}
which contains the so-called phonon-assisted density matrix
\begin{align}
\Xi^{\lambda\lambda'}_{{\bf P},{\bf q},j}
\equiv
&
\langle
b^{\dag}_{{\bf q},j}
a^{\dag}_{\lambda',{\bf P}-{\bf q}}a_{\lambda,{\bf P}}
\rangle
\notag\\
&
-
\langle
b^{\dag}_{{\bf q},j}
\rangle
\langle
a^{\dag}_{\lambda',{\bf P}-{\bf q}}a_{\lambda,{\bf P}}
\rangle.
\label{EQ:phonon_density_matrix}
\end{align} 
Here, we assume that the system is homogeneous and that the phonons are in thermal states so that $\langle b^{\dag}_{{\bf q},j}\rangle=0$. The identity $(G^{-{\bf q},j }_{{\bf P}-{\bf q}})^{\dag}=G^{{\bf q},j}_{\bf P}$ from the hermiticity of the Hamiltonian has been used. By considering the Heisenberg equations of motion for the operators in $\Xi^{\lambda\lambda'}_{{\bf P},{\bf q},j}$, we obtain the following matrix equation:
\begin{align}
i\hbar\partial_t
\Xi_{{\bf P},{\bf q},j}
=
&
(-
\hbar\Omega_{{\bf q},j}-i\Gamma_{\rm e-ph})
\Xi_{{\bf P},{\bf q},j}
\notag\\
&
+
h_{{\bf P}+\frac{e}{\hbar}{\bf A}(t)}
\Xi_{{\bf P},{\bf q},j}
-
\Xi_{{\bf P},{\bf q},j}
h_{{\bf P}-{\bf q}+\frac{e}{\hbar}{\bf A}(t)}
\notag\\
&
+
N_{{\bf q},j}
(
G^{{\bf q},j}_{{\bf P}+\frac{e}{\hbar}{\bf A}(t)}
\rho_{{\bf P}-{\bf q}}
-
\rho_{{\bf P}}
G^{{\bf q},j}_{{\bf P}+\frac{e}{\hbar}{\bf A}(t)}
)
\notag\\
&
-
\rho_{\bf P}
G^{{\bf q},j}_{{\bf P}+\frac{e}{\hbar}{\bf A}(t)}
(
{\bf 1}
-
\rho_{\bf P-q}
),
\label{EQ:phonon_density_matrix_dyanmics}
\end{align}
where we have introduced a phenomenological dephasing constant $\Gamma_{\rm e-ph}$ to account for the effects from the four-point correlation terms such as $\langle a^{\dag}_{\lambda'',{\bf P}''}a_{\lambda''',{\bf P}''-{\bf q}}a^{\dag}_{\lambda',{\bf P}-{\bf q}}a_{\lambda,{\bf P}}\rangle$ and truncated the dynamic equations into a closed set~\cite{schilp1994electron}. By using the unitary matrix $U_{\bf k}$, Eq.~(\ref{EQ:density_matrix_dynamics}) can be transformed into an equation governing the dynamics of the density matrix ${\tilde\rho}^{\prime}_{\bf k}$:
\begin{align}
i\hbar\partial_t{\tilde \rho}'_{\bf k}
-
&
ie{\bf E}(t)
\cdot
\partial_{\bf k}{\tilde\rho}'_{\bf k}
=
[
\Lambda_{\bf k}
+e{\bf E}(t)\cdot{\bf R}_{\bf k}
,
{\tilde\rho}'_{\bf k}
]
\notag\\
&
+
\sum_{{\bf q},j}
G^{\prime{\bf q},j}_{\bf k}
[
{\tilde\Xi}^{\prime\dag}_{{\bf k},{\bf q},j}
+
{\tilde\Xi}'_{{\bf k-q},-{\bf q},j}
]
\notag\\
&
-
\sum_{{\bf q},j}
[
{\tilde\Xi}'_{{\bf k},{\bf q},j}
+
{\tilde\Xi}^{\prime\dag}_{{\bf k-q},-{\bf q},j}
]
(G^{{\prime\bf q},j }_{\bf k})^{\dag},
\label{EQ:density_matrix_dynamics2}
\end{align}
where the electron-hole coupling constant $G^{{\bf q},j}_{\lambda\lambda',\bf k}$ is transformed into $G^{\prime{\bf q},j}_{\bf k}\equiv U_{\bf k}G^{{\bf q},j}_{\bf k}U^{\dag}_{\bf k-q}$. The transformed phonon-assisted density matrix ${\tilde\Xi}^{\prime\lambda\lambda'}_{{\bf k},{\bf q},j}=\langle b^{\dag}_{{\bf q},j}{\tilde a}^{\dag}_{\lambda',{\bf k}-{\bf q}}{\tilde a}_{\lambda,{\bf k}}\rangle-\langle b^{\dag}_{{\bf q},j}\rangle\langle{\tilde a}^{\dag}_{\lambda',{\bf k}-{\bf q}}{\tilde a}_{\lambda,{\bf k}}\rangle$ satisfies the following equation from Eq.~(\ref{EQ:phonon_density_matrix_dyanmics}):
\begin{align}
&
i\hbar\partial_t
{\tilde\Xi}^{\prime}_{{\bf k},{\bf q},j}
-
ie{\bf E}(t)
\cdot
\partial_{\bf k}{\tilde\Xi}^{\prime}_{{\bf k},{\bf q},j}
\notag\\
=
&
e{\bf E}(t)
\cdot
({\bf R}_{\bf k}{\tilde\Xi}'_{{\bf k},{\bf q},j}
-
{\tilde\Xi}'_{{\bf k},{\bf q},j}{\bf R}_{\bf k-q})
\notag\\
&
+
\Lambda_{\bf k}{\tilde\Xi}'_{{\bf k},{\bf q},j}
-
{\tilde\Xi}'_{{\bf k},{\bf q},j}
\Lambda_{\bf k-q}
\notag\\
&
+
(-
\hbar\Omega_{{\bf q},j}
-i\Gamma_{\rm e-ph}
)
{\tilde\Xi}'_{{\bf k},{\bf q},j}
\notag\\
&
+
N_{{\bf q},j}
(
G^{\prime{\bf q},j}_{\bf k}
{\tilde\rho}^{\prime}_{{\bf k}-{\bf q}}
-
{\tilde\rho}^{\prime}_{{\bf k}}
G^{\prime{\bf q},j}_{\bf k}
)
\notag\\
&
-
{\tilde\rho}^{\prime}_{\bf k}
G^{\prime{\bf q},j}_{\bf k}
(
{\bf 1}
-
{\tilde\rho}^{\prime}_{\bf k-q}
).
\label{EQ:phonon_density_matrix_dyanmics2}
\end{align}

In HSG from an insulator, because of the relatively weak NIR excitation, one can ignore the carrier occupations in the conduction bands and take ${\tilde \rho}^{\prime \lambda\lambda}_{\bf k}=1$ ( ${\tilde \rho}^{\prime \lambda\lambda}_{\bf k}=0$) for the valence (conduction) band. In this case, from Eq.~(\ref{EQ:density_matrix_dynamics2}), for a parabolic two-band model with a constant dipole vector ${\bf d}^{cv}$ and negligible Berry curvature effects under a linearly polarized THz field, the density matrix element ${\tilde \rho}^{\prime cv}_{\bf k}$ that determines the interband polarization $\mathbb P$ satisfies the following equation:
\begin{align}
&
i\hbar\partial_t{\tilde \rho}^{\prime cv}_{\bf k}
-
ie{\bf E}_{\rm THz}(t)
\cdot
\partial_{\bf k}{\tilde\rho}^{\prime cv}_{\bf k}
\notag\\
&
=
(E_{c,\bf k}-E_{v,\bf k})
{\tilde\rho}^{\prime cv}_{\bf k}
-{\bf d}^{cv}\cdot{\bf E}_{\rm NIR}e^{-i\omega_{\rm NIR}t}
\notag\\
&
+
\sum_{{\bf q},j}
G^{\prime{\bf q},j}_{cc,\bf k}
[
{\tilde\Xi}^{\prime vc *}_{{\bf k},{\bf q},j}
+
{\tilde\Xi}^{\prime cv}_{{\bf k-q},-{\bf q},j}
]
\notag\\
&
-
\sum_{{\bf q},j}
[
{\tilde\Xi}^{\prime cv}_{{\bf k},{\bf q},j}
+
{\tilde\Xi}^{\prime vc *}_{{\bf k-q},-{\bf q},j}
]
G^{{\prime\bf q},j *}_{vv,\bf k},
\label{EQ:twoband_density_matrix_dyanmics2}
\end{align}
where the phonon-assisted density matrix elements ${\tilde\Xi}^{\prime vc *}_{{\bf k},{\bf q},j}$ and ${\tilde\Xi}^{\prime cv}_{{\bf k-q},-{\bf q},j}$ (${\tilde\Xi}^{\prime cv}_{{\bf k},{\bf q},j}$ and ${\tilde\Xi}^{\prime vc *}_{{\bf k-q},-{\bf q},j}$) describe a phonon emission process and a phonon absorption process in the conduction (valence) band, respectively. We have ignored the NIR-laser field in the carrier acceleration described by the term containing the $k$-gradient of ${\tilde\rho}^{\prime cv}_{\bf k}$. For the initial creation of electron-hole pairs, we include only the NIR-laser field ${\bf E}_{\rm NIR}\exp{(-i\omega_{\rm NIR}t)}$ under the rotating-wave approximation in the coupling between the dipole vector ${\bf d}^{cv}$ and the electric field. For the electron-phonon coupling constant $G^{{\prime\bf q},j *}_{\lambda\lambda',\bf k}$, the off-diagonal elements are assumed to be negligible. Explicitly, the phonon-assisted density matrix elements in Eq.~(\ref{EQ:twoband_density_matrix_dyanmics2}) satisfy the following equations:
\begin{align}
&
i\hbar\partial_t
{\tilde\Xi}^{\prime vc *}_{{\bf k},{\bf q},j}
-
ie{\bf E}_{\rm THz}(t)
\cdot
\partial_{\bf k}{\tilde\Xi}^{\prime vc *}_{{\bf k},{\bf q},j}
\notag\\
=
&
(E_{c,\bf k-q}-E_{v,\bf k}+\hbar\Omega_{{\bf q},j}-i\Gamma_{\rm e-ph}){\tilde\Xi}^{\prime vc *}_{{\bf k},{\bf q},j}
\notag\\
&
-
(N_{{\bf q},j}+1)
(
G^{\prime{\bf q},j *}_{vv,\bf k}
{\tilde\rho}^{\prime cv}_{{\bf k}-{\bf q}}
-
{\tilde\rho}^{\prime cv}_{{\bf k}}
G^{\prime{\bf q},j *}_{cc,\bf k}
),
\label{EQ:twoband_xi_dyanmics1}
\\
&
i\hbar\partial_t
{\tilde\Xi}^{\prime cv}_{{\bf k-q},{-\bf q},j}
-
ie{\bf E}_{\rm THz}(t)
\cdot
\partial_{\bf k}{\tilde\Xi}^{\prime cv}_{{\bf k-q},{-\bf q},j}
\notag\\
=
&
(E_{c,\bf k-q}-E_{v,\bf k}-\hbar\Omega_{{\bf q},j}-i\Gamma_{\rm e-ph}
)
{\tilde\Xi}^{\prime cv}_{{\bf k-q},-{\bf q},j}
\notag\\
&
+
N_{{\bf q},j}
(
G^{\prime{\bf q},j *}_{cc,\bf k}
{\tilde\rho}^{\prime cv}_{\bf k}
-
{\tilde\rho}^{\prime cv}_{{\bf k}-{\bf q}}
G^{\prime{\bf q},j*}_{vv,\bf k}
),
\label{EQ:twoband_xi_dyanmics2}
\\
&
i\hbar\partial_t
{\tilde\Xi}^{\prime cv}_{{\bf k},{\bf q},j}
-
ie{\bf E}_{\rm THz}(t)
\cdot
\partial_{\bf k}{\tilde\Xi}^{\prime cv}_{{\bf k},{\bf q},j}
\notag\\
=
&
(E_{c,\bf k}-E_{v,\bf k-q}-\hbar\Omega_{{\bf q},j}-i\Gamma_{\rm e-ph}
)
{\tilde\Xi}^{\prime cv}_{{\bf k},{\bf q},j}
\notag\\
&
+
N_{{\bf q},j}
(
G^{\prime{\bf q},j}_{cc,\bf k}
{\tilde\rho}^{\prime cv}_{{\bf k}-{\bf q}}
-
{\tilde\rho}^{\prime cv}_{{\bf k}}
G^{\prime{\bf q},j}_{vv,\bf k}
),
\label{EQ:twoband_xi_dyanmics3}
\\
&i\hbar\partial_t
{\tilde\Xi}^{\prime vc *}_{{\bf k-q},-{\bf q},j}
-
ie{\bf E}_{\rm THz}(t)
\cdot
\partial_{\bf k}{\tilde\Xi}^{\prime vc *}_{{\bf k-q},-{\bf q},j}
\notag\\
=
&
(E_{c,\bf k}-E_{v,\bf k-q}+\hbar\Omega_{{\bf q},j}-i\Gamma_{\rm e-ph}){\tilde\Xi}^{\prime vc *}_{{\bf k-q},-{\bf q},j}
\notag\\
&
-
(N_{{\bf q},j}+1)
(
G^{\prime{\bf q},j }_{vv,\bf k}
{\tilde\rho}^{\prime cv}_{\bf k}
-
{\tilde\rho}^{\prime cv}_{{\bf k-q}}
G^{\prime{\bf q},j}_{cc,\bf k}
),
\label{EQ:twoband_xi_dyanmics4}
\end{align}
where the Berry connection is ignored and only the THz electric field ${\bf E}_{\rm THz}(t)$ is included to describe the carrier acceleration. Assuming that the coupling constant $G^{\prime{\bf q},j}_{\lambda\lambda,{\bf k}(t)}$ ($\lambda=c,v$) and the quantity ${\tilde\rho}^{\prime cv}_{{\bf k}(t)}(t)\exp\{\frac{i}{\hbar}\int_{-\infty}^{t}dt''[E_{c,{\bf k}(t'')}-E_{v,{\bf k}(t'')}]\}$ are slowly varying, we make the Markovian approximation and convert the terms in the third and fourth lines of Eq.~(\ref {EQ:twoband_density_matrix_dyanmics2}) into terms proportional the the density-matrix elements ${\tilde\rho}^{\prime cv}_{{\bf k}}$ and ${\tilde\rho}^{\prime cv}_{{\bf k}-{\bf q}}$:
\begin{align}
&
\sum_{{\bf q},j}
G^{\prime{\bf q},j}_{cc,\bf k}
[
{\tilde\Xi}^{\prime vc *}_{{\bf k},{\bf q},j}
+
{\tilde\Xi}^{\prime cv}_{{\bf k-q},-{\bf q},j}
]
\notag\\
&
-
\sum_{{\bf q},j}
[
{\tilde\Xi}^{\prime cv}_{{\bf k},{\bf q},j}
+
{\tilde\Xi}^{\prime vc *}_{{\bf k-q},-{\bf q},j}
]
(G^{{\prime\bf q},j *}_{vv,\bf k}),
\notag\\
\rightarrow
&
-i[{\mathcal Q}_{\bf P}(t)
{\tilde\rho}^{\prime cv}_{{\bf k}}
+
\sum_{{\bf q}} 
{\mathcal W}'_{\bf P,q}(t)
{\tilde\rho}^{\prime cv}_{{\bf k}-{\bf q}}].
\label{EQ:markovian_app}
\end{align}
Here, we have introduced the coefficients ${\mathcal Q}_{\bf P}(t)$ and ${\mathcal W}'_{\bf P,q}(t)$, which are defined as
\begin{widetext}
\begin{align}
&
{\mathcal Q}_{\bf P}(t)
\equiv
\sum_{{\bf q},j} 
|G^{\prime{\bf q},j }_{cc,\bf k}|^2
[
(N_{{\bf q},j}+1)
{\mathcal G}^{c,(+)}_{{\bf P},{\bf q}}(t)
+
N_{{\bf q},j}
{\mathcal G}^{c,(-)}_{{\bf P},{\bf q}}(t)]
+
|G^{\prime{\bf q},j}_{vv,\bf k}|^2
[(N_{{\bf q},j}+1)
{\mathcal G}^{v,(-)}_{{\bf P-q},-{\bf q}}(t)
+
N_{{\bf q},j}
{\mathcal G}^{v,(-)}_{{\bf P-q},-{\bf q}}(t)],
\label{EQ:constructive_q}
\\
&
{\mathcal W}'_{\bf P,q}(t)
\equiv
\sum_{j} 
G^{\prime{\bf q},j }_{cc,\bf k}G^{\prime{\bf q},j *}_{vv,\bf k}
\{
(N_{{\bf q},j}+1)
[{\mathcal G}^{c,(+)}_{{\bf P-q},-{\bf q}}(t)
-
{\mathcal G}^{v,(+)}_{{\bf P},{\bf q}}(t)]
+
N_{{\bf q},j}
[{\mathcal G}^{c,(-)}_{{\bf P-q},-{\bf q}}(t)
-
{\mathcal G}^{v,(-)}_{{\bf P},{\bf q}}(t)]
\},
\label{EQ:destructive_w}
\end{align}
\end{widetext}
where
\begin{align}
{\mathcal G}^{\lambda, (\eta)}_{{\bf P},{\bf q}}(t)
=
&
\frac{1}{\hbar}
\int_{-\infty}^t
dt'
\exp
\{
-\frac{i}{\hbar}
\int_{t'}^{t}
dt''
[
E_{\lambda,{\bf k}(t'')-{\bf q}}
\notag\\
&
-E_{\lambda,{\bf k}(t'')}
+
\eta
\hbar\Omega_{{\bf q},j}
-i\Gamma_{\rm e-ph}
]
\},
\label{EQ:factor_g}
\end{align}
with $\eta=\pm1$. Following Ref.~\cite{renbao2002electron}, we assume that the summation over $\bf q$ in the third line of Eq.~(\ref{EQ:markovian_app}) results in a negligible contribution. Thus Eq.~(\ref{EQ:twoband_density_matrix_dyanmics2}) reduces to
\begin{align}
&
i\hbar\partial_t{\tilde \rho}^{\prime cv}_{\bf k}
-
ie{\bf E}_{\rm THz}(t)
\cdot
\partial_{\bf k}{\tilde\rho}^{\prime cv}_{\bf k}
\notag\\
=
&
[E_{c,\bf k}-E_{v,\bf k}-i{\mathcal Q}_{\bf P}(t)]
{\tilde\rho}^{\prime cv}_{\bf k}
\notag\\
&
-{\bf d}^{cv}\cdot{\bf E}_{\rm NIR}e^{-i\omega_{\rm NIR}t}.
\label{EQ:twoband_density_matrix_dyanmics3}
\end{align}
We can see that the real part of ${\mathcal Q}_{\bf P}(t)$ describes dephasing, while the imaginary part of ${\mathcal Q}_{\bf P}(t)$ describes renormalization of the electron-hole energy. Consider electrons and holes driven by a linearly polarized THz field ${\bf E}_{\rm THz}(t)=\hat{x}F_{\rm THz}\cos(\omega_{\rm THz}t)$ in parabolic bands with effective masses $m_c$ and $m_v$, respectively. By using the identity with the Bessel functions of the first kind, $J_n$, $\exp(iz\cos x)=\sum_nJ_n(z)i^n \exp{(inx)}$, Eq.~(\ref{EQ:factor_g}) has the explicit forms:
\begin{align}
{\mathcal G}^{c, (\eta)}_{{\bf P},{\bf q}}(t)
=
\sum_{n,n',{\bf q},j}
i
e^{i(n-n')\omega t}
i^{n-n'}
\notag\\
\times
\frac{
J_n(
\frac{eF_{\rm THz}q_x}{m_{c}\omega_{\rm THz}^2}
)
J_{n'}(
\frac{eF_{\rm THz}q_x}{m_{c}\omega_{\rm THz}^2}
)
}
{E_{c,{\bf P}}
-
E_{c,{\bf P}-{\bf q}}
-\eta\hbar\Omega_{{\bf q},j}
+n'\hbar\omega_{\rm THz}
+i\Gamma_{\rm e-ph}},
\label{EQ:factor_g2c}
\\
{\mathcal G}^{v, (\eta)}_{{\bf P-q},-{\bf q}}(t)
=
\sum_{n,n',{\bf q},j}
i
e^{i(n-n')\omega t}
i^{n-n'}
\notag\\
\times
\frac{
J_n(
\frac{eF_{\rm THz}q_x}{m_{c}\omega_{\rm THz}^2}
)
J_{n'}(
\frac{eF_{\rm THz}q_x}{m_{c}\omega_{\rm THz}^2}
)
}
{E_{v,{\bf P-q}}
-
E_{v,{\bf P}}
-\eta\hbar\Omega_{{\bf q},j}
+n'\hbar\omega_{\rm THz}
+i\Gamma_{\rm e-ph}},
\label{EQ:factor_g2v}
\end{align}
where $E_{c,{\bf P}}=\hbar^2P^2/(2m_c)$ and $E_{v,{\bf P}}=-\hbar^2P^2/(2m_v)$ are the energy dispersion relations of the conduction and valence bands, respectively. By ignoring the time-dependent oscillating terms in ${\mathcal G}^{\lambda, (\eta)}_{{\bf P},{\bf q}}(t)$ and assuming that the coupling constant $G^{\prime{\bf q},j}_{\lambda\lambda,{\bf k}(t)}$ does not depend on $\bf k$ ($G^{\prime{\bf q},j}_{\lambda\lambda,{\bf k}(t)}\rightarrow G^{\prime{\bf q},j}_{\lambda\lambda}$), the function ${\mathcal Q}_{\bf P}(t)$ becomes a time-independent factor:
\begin{widetext}
\begin{align}
{\mathcal Q}_{\bf P}
&\approx
\sum_{n,{\bf q},j}
i
(N_{{\bf q},j}+1)
[
\frac{
|J_n(
\frac{eF_{\rm THz}q_x}{m_{c}\omega_{\rm THz}^2}
)|^2
|G^{\prime{\bf q},j}_{cc}|^2}
{E_{c,{\bf P}}
-
E_{c,{\bf P}-{\bf q}}
-\hbar\Omega_{{\bf q},j}
+n\hbar\omega_{\rm THz}
+i\Gamma_{\rm e-ph}}
+
\frac{
|J_n(
\frac{eF_{\rm THz}q_x}{m_{v}\omega_{\rm THz}^2}
)|^2
|G^{\prime{\bf q},j}_{vv}|^2}
{E_{v,{\bf P}-{\bf q}}
-
E_{v,{\bf P}}
-\hbar\Omega_{{\bf q},j}
+n\hbar\omega_{\rm THz}+i\Gamma_{\rm e-ph}}
]
\notag\\
&
+
\sum_{n,{\bf q},j}
i
N_{{\bf q},j}
[
\frac{
|J_n(
\frac{eF_{\rm THz}q_x}{m_{c}\omega_{\rm THz}^2}
)|^2
|G^{\prime{\bf q},j}_{cc}|^2}
{E_{c,{\bf P}}
-
E_{c,{\bf P}-{\bf q}}
+\hbar\Omega_{{\bf q},j}
+n\hbar\omega_{\rm THz}
+i\Gamma_{\rm e-ph}}
+
\frac{
|J_n(
\frac{eF_{\rm THz}q_x}{m_{v}\omega_{\rm THz}^2}
)|^2
|G^{\prime{\bf q},j}_{vv}|^2}
{E_{v,{\bf P}-{\bf q}}
-
E_{v,{\bf P}}
+\hbar\Omega_{{\bf q},j}
+n\hbar\omega_{\rm THz}
+i\Gamma_{\rm e-ph}}
].
\label{EQ:constant_Q}
\end{align}
\end{widetext}
Note that the real part of ${\mathcal Q}_{\bf P}$ is positive, corresponding to dephasing of the electron-hole coherences. In the limit of zero THz field and small $\Gamma_{\rm e-ph}$, the real part of Eq.~(\ref{EQ:constant_Q}) reduces to the result given by Fermi's golden rule.


\begin{thebibliography}{75}%
\makeatletter
\providecommand \@ifxundefined [1]{%
 \@ifx{#1\undefined}
}%
\providecommand \@ifnum [1]{%
 \ifnum #1\expandafter \@firstoftwo
 \else \expandafter \@secondoftwo
 \fi
}%
\providecommand \@ifx [1]{%
 \ifx #1\expandafter \@firstoftwo
 \else \expandafter \@secondoftwo
 \fi
}%
\providecommand \natexlab [1]{#1}%
\providecommand \enquote  [1]{``#1''}%
\providecommand \bibnamefont  [1]{#1}%
\providecommand \bibfnamefont [1]{#1}%
\providecommand \citenamefont [1]{#1}%
\providecommand \href@noop [0]{\@secondoftwo}%
\providecommand \href [0]{\begingroup \@sanitize@url \@href}%
\providecommand \@href[1]{\@@startlink{#1}\@@href}%
\providecommand \@@href[1]{\endgroup#1\@@endlink}%
\providecommand \@sanitize@url [0]{\catcode `\\12\catcode `\$12\catcode
  `\&12\catcode `\#12\catcode `\^12\catcode `\_12\catcode `\%12\relax}%
\providecommand \@@startlink[1]{}%
\providecommand \@@endlink[0]{}%
\providecommand \url  [0]{\begingroup\@sanitize@url \@url }%
\providecommand \@url [1]{\endgroup\@href {#1}{\urlprefix }}%
\providecommand \urlprefix  [0]{URL }%
\providecommand \Eprint [0]{\href }%
\providecommand \doibase [0]{https://doi.org/}%
\providecommand \selectlanguage [0]{\@gobble}%
\providecommand \bibinfo  [0]{\@secondoftwo}%
\providecommand \bibfield  [0]{\@secondoftwo}%
\providecommand \translation [1]{[#1]}%
\providecommand \BibitemOpen [0]{}%
\providecommand \bibitemStop [0]{}%
\providecommand \bibitemNoStop [0]{.\EOS\space}%
\providecommand \EOS [0]{\spacefactor3000\relax}%
\providecommand \BibitemShut  [1]{\csname bibitem#1\endcsname}%
\let\auto@bib@innerbib\@empty
\bibitem [{\citenamefont {Laughlin}\ and\ \citenamefont
  {Pines}(2000)}]{laughlin2000theory}%
  \BibitemOpen
  \bibfield  {author} {\bibinfo {author} {\bibfnamefont {R.~B.}\ \bibnamefont
  {Laughlin}}\ and\ \bibinfo {author} {\bibfnamefont {D.}~\bibnamefont
  {Pines}},\ }\bibfield  {title} {\bibinfo {title} {The theory of everything},\
  }\href@noop {} {\bibfield  {journal} {\bibinfo  {journal} {Proceedings of the
  national academy of sciences}\ }\textbf {\bibinfo {volume} {97}},\ \bibinfo
  {pages} {28} (\bibinfo {year} {2000})}\BibitemShut {NoStop}%
\bibitem [{\citenamefont {Powell}(2009)}]{powell2009introduction}%
  \BibitemOpen
  \bibfield  {author} {\bibinfo {author} {\bibfnamefont {B.~J.}\ \bibnamefont
  {Powell}},\ }\bibfield  {title} {\bibinfo {title} {An introduction to
  effective low-energy {H}amiltonians in condensed matter physics and
  chemistry},\ }\href@noop {} {\bibfield  {journal} {\bibinfo  {journal}
  {arXiv:0906.1640}\ } (\bibinfo {year} {2009})}\BibitemShut {NoStop}%
\bibitem [{\citenamefont {Ashcroft}\ and\ \citenamefont
  {Mermin}(1976)}]{ashcroft1976solid}%
  \BibitemOpen
  \bibfield  {author} {\bibinfo {author} {\bibfnamefont {N.~W.}\ \bibnamefont
  {Ashcroft}}\ and\ \bibinfo {author} {\bibfnamefont {N.~D.}\ \bibnamefont
  {Mermin}},\ }\href@noop {} {\emph {\bibinfo {title} {Solid State Physics}}}\
  (\bibinfo  {publisher} {Saunders College Publishing},\ \bibinfo {year}
  {1976})\BibitemShut {NoStop}%
\bibitem [{\citenamefont {Born}\ and\ \citenamefont
  {Oppenheimer}(1927)}]{born1927zur}%
  \BibitemOpen
  \bibfield  {author} {\bibinfo {author} {\bibfnamefont {M.}~\bibnamefont
  {Born}}\ and\ \bibinfo {author} {\bibfnamefont {R.}~\bibnamefont
  {Oppenheimer}},\ }\bibfield  {title} {\bibinfo {title} {Zur quantentheorie
  der molekeln},\ }\href@noop {} {\bibfield  {journal} {\bibinfo  {journal}
  {Annalen der Physik}\ }\textbf {\bibinfo {volume} {389}},\ \bibinfo {pages}
  {457} (\bibinfo {year} {1927})}\BibitemShut {NoStop}%
\bibitem [{\citenamefont {Cohen}\ and\ \citenamefont
  {Heine}(1970)}]{cohen1970fitting}%
  \BibitemOpen
  \bibfield  {author} {\bibinfo {author} {\bibfnamefont {M.~L.}\ \bibnamefont
  {Cohen}}\ and\ \bibinfo {author} {\bibfnamefont {V.}~\bibnamefont {Heine}},\
  }\bibfield  {title} {\bibinfo {title} {The fitting of pseudopotentials to
  experimental data and their subsequent application},\ }in\ \href@noop {}
  {\emph {\bibinfo {booktitle} {Solid state physics}}},\ Vol.~\bibinfo {volume}
  {24}\ (\bibinfo  {publisher} {Elsevier},\ \bibinfo {year} {1970})\ pp.\
  \bibinfo {pages} {37--248}\BibitemShut {NoStop}%
\bibitem [{\citenamefont {Willatzen}\ and\ \citenamefont
  {Voon}(2009)}]{willatzen2009kp}%
  \BibitemOpen
  \bibfield  {author} {\bibinfo {author} {\bibfnamefont {M.}~\bibnamefont
  {Willatzen}}\ and\ \bibinfo {author} {\bibfnamefont {L.~C. L.~Y.}\
  \bibnamefont {Voon}},\ }\href@noop {} {\emph {\bibinfo {title} {The $k\cdot
  p$ method: electronic properties of semiconductors}}}\ (\bibinfo  {publisher}
  {Springer},\ \bibinfo {year} {2009})\BibitemShut {NoStop}%
\bibitem [{\citenamefont {Hubbard}(1963)}]{hubbard1963electron}%
  \BibitemOpen
  \bibfield  {author} {\bibinfo {author} {\bibfnamefont {J.}~\bibnamefont
  {Hubbard}},\ }\bibfield  {title} {\bibinfo {title} {Electron correlations in
  narrow energy bands},\ }\href@noop {} {\bibfield  {journal} {\bibinfo
  {journal} {Proceedings of the Royal Society of London. Series A. Mathematical
  and Physical Sciences}\ }\textbf {\bibinfo {volume} {276}},\ \bibinfo {pages}
  {238} (\bibinfo {year} {1963})}\BibitemShut {NoStop}%
\bibitem [{\citenamefont {Arovas}\ \emph {et~al.}(2022)\citenamefont {Arovas},
  \citenamefont {Berg}, \citenamefont {Kivelson},\ and\ \citenamefont
  {Raghu}}]{arovas2022hubbard}%
  \BibitemOpen
  \bibfield  {author} {\bibinfo {author} {\bibfnamefont {D.~P.}\ \bibnamefont
  {Arovas}}, \bibinfo {author} {\bibfnamefont {E.}~\bibnamefont {Berg}},
  \bibinfo {author} {\bibfnamefont {S.~A.}\ \bibnamefont {Kivelson}},\ and\
  \bibinfo {author} {\bibfnamefont {S.}~\bibnamefont {Raghu}},\ }\bibfield
  {title} {\bibinfo {title} {The {H}ubbard model},\ }\href@noop {} {\bibfield
  {journal} {\bibinfo  {journal} {Annual review of condensed matter physics}\
  }\textbf {\bibinfo {volume} {13}},\ \bibinfo {pages} {239} (\bibinfo {year}
  {2022})}\BibitemShut {NoStop}%
\bibitem [{\citenamefont {Mott}(1968)}]{mott1968metal}%
  \BibitemOpen
  \bibfield  {author} {\bibinfo {author} {\bibfnamefont {N.~F.}\ \bibnamefont
  {Mott}},\ }\bibfield  {title} {\bibinfo {title} {Metal-insulator
  transition},\ }\href@noop {} {\bibfield  {journal} {\bibinfo  {journal} {Rev.
  Mod. Phys.}\ }\textbf {\bibinfo {volume} {40}},\ \bibinfo {pages} {677}
  (\bibinfo {year} {1968})}\BibitemShut {NoStop}%
\bibitem [{\citenamefont {Imada}\ \emph {et~al.}(1998)\citenamefont {Imada},
  \citenamefont {Fujimori},\ and\ \citenamefont {Tokura}}]{imada1998metal}%
  \BibitemOpen
  \bibfield  {author} {\bibinfo {author} {\bibfnamefont {M.}~\bibnamefont
  {Imada}}, \bibinfo {author} {\bibfnamefont {A.}~\bibnamefont {Fujimori}},\
  and\ \bibinfo {author} {\bibfnamefont {Y.}~\bibnamefont {Tokura}},\
  }\bibfield  {title} {\bibinfo {title} {Metal-insulator transitions},\
  }\href@noop {} {\bibfield  {journal} {\bibinfo  {journal} {Rev. Mod. Phys.}\
  }\textbf {\bibinfo {volume} {70}},\ \bibinfo {pages} {1039} (\bibinfo {year}
  {1998})}\BibitemShut {NoStop}%
\bibitem [{\citenamefont {Lee}\ \emph {et~al.}(2006)\citenamefont {Lee},
  \citenamefont {Nagaosa},\ and\ \citenamefont {Wen}}]{lee2006doping}%
  \BibitemOpen
  \bibfield  {author} {\bibinfo {author} {\bibfnamefont {P.~A.}\ \bibnamefont
  {Lee}}, \bibinfo {author} {\bibfnamefont {N.}~\bibnamefont {Nagaosa}},\ and\
  \bibinfo {author} {\bibfnamefont {X.-G.}\ \bibnamefont {Wen}},\ }\bibfield
  {title} {\bibinfo {title} {Doping a {M}ott insulator: {P}hysics of
  high-temperature superconductivity},\ }\href@noop {} {\bibfield  {journal}
  {\bibinfo  {journal} {Rev. Mod. Phys.}\ }\textbf {\bibinfo {volume} {78}},\
  \bibinfo {pages} {17} (\bibinfo {year} {2006})}\BibitemShut {NoStop}%
\bibitem [{\citenamefont {Fradkin}\ \emph {et~al.}(2015)\citenamefont
  {Fradkin}, \citenamefont {Kivelson},\ and\ \citenamefont
  {Tranquada}}]{fradkin2015colloquium}%
  \BibitemOpen
  \bibfield  {author} {\bibinfo {author} {\bibfnamefont {E.}~\bibnamefont
  {Fradkin}}, \bibinfo {author} {\bibfnamefont {S.~A.}\ \bibnamefont
  {Kivelson}},\ and\ \bibinfo {author} {\bibfnamefont {J.~M.}\ \bibnamefont
  {Tranquada}},\ }\bibfield  {title} {\bibinfo {title} {Colloquium: {T}heory of
  intertwined orders in high temperature superconductors},\ }\href@noop {}
  {\bibfield  {journal} {\bibinfo  {journal} {Rev. Mod. Phys.}\ }\textbf
  {\bibinfo {volume} {87}},\ \bibinfo {pages} {457} (\bibinfo {year}
  {2015})}\BibitemShut {NoStop}%
\bibitem [{\citenamefont {Joci{\'c}}\ and\ \citenamefont
  {Vukmirovi{\'c}}(2020)}]{jocic2020ab}%
  \BibitemOpen
  \bibfield  {author} {\bibinfo {author} {\bibfnamefont {M.}~\bibnamefont
  {Joci{\'c}}}\ and\ \bibinfo {author} {\bibfnamefont {N.}~\bibnamefont
  {Vukmirovi{\'c}}},\ }\bibfield  {title} {\bibinfo {title} {Ab initio
  construction of symmetry-adapted $k\cdot p$ {H}amiltonians for the electronic
  structure of semiconductors},\ }\href@noop {} {\bibfield  {journal} {\bibinfo
   {journal} {Phys. Rev. B}\ }\textbf {\bibinfo {volume} {102}},\ \bibinfo
  {pages} {085121} (\bibinfo {year} {2020})}\BibitemShut {NoStop}%
\bibitem [{\citenamefont {Bloch}(1929)}]{bloch1929quantenmechanik}%
  \BibitemOpen
  \bibfield  {author} {\bibinfo {author} {\bibfnamefont {F.}~\bibnamefont
  {Bloch}},\ }\bibfield  {title} {\bibinfo {title} {{\"U}ber die
  quantenmechanik der elektronen in kristallgittern},\ }\href@noop {}
  {\bibfield  {journal} {\bibinfo  {journal} {Zeitschrift f{\"u}r physik}\
  }\textbf {\bibinfo {volume} {52}},\ \bibinfo {pages} {555} (\bibinfo {year}
  {1929})}\BibitemShut {NoStop}%
\bibitem [{\citenamefont {Sobota}\ \emph {et~al.}(2021)\citenamefont {Sobota},
  \citenamefont {He},\ and\ \citenamefont {Shen}}]{sobota2021angle}%
  \BibitemOpen
  \bibfield  {author} {\bibinfo {author} {\bibfnamefont {J.~A.}\ \bibnamefont
  {Sobota}}, \bibinfo {author} {\bibfnamefont {Y.}~\bibnamefont {He}},\ and\
  \bibinfo {author} {\bibfnamefont {Z.-X.}\ \bibnamefont {Shen}},\ }\bibfield
  {title} {\bibinfo {title} {Angle-resolved photoemission studies of quantum
  materials},\ }\href@noop {} {\bibfield  {journal} {\bibinfo  {journal} {Rev.
  Mod. Phys.}\ }\textbf {\bibinfo {volume} {93}},\ \bibinfo {pages} {025006}
  (\bibinfo {year} {2021})}\BibitemShut {NoStop}%
\bibitem [{\citenamefont {Zhang}\ \emph {et~al.}(2022)\citenamefont {Zhang},
  \citenamefont {Pincelli}, \citenamefont {Jozwiak}, \citenamefont {Kondo},
  \citenamefont {Ernstorfer}, \citenamefont {Sato},\ and\ \citenamefont
  {Zhou}}]{zhang2022angle}%
  \BibitemOpen
  \bibfield  {author} {\bibinfo {author} {\bibfnamefont {H.}~\bibnamefont
  {Zhang}}, \bibinfo {author} {\bibfnamefont {T.}~\bibnamefont {Pincelli}},
  \bibinfo {author} {\bibfnamefont {C.}~\bibnamefont {Jozwiak}}, \bibinfo
  {author} {\bibfnamefont {T.}~\bibnamefont {Kondo}}, \bibinfo {author}
  {\bibfnamefont {R.}~\bibnamefont {Ernstorfer}}, \bibinfo {author}
  {\bibfnamefont {T.}~\bibnamefont {Sato}},\ and\ \bibinfo {author}
  {\bibfnamefont {S.}~\bibnamefont {Zhou}},\ }\bibfield  {title} {\bibinfo
  {title} {Angle-resolved photoemission spectroscopy},\ }\href@noop {}
  {\bibfield  {journal} {\bibinfo  {journal} {Nat. Rev. Methods Primers}\
  }\textbf {\bibinfo {volume} {2}},\ \bibinfo {pages} {54} (\bibinfo {year}
  {2022})}\BibitemShut {NoStop}%
\bibitem [{\citenamefont {Sch{\"u}ler}\ \emph {et~al.}(2022)\citenamefont
  {Sch{\"u}ler}, \citenamefont {Pincelli}, \citenamefont {Dong}, \citenamefont
  {Devereaux}, \citenamefont {Wolf}, \citenamefont {Rettig}, \citenamefont
  {Ernstorfer},\ and\ \citenamefont {Beaulieu}}]{schuler2022polarization}%
  \BibitemOpen
  \bibfield  {author} {\bibinfo {author} {\bibfnamefont {M.}~\bibnamefont
  {Sch{\"u}ler}}, \bibinfo {author} {\bibfnamefont {T.}~\bibnamefont
  {Pincelli}}, \bibinfo {author} {\bibfnamefont {S.}~\bibnamefont {Dong}},
  \bibinfo {author} {\bibfnamefont {T.~P.}\ \bibnamefont {Devereaux}}, \bibinfo
  {author} {\bibfnamefont {M.}~\bibnamefont {Wolf}}, \bibinfo {author}
  {\bibfnamefont {L.}~\bibnamefont {Rettig}}, \bibinfo {author} {\bibfnamefont
  {R.}~\bibnamefont {Ernstorfer}},\ and\ \bibinfo {author} {\bibfnamefont
  {S.}~\bibnamefont {Beaulieu}},\ }\bibfield  {title} {\bibinfo {title}
  {Polarization-modulated angle-resolved photoemission spectroscopy: toward
  circular dichroism without circular photons and {B}loch wave-function
  reconstruction},\ }\href@noop {} {\bibfield  {journal} {\bibinfo  {journal}
  {Phys. Rev. X}\ }\textbf {\bibinfo {volume} {12}},\ \bibinfo {pages} {011019}
  (\bibinfo {year} {2022})}\BibitemShut {NoStop}%
\bibitem [{\citenamefont {Ghimire}\ \emph {et~al.}(2011)\citenamefont
  {Ghimire}, \citenamefont {DiChiara}, \citenamefont {Sistrunk}, \citenamefont
  {Agostini}, \citenamefont {DiMauro},\ and\ \citenamefont
  {Reis}}]{ghimire2011observation}%
  \BibitemOpen
  \bibfield  {author} {\bibinfo {author} {\bibfnamefont {S.}~\bibnamefont
  {Ghimire}}, \bibinfo {author} {\bibfnamefont {A.~D.}\ \bibnamefont
  {DiChiara}}, \bibinfo {author} {\bibfnamefont {E.}~\bibnamefont {Sistrunk}},
  \bibinfo {author} {\bibfnamefont {P.}~\bibnamefont {Agostini}}, \bibinfo
  {author} {\bibfnamefont {L.~F.}\ \bibnamefont {DiMauro}},\ and\ \bibinfo
  {author} {\bibfnamefont {D.~A.}\ \bibnamefont {Reis}},\ }\bibfield  {title}
  {\bibinfo {title} {Observation of high-order harmonic generation in a bulk
  crystal},\ }\href@noop {} {\bibfield  {journal} {\bibinfo  {journal} {Nat.
  Phys.}\ }\textbf {\bibinfo {volume} {7}},\ \bibinfo {pages} {138} (\bibinfo
  {year} {2011})}\BibitemShut {NoStop}%
\bibitem [{\citenamefont {Schubert}\ \emph {et~al.}(2014)\citenamefont
  {Schubert}, \citenamefont {Hohenleutner}, \citenamefont {Langer},
  \citenamefont {Urbanek}, \citenamefont {Lange}, \citenamefont {Huttner},
  \citenamefont {Golde}, \citenamefont {Meier}, \citenamefont {Kira},
  \citenamefont {Koch} \emph {et~al.}}]{schubert2014sub}%
  \BibitemOpen
  \bibfield  {author} {\bibinfo {author} {\bibfnamefont {O.}~\bibnamefont
  {Schubert}}, \bibinfo {author} {\bibfnamefont {M.}~\bibnamefont
  {Hohenleutner}}, \bibinfo {author} {\bibfnamefont {F.}~\bibnamefont
  {Langer}}, \bibinfo {author} {\bibfnamefont {B.}~\bibnamefont {Urbanek}},
  \bibinfo {author} {\bibfnamefont {C.}~\bibnamefont {Lange}}, \bibinfo
  {author} {\bibfnamefont {U.}~\bibnamefont {Huttner}}, \bibinfo {author}
  {\bibfnamefont {D.}~\bibnamefont {Golde}}, \bibinfo {author} {\bibfnamefont
  {T.}~\bibnamefont {Meier}}, \bibinfo {author} {\bibfnamefont
  {M.}~\bibnamefont {Kira}}, \bibinfo {author} {\bibfnamefont {S.~W.}\
  \bibnamefont {Koch}}, \emph {et~al.},\ }\bibfield  {title} {\bibinfo {title}
  {Sub-cycle control of terahertz high-harmonic generation by dynamical {B}loch
  oscillations},\ }\href@noop {} {\bibfield  {journal} {\bibinfo  {journal}
  {Nat. Photon.}\ }\textbf {\bibinfo {volume} {8}},\ \bibinfo {pages} {119}
  (\bibinfo {year} {2014})}\BibitemShut {NoStop}%
\bibitem [{\citenamefont {Hohenleutner}\ \emph {et~al.}(2015)\citenamefont
  {Hohenleutner}, \citenamefont {Langer}, \citenamefont {Schubert},
  \citenamefont {Knorr}, \citenamefont {Huttner}, \citenamefont {Koch},
  \citenamefont {Kira},\ and\ \citenamefont {Huber}}]{hohenleutner2015real}%
  \BibitemOpen
  \bibfield  {author} {\bibinfo {author} {\bibfnamefont {M.}~\bibnamefont
  {Hohenleutner}}, \bibinfo {author} {\bibfnamefont {F.}~\bibnamefont
  {Langer}}, \bibinfo {author} {\bibfnamefont {O.}~\bibnamefont {Schubert}},
  \bibinfo {author} {\bibfnamefont {M.}~\bibnamefont {Knorr}}, \bibinfo
  {author} {\bibfnamefont {U.}~\bibnamefont {Huttner}}, \bibinfo {author}
  {\bibfnamefont {S.~W.}\ \bibnamefont {Koch}}, \bibinfo {author}
  {\bibfnamefont {M.}~\bibnamefont {Kira}},\ and\ \bibinfo {author}
  {\bibfnamefont {R.}~\bibnamefont {Huber}},\ }\bibfield  {title} {\bibinfo
  {title} {Real-time observation of interfering crystal electrons in
  high-harmonic generation},\ }\href@noop {} {\bibfield  {journal} {\bibinfo
  {journal} {Nature}\ }\textbf {\bibinfo {volume} {523}},\ \bibinfo {pages}
  {572} (\bibinfo {year} {2015})}\BibitemShut {NoStop}%
\bibitem [{\citenamefont {Liu}\ \emph {et~al.}(2017)\citenamefont {Liu},
  \citenamefont {Li}, \citenamefont {You}, \citenamefont {Ghimire},
  \citenamefont {Heinz},\ and\ \citenamefont {Reis}}]{liu2017high}%
  \BibitemOpen
  \bibfield  {author} {\bibinfo {author} {\bibfnamefont {H.}~\bibnamefont
  {Liu}}, \bibinfo {author} {\bibfnamefont {Y.}~\bibnamefont {Li}}, \bibinfo
  {author} {\bibfnamefont {Y.~S.}\ \bibnamefont {You}}, \bibinfo {author}
  {\bibfnamefont {S.}~\bibnamefont {Ghimire}}, \bibinfo {author} {\bibfnamefont
  {T.~F.}\ \bibnamefont {Heinz}},\ and\ \bibinfo {author} {\bibfnamefont
  {D.~A.}\ \bibnamefont {Reis}},\ }\bibfield  {title} {\bibinfo {title}
  {High-harmonic generation from an atomically thin semiconductor},\
  }\href@noop {} {\bibfield  {journal} {\bibinfo  {journal} {Nat. Phys.}\
  }\textbf {\bibinfo {volume} {13}},\ \bibinfo {pages} {262} (\bibinfo {year}
  {2017})}\BibitemShut {NoStop}%
\bibitem [{\citenamefont {Liu}\ and\ \citenamefont {Zhu}(2007)}]{liu2007high}%
  \BibitemOpen
  \bibfield  {author} {\bibinfo {author} {\bibfnamefont {R.-B.}\ \bibnamefont
  {Liu}}\ and\ \bibinfo {author} {\bibfnamefont {B.-F.}\ \bibnamefont {Zhu}},\
  }\bibfield  {title} {\bibinfo {title} {High-order {TH}z-sideband generation
  in semiconductors},\ }in\ \href@noop {} {\emph {\bibinfo {booktitle} {AIP
  Conf. Proc.}}},\ Vol.\ \bibinfo {volume} {893}\ (\bibinfo {organization}
  {American Institute of Physics},\ \bibinfo {year} {2007})\ pp.\ \bibinfo
  {pages} {1455--1456}\BibitemShut {NoStop}%
\bibitem [{\citenamefont {Zaks}\ \emph {et~al.}(2012)\citenamefont {Zaks},
  \citenamefont {Liu},\ and\ \citenamefont {Sherwin}}]{zaks2012experimental}%
  \BibitemOpen
  \bibfield  {author} {\bibinfo {author} {\bibfnamefont {B.}~\bibnamefont
  {Zaks}}, \bibinfo {author} {\bibfnamefont {R.-B.}\ \bibnamefont {Liu}},\ and\
  \bibinfo {author} {\bibfnamefont {M.~S.}\ \bibnamefont {Sherwin}},\
  }\bibfield  {title} {\bibinfo {title} {Experimental observation of
  electron--hole recollisions},\ }\href@noop {} {\bibfield  {journal} {\bibinfo
   {journal} {Nature}\ }\textbf {\bibinfo {volume} {483}},\ \bibinfo {pages}
  {580} (\bibinfo {year} {2012})}\BibitemShut {NoStop}%
\bibitem [{\citenamefont {Golde}\ \emph {et~al.}(2008)\citenamefont {Golde},
  \citenamefont {Meier},\ and\ \citenamefont {Koch}}]{golde2008high}%
  \BibitemOpen
  \bibfield  {author} {\bibinfo {author} {\bibfnamefont {D.}~\bibnamefont
  {Golde}}, \bibinfo {author} {\bibfnamefont {T.}~\bibnamefont {Meier}},\ and\
  \bibinfo {author} {\bibfnamefont {S.~W.}\ \bibnamefont {Koch}},\ }\bibfield
  {title} {\bibinfo {title} {High harmonics generated in semiconductor
  nanostructures by the coupled dynamics of optical inter-and intraband
  excitations},\ }\href@noop {} {\bibfield  {journal} {\bibinfo  {journal}
  {Phys. Rev. B}\ }\textbf {\bibinfo {volume} {77}},\ \bibinfo {pages} {075330}
  (\bibinfo {year} {2008})}\BibitemShut {NoStop}%
\bibitem [{\citenamefont {Golde}\ \emph {et~al.}(2009)\citenamefont {Golde},
  \citenamefont {Meier},\ and\ \citenamefont {Koch}}]{golde2009microscopic}%
  \BibitemOpen
  \bibfield  {author} {\bibinfo {author} {\bibfnamefont {D.}~\bibnamefont
  {Golde}}, \bibinfo {author} {\bibfnamefont {T.}~\bibnamefont {Meier}},\ and\
  \bibinfo {author} {\bibfnamefont {S.~W.}\ \bibnamefont {Koch}},\ }\bibfield
  {title} {\bibinfo {title} {Microscopic analysis of high-harmonic generation
  in semiconductor nanostructures},\ }\href@noop {} {\bibfield  {journal}
  {\bibinfo  {journal} {Phys. Status Solidi C}\ }\textbf {\bibinfo {volume}
  {6}},\ \bibinfo {pages} {420} (\bibinfo {year} {2009})}\BibitemShut {NoStop}%
\bibitem [{\citenamefont {Avetissian}\ \emph {et~al.}(2022)\citenamefont
  {Avetissian}, \citenamefont {Avetisyan}, \citenamefont {Avchyan},\ and\
  \citenamefont {Mkrtchian}}]{avetissian2022high}%
  \BibitemOpen
  \bibfield  {author} {\bibinfo {author} {\bibfnamefont {H.~K.}\ \bibnamefont
  {Avetissian}}, \bibinfo {author} {\bibfnamefont {V.~N.}\ \bibnamefont
  {Avetisyan}}, \bibinfo {author} {\bibfnamefont {B.~R.}\ \bibnamefont
  {Avchyan}},\ and\ \bibinfo {author} {\bibfnamefont {G.~F.}\ \bibnamefont
  {Mkrtchian}},\ }\bibfield  {title} {\bibinfo {title} {High-order harmonic
  generation in three-dimensional {W}eyl semimetals with broken time-reversal
  symmetry},\ }\href@noop {} {\bibfield  {journal} {\bibinfo  {journal} {Phys.
  Rev. A}\ }\textbf {\bibinfo {volume} {106}},\ \bibinfo {pages} {033107}
  (\bibinfo {year} {2022})}\BibitemShut {NoStop}%
\bibitem [{\citenamefont {Lindberg}\ and\ \citenamefont
  {Koch}(1988)}]{lindberg1988effective}%
  \BibitemOpen
  \bibfield  {author} {\bibinfo {author} {\bibfnamefont {M.}~\bibnamefont
  {Lindberg}}\ and\ \bibinfo {author} {\bibfnamefont {S.~W.}\ \bibnamefont
  {Koch}},\ }\bibfield  {title} {\bibinfo {title} {Effective {B}loch equations
  for semiconductors},\ }\href@noop {} {\bibfield  {journal} {\bibinfo
  {journal} {Phys. Rev. B}\ }\textbf {\bibinfo {volume} {38}},\ \bibinfo
  {pages} {3342} (\bibinfo {year} {1988})}\BibitemShut {NoStop}%
\bibitem [{\citenamefont {Vampa}\ \emph
  {et~al.}(2015{\natexlab{a}})\citenamefont {Vampa}, \citenamefont {Hammond},
  \citenamefont {Thir{\'e}}, \citenamefont {Schmidt}, \citenamefont
  {L{\'e}gar{\'e}}, \citenamefont {McDonald}, \citenamefont {Brabec},
  \citenamefont {Klug},\ and\ \citenamefont {Corkum}}]{vampa2015all}%
  \BibitemOpen
  \bibfield  {author} {\bibinfo {author} {\bibfnamefont {G.}~\bibnamefont
  {Vampa}}, \bibinfo {author} {\bibfnamefont {T.~J.}\ \bibnamefont {Hammond}},
  \bibinfo {author} {\bibfnamefont {N.}~\bibnamefont {Thir{\'e}}}, \bibinfo
  {author} {\bibfnamefont {B.~E.}\ \bibnamefont {Schmidt}}, \bibinfo {author}
  {\bibfnamefont {F.}~\bibnamefont {L{\'e}gar{\'e}}}, \bibinfo {author}
  {\bibfnamefont {C.~R.}\ \bibnamefont {McDonald}}, \bibinfo {author}
  {\bibfnamefont {T.}~\bibnamefont {Brabec}}, \bibinfo {author} {\bibfnamefont
  {D.~D.}\ \bibnamefont {Klug}},\ and\ \bibinfo {author} {\bibfnamefont
  {P.~B.}\ \bibnamefont {Corkum}},\ }\bibfield  {title} {\bibinfo {title}
  {All-optical reconstruction of crystal band structure},\ }\href@noop {}
  {\bibfield  {journal} {\bibinfo  {journal} {Phys. Rev. Lett.}\ }\textbf
  {\bibinfo {volume} {115}},\ \bibinfo {pages} {193603} (\bibinfo {year}
  {2015}{\natexlab{a}})}\BibitemShut {NoStop}%
\bibitem [{\citenamefont {Vampa}\ \emph
  {et~al.}(2015{\natexlab{b}})\citenamefont {Vampa}, \citenamefont {McDonald},
  \citenamefont {Orlando}, \citenamefont {Corkum},\ and\ \citenamefont
  {Brabec}}]{vampa2015semiclassical}%
  \BibitemOpen
  \bibfield  {author} {\bibinfo {author} {\bibfnamefont {G.}~\bibnamefont
  {Vampa}}, \bibinfo {author} {\bibfnamefont {C.~R.}\ \bibnamefont {McDonald}},
  \bibinfo {author} {\bibfnamefont {G.}~\bibnamefont {Orlando}}, \bibinfo
  {author} {\bibfnamefont {P.~B.}\ \bibnamefont {Corkum}},\ and\ \bibinfo
  {author} {\bibfnamefont {T.}~\bibnamefont {Brabec}},\ }\bibfield  {title}
  {\bibinfo {title} {Semiclassical analysis of high harmonic generation in bulk
  crystals},\ }\href@noop {} {\bibfield  {journal} {\bibinfo  {journal} {Phys.
  Rev. B}\ }\textbf {\bibinfo {volume} {91}},\ \bibinfo {pages} {064302}
  (\bibinfo {year} {2015}{\natexlab{b}})}\BibitemShut {NoStop}%
\bibitem [{\citenamefont {Li}\ \emph {et~al.}(2020)\citenamefont {Li},
  \citenamefont {Lan}, \citenamefont {He}, \citenamefont {Cao}, \citenamefont
  {Zhang},\ and\ \citenamefont {Lu}}]{li2020determination}%
  \BibitemOpen
  \bibfield  {author} {\bibinfo {author} {\bibfnamefont {L.}~\bibnamefont
  {Li}}, \bibinfo {author} {\bibfnamefont {P.}~\bibnamefont {Lan}}, \bibinfo
  {author} {\bibfnamefont {L.}~\bibnamefont {He}}, \bibinfo {author}
  {\bibfnamefont {W.}~\bibnamefont {Cao}}, \bibinfo {author} {\bibfnamefont
  {Q.}~\bibnamefont {Zhang}},\ and\ \bibinfo {author} {\bibfnamefont
  {P.}~\bibnamefont {Lu}},\ }\bibfield  {title} {\bibinfo {title}
  {Determination of electron band structure using temporal interferometry},\
  }\href@noop {} {\bibfield  {journal} {\bibinfo  {journal} {Phys. Rev. Lett.}\
  }\textbf {\bibinfo {volume} {124}},\ \bibinfo {pages} {157403} (\bibinfo
  {year} {2020})}\BibitemShut {NoStop}%
\bibitem [{\citenamefont {Chen}\ \emph {et~al.}(2021)\citenamefont {Chen},
  \citenamefont {Xia},\ and\ \citenamefont {Fu}}]{chen2021reconstruction}%
  \BibitemOpen
  \bibfield  {author} {\bibinfo {author} {\bibfnamefont {J.}~\bibnamefont
  {Chen}}, \bibinfo {author} {\bibfnamefont {Q.}~\bibnamefont {Xia}},\ and\
  \bibinfo {author} {\bibfnamefont {L.}~\bibnamefont {Fu}},\ }\bibfield
  {title} {\bibinfo {title} {Reconstruction of crystal band structure by
  spectral caustics in high-order harmonic generation},\ }\href@noop {}
  {\bibfield  {journal} {\bibinfo  {journal} {Phys. Rev. A}\ }\textbf {\bibinfo
  {volume} {104}},\ \bibinfo {pages} {063109} (\bibinfo {year}
  {2021})}\BibitemShut {NoStop}%
\bibitem [{\citenamefont {Luu}\ \emph {et~al.}(2015)\citenamefont {Luu},
  \citenamefont {Garg}, \citenamefont {Kruchinin}, \citenamefont {Moulet},
  \citenamefont {Hassan},\ and\ \citenamefont {Goulielmakis}}]{luu2015extreme}%
  \BibitemOpen
  \bibfield  {author} {\bibinfo {author} {\bibfnamefont {T.~T.}\ \bibnamefont
  {Luu}}, \bibinfo {author} {\bibfnamefont {M.}~\bibnamefont {Garg}}, \bibinfo
  {author} {\bibfnamefont {S.~Y.}\ \bibnamefont {Kruchinin}}, \bibinfo {author}
  {\bibfnamefont {A.}~\bibnamefont {Moulet}}, \bibinfo {author} {\bibfnamefont
  {M.~T.}\ \bibnamefont {Hassan}},\ and\ \bibinfo {author} {\bibfnamefont
  {E.}~\bibnamefont {Goulielmakis}},\ }\bibfield  {title} {\bibinfo {title}
  {Extreme ultraviolet high-harmonic spectroscopy of solids},\ }\href@noop {}
  {\bibfield  {journal} {\bibinfo  {journal} {Nature}\ }\textbf {\bibinfo
  {volume} {521}},\ \bibinfo {pages} {498} (\bibinfo {year}
  {2015})}\BibitemShut {NoStop}%
\bibitem [{\citenamefont {Lanin}\ \emph {et~al.}(2017)\citenamefont {Lanin},
  \citenamefont {Stepanov}, \citenamefont {Fedotov},\ and\ \citenamefont
  {Zheltikov}}]{lanin2017mapping}%
  \BibitemOpen
  \bibfield  {author} {\bibinfo {author} {\bibfnamefont {A.~A.}\ \bibnamefont
  {Lanin}}, \bibinfo {author} {\bibfnamefont {E.~A.}\ \bibnamefont {Stepanov}},
  \bibinfo {author} {\bibfnamefont {A.~B.}\ \bibnamefont {Fedotov}},\ and\
  \bibinfo {author} {\bibfnamefont {A.~M.}\ \bibnamefont {Zheltikov}},\
  }\bibfield  {title} {\bibinfo {title} {Mapping the electron band structure by
  intraband high-harmonic generation in solids},\ }\href@noop {} {\bibfield
  {journal} {\bibinfo  {journal} {Optica}\ }\textbf {\bibinfo {volume} {4}},\
  \bibinfo {pages} {516} (\bibinfo {year} {2017})}\BibitemShut {NoStop}%
\bibitem [{\citenamefont {Lv}\ \emph {et~al.}(2021)\citenamefont {Lv},
  \citenamefont {Xu}, \citenamefont {Han}, \citenamefont {Zhang}, \citenamefont
  {Han}, \citenamefont {Zhou}, \citenamefont {Yao}, \citenamefont {Liu},
  \citenamefont {Lu}, \citenamefont {Weng} \emph {et~al.}}]{lv2021high}%
  \BibitemOpen
  \bibfield  {author} {\bibinfo {author} {\bibfnamefont {Y.-Y.}\ \bibnamefont
  {Lv}}, \bibinfo {author} {\bibfnamefont {J.}~\bibnamefont {Xu}}, \bibinfo
  {author} {\bibfnamefont {S.}~\bibnamefont {Han}}, \bibinfo {author}
  {\bibfnamefont {C.}~\bibnamefont {Zhang}}, \bibinfo {author} {\bibfnamefont
  {Y.}~\bibnamefont {Han}}, \bibinfo {author} {\bibfnamefont {J.}~\bibnamefont
  {Zhou}}, \bibinfo {author} {\bibfnamefont {S.-H.}\ \bibnamefont {Yao}},
  \bibinfo {author} {\bibfnamefont {X.-P.}\ \bibnamefont {Liu}}, \bibinfo
  {author} {\bibfnamefont {M.-H.}\ \bibnamefont {Lu}}, \bibinfo {author}
  {\bibfnamefont {H.}~\bibnamefont {Weng}}, \emph {et~al.},\ }\bibfield
  {title} {\bibinfo {title} {High-harmonic generation in {W}eyl semimetal
  $\beta$-{$\rm WP_2$} crystals},\ }\href@noop {} {\bibfield  {journal}
  {\bibinfo  {journal} {Nat. Commun.}\ }\textbf {\bibinfo {volume} {12}},\
  \bibinfo {pages} {6437} (\bibinfo {year} {2021})}\BibitemShut {NoStop}%
\bibitem [{\citenamefont {Parks}\ and\ \citenamefont
  {Kolesik}(2025)}]{parks2025full}%
  \BibitemOpen
  \bibfield  {author} {\bibinfo {author} {\bibfnamefont {A.~M.}\ \bibnamefont
  {Parks}}\ and\ \bibinfo {author} {\bibfnamefont {M.}~\bibnamefont
  {Kolesik}},\ }\bibfield  {title} {\bibinfo {title} {Full brillouin zone,
  multi-band reconstruction of the electronic structure from high-harmonic
  spectra},\ }\href@noop {} {\bibfield  {journal} {\bibinfo  {journal} {Opt.
  Express}\ }\textbf {\bibinfo {volume} {33}},\ \bibinfo {pages} {13986}
  (\bibinfo {year} {2025})}\BibitemShut {NoStop}%
\bibitem [{\citenamefont {Zaks}\ \emph {et~al.}(2013)\citenamefont {Zaks},
  \citenamefont {Banks},\ and\ \citenamefont {Sherwin}}]{zaks2013high}%
  \BibitemOpen
  \bibfield  {author} {\bibinfo {author} {\bibfnamefont {B.}~\bibnamefont
  {Zaks}}, \bibinfo {author} {\bibfnamefont {H.}~\bibnamefont {Banks}},\ and\
  \bibinfo {author} {\bibfnamefont {M.~S.}\ \bibnamefont {Sherwin}},\
  }\bibfield  {title} {\bibinfo {title} {High-order sideband generation in bulk
  gaas},\ }\href@noop {} {\bibfield  {journal} {\bibinfo  {journal} {Appl.
  Phys. Lett.}\ }\textbf {\bibinfo {volume} {102}},\ \bibinfo {pages} {012104}
  (\bibinfo {year} {2013})}\BibitemShut {NoStop}%
\bibitem [{\citenamefont {Banks}\ \emph {et~al.}(2013)\citenamefont {Banks},
  \citenamefont {Zaks}, \citenamefont {Yang}, \citenamefont {Mack},
  \citenamefont {Gossard}, \citenamefont {Liu},\ and\ \citenamefont
  {Sherwin}}]{banks2013terahertz}%
  \BibitemOpen
  \bibfield  {author} {\bibinfo {author} {\bibfnamefont {H.}~\bibnamefont
  {Banks}}, \bibinfo {author} {\bibfnamefont {B.}~\bibnamefont {Zaks}},
  \bibinfo {author} {\bibfnamefont {F.}~\bibnamefont {Yang}}, \bibinfo {author}
  {\bibfnamefont {S.}~\bibnamefont {Mack}}, \bibinfo {author} {\bibfnamefont
  {A.~C.}\ \bibnamefont {Gossard}}, \bibinfo {author} {\bibfnamefont
  {R.}~\bibnamefont {Liu}},\ and\ \bibinfo {author} {\bibfnamefont {M.~S.}\
  \bibnamefont {Sherwin}},\ }\bibfield  {title} {\bibinfo {title} {Terahertz
  electron-hole recollisions in {GaAs/AlGaAs} quantum wells: robustness to
  scattering by optical phonons and thermal fluctuations},\ }\href@noop {}
  {\bibfield  {journal} {\bibinfo  {journal} {Phys. Rev. Lett.}\ }\textbf
  {\bibinfo {volume} {111}},\ \bibinfo {pages} {267402} (\bibinfo {year}
  {2013})}\BibitemShut {NoStop}%
\bibitem [{\citenamefont {Langer}\ \emph {et~al.}(2016)\citenamefont {Langer},
  \citenamefont {Hohenleutner}, \citenamefont {Schmid}, \citenamefont
  {P{\"o}llmann}, \citenamefont {Nagler}, \citenamefont {Korn}, \citenamefont
  {Sch{\"u}ller}, \citenamefont {Sherwin}, \citenamefont {Huttner},
  \citenamefont {Steiner} \emph {et~al.}}]{langer2016lightwave}%
  \BibitemOpen
  \bibfield  {author} {\bibinfo {author} {\bibfnamefont {F.}~\bibnamefont
  {Langer}}, \bibinfo {author} {\bibfnamefont {M.}~\bibnamefont
  {Hohenleutner}}, \bibinfo {author} {\bibfnamefont {C.~P.}\ \bibnamefont
  {Schmid}}, \bibinfo {author} {\bibfnamefont {C.}~\bibnamefont
  {P{\"o}llmann}}, \bibinfo {author} {\bibfnamefont {P.}~\bibnamefont
  {Nagler}}, \bibinfo {author} {\bibfnamefont {T.}~\bibnamefont {Korn}},
  \bibinfo {author} {\bibfnamefont {C.}~\bibnamefont {Sch{\"u}ller}}, \bibinfo
  {author} {\bibfnamefont {M.~S.}\ \bibnamefont {Sherwin}}, \bibinfo {author}
  {\bibfnamefont {U.}~\bibnamefont {Huttner}}, \bibinfo {author} {\bibfnamefont
  {J.~T.}\ \bibnamefont {Steiner}}, \emph {et~al.},\ }\bibfield  {title}
  {\bibinfo {title} {Lightwave-driven quasiparticle collisions on a subcycle
  timescale},\ }\href@noop {} {\bibfield  {journal} {\bibinfo  {journal}
  {Nature}\ }\textbf {\bibinfo {volume} {533}},\ \bibinfo {pages} {225}
  (\bibinfo {year} {2016})}\BibitemShut {NoStop}%
\bibitem [{\citenamefont {Banks}\ \emph {et~al.}(2017)\citenamefont {Banks},
  \citenamefont {Wu}, \citenamefont {Valovcin}, \citenamefont {Mack},
  \citenamefont {Gossard}, \citenamefont {Pfeiffer}, \citenamefont {Liu},\ and\
  \citenamefont {Sherwin}}]{banks2017dynamical}%
  \BibitemOpen
  \bibfield  {author} {\bibinfo {author} {\bibfnamefont {H.~B.}\ \bibnamefont
  {Banks}}, \bibinfo {author} {\bibfnamefont {Q.}~\bibnamefont {Wu}}, \bibinfo
  {author} {\bibfnamefont {D.~C.}\ \bibnamefont {Valovcin}}, \bibinfo {author}
  {\bibfnamefont {S.}~\bibnamefont {Mack}}, \bibinfo {author} {\bibfnamefont
  {A.~C.}\ \bibnamefont {Gossard}}, \bibinfo {author} {\bibfnamefont
  {L.}~\bibnamefont {Pfeiffer}}, \bibinfo {author} {\bibfnamefont {R.-B.}\
  \bibnamefont {Liu}},\ and\ \bibinfo {author} {\bibfnamefont {M.~S.}\
  \bibnamefont {Sherwin}},\ }\bibfield  {title} {\bibinfo {title} {Dynamical
  birefringence: electron-hole recollisions as probes of {B}erry curvature},\
  }\href@noop {} {\bibfield  {journal} {\bibinfo  {journal} {Phys. Rev. X}\
  }\textbf {\bibinfo {volume} {7}},\ \bibinfo {pages} {041042} (\bibinfo {year}
  {2017})}\BibitemShut {NoStop}%
\bibitem [{\citenamefont {Valovcin}\ \emph {et~al.}(2018)\citenamefont
  {Valovcin}, \citenamefont {Banks}, \citenamefont {Mack}, \citenamefont
  {Gossard}, \citenamefont {West}, \citenamefont {Pfeiffer},\ and\
  \citenamefont {Sherwin}}]{valovcin2018optical}%
  \BibitemOpen
  \bibfield  {author} {\bibinfo {author} {\bibfnamefont {D.~C.}\ \bibnamefont
  {Valovcin}}, \bibinfo {author} {\bibfnamefont {H.~B.}\ \bibnamefont {Banks}},
  \bibinfo {author} {\bibfnamefont {S.}~\bibnamefont {Mack}}, \bibinfo {author}
  {\bibfnamefont {A.~C.}\ \bibnamefont {Gossard}}, \bibinfo {author}
  {\bibfnamefont {K.}~\bibnamefont {West}}, \bibinfo {author} {\bibfnamefont
  {L.}~\bibnamefont {Pfeiffer}},\ and\ \bibinfo {author} {\bibfnamefont
  {M.~S.}\ \bibnamefont {Sherwin}},\ }\bibfield  {title} {\bibinfo {title}
  {Optical frequency combs from high-order sideband generation},\ }\href@noop
  {} {\bibfield  {journal} {\bibinfo  {journal} {Opt. Express}\ }\textbf
  {\bibinfo {volume} {26}},\ \bibinfo {pages} {29807} (\bibinfo {year}
  {2018})}\BibitemShut {NoStop}%
\bibitem [{\citenamefont {Langer}\ \emph {et~al.}(2018)\citenamefont {Langer},
  \citenamefont {Schmid}, \citenamefont {Schlauderer}, \citenamefont {Gmitra},
  \citenamefont {Fabian}, \citenamefont {Nagler}, \citenamefont {Sch{\"u}ller},
  \citenamefont {Korn}, \citenamefont {Hawkins}, \citenamefont {Steiner} \emph
  {et~al.}}]{langer2018lightwave}%
  \BibitemOpen
  \bibfield  {author} {\bibinfo {author} {\bibfnamefont {F.}~\bibnamefont
  {Langer}}, \bibinfo {author} {\bibfnamefont {C.~P.}\ \bibnamefont {Schmid}},
  \bibinfo {author} {\bibfnamefont {S.}~\bibnamefont {Schlauderer}}, \bibinfo
  {author} {\bibfnamefont {M.}~\bibnamefont {Gmitra}}, \bibinfo {author}
  {\bibfnamefont {J.}~\bibnamefont {Fabian}}, \bibinfo {author} {\bibfnamefont
  {P.}~\bibnamefont {Nagler}}, \bibinfo {author} {\bibfnamefont
  {C.}~\bibnamefont {Sch{\"u}ller}}, \bibinfo {author} {\bibfnamefont
  {T.}~\bibnamefont {Korn}}, \bibinfo {author} {\bibfnamefont {P.~G.}\
  \bibnamefont {Hawkins}}, \bibinfo {author} {\bibfnamefont {J.~T.}\
  \bibnamefont {Steiner}}, \emph {et~al.},\ }\bibfield  {title} {\bibinfo
  {title} {Lightwave valleytronics in a monolayer of tungsten diselenide},\
  }\href@noop {} {\bibfield  {journal} {\bibinfo  {journal} {Nature}\ }\textbf
  {\bibinfo {volume} {557}},\ \bibinfo {pages} {76} (\bibinfo {year}
  {2018})}\BibitemShut {NoStop}%
\bibitem [{\citenamefont {Borsch}\ \emph {et~al.}(2020)\citenamefont {Borsch},
  \citenamefont {Schmid}, \citenamefont {Weigl}, \citenamefont {Schlauderer},
  \citenamefont {Hofmann}, \citenamefont {Lange}, \citenamefont {Steiner},
  \citenamefont {Koch}, \citenamefont {Huber},\ and\ \citenamefont
  {Kira}}]{borsch2020super}%
  \BibitemOpen
  \bibfield  {author} {\bibinfo {author} {\bibfnamefont {M.}~\bibnamefont
  {Borsch}}, \bibinfo {author} {\bibfnamefont {C.~P.}\ \bibnamefont {Schmid}},
  \bibinfo {author} {\bibfnamefont {L.}~\bibnamefont {Weigl}}, \bibinfo
  {author} {\bibfnamefont {S.}~\bibnamefont {Schlauderer}}, \bibinfo {author}
  {\bibfnamefont {N.}~\bibnamefont {Hofmann}}, \bibinfo {author} {\bibfnamefont
  {C.}~\bibnamefont {Lange}}, \bibinfo {author} {\bibfnamefont {J.~T.}\
  \bibnamefont {Steiner}}, \bibinfo {author} {\bibfnamefont {S.~W.}\
  \bibnamefont {Koch}}, \bibinfo {author} {\bibfnamefont {R.}~\bibnamefont
  {Huber}},\ and\ \bibinfo {author} {\bibfnamefont {M.}~\bibnamefont {Kira}},\
  }\bibfield  {title} {\bibinfo {title} {Super-resolution lightwave tomography
  of electronic bands in quantum materials},\ }\href@noop {} {\bibfield
  {journal} {\bibinfo  {journal} {Science}\ }\textbf {\bibinfo {volume}
  {370}},\ \bibinfo {pages} {1204} (\bibinfo {year} {2020})}\BibitemShut
  {NoStop}%
\bibitem [{\citenamefont {Nagai}\ \emph {et~al.}(2020)\citenamefont {Nagai},
  \citenamefont {Uchida}, \citenamefont {Yoshikawa}, \citenamefont {Endo},
  \citenamefont {Miyata},\ and\ \citenamefont {Tanaka}}]{nagai2020dynamical}%
  \BibitemOpen
  \bibfield  {author} {\bibinfo {author} {\bibfnamefont {K.}~\bibnamefont
  {Nagai}}, \bibinfo {author} {\bibfnamefont {K.}~\bibnamefont {Uchida}},
  \bibinfo {author} {\bibfnamefont {N.}~\bibnamefont {Yoshikawa}}, \bibinfo
  {author} {\bibfnamefont {T.}~\bibnamefont {Endo}}, \bibinfo {author}
  {\bibfnamefont {Y.}~\bibnamefont {Miyata}},\ and\ \bibinfo {author}
  {\bibfnamefont {K.}~\bibnamefont {Tanaka}},\ }\bibfield  {title} {\bibinfo
  {title} {Dynamical symmetry of strongly light-driven electronic system in
  crystalline solids},\ }\href@noop {} {\bibfield  {journal} {\bibinfo
  {journal} {Communications Physics}\ }\textbf {\bibinfo {volume} {3}},\
  \bibinfo {pages} {137} (\bibinfo {year} {2020})}\BibitemShut {NoStop}%
\bibitem [{\citenamefont {Costello}\ \emph {et~al.}(2021)\citenamefont
  {Costello}, \citenamefont {O'Hara}, \citenamefont {Wu}, \citenamefont
  {Valovcin}, \citenamefont {Pfeiffer}, \citenamefont {West},\ and\
  \citenamefont {Sherwin}}]{costello2021reconstruction}%
  \BibitemOpen
  \bibfield  {author} {\bibinfo {author} {\bibfnamefont {J.~B.}\ \bibnamefont
  {Costello}}, \bibinfo {author} {\bibfnamefont {S.~D.}\ \bibnamefont
  {O'Hara}}, \bibinfo {author} {\bibfnamefont {Q.}~\bibnamefont {Wu}}, \bibinfo
  {author} {\bibfnamefont {D.~C.}\ \bibnamefont {Valovcin}}, \bibinfo {author}
  {\bibfnamefont {L.~N.}\ \bibnamefont {Pfeiffer}}, \bibinfo {author}
  {\bibfnamefont {K.~W.}\ \bibnamefont {West}},\ and\ \bibinfo {author}
  {\bibfnamefont {M.~S.}\ \bibnamefont {Sherwin}},\ }\bibfield  {title}
  {\bibinfo {title} {Reconstruction of {B}loch wavefunctions of holes in a
  semiconductor},\ }\href@noop {} {\bibfield  {journal} {\bibinfo  {journal}
  {Nature}\ }\textbf {\bibinfo {volume} {599}},\ \bibinfo {pages} {57}
  (\bibinfo {year} {2021})}\BibitemShut {NoStop}%
\bibitem [{\citenamefont {Freudenstein}\ \emph {et~al.}(2022)\citenamefont
  {Freudenstein}, \citenamefont {Borsch}, \citenamefont {Meierhofer},
  \citenamefont {Afanasiev}, \citenamefont {Schmid}, \citenamefont {Sandner},
  \citenamefont {Liebich}, \citenamefont {Girnghuber}, \citenamefont {Knorr},
  \citenamefont {Kira} \emph {et~al.}}]{freudenstein2022attosecond}%
  \BibitemOpen
  \bibfield  {author} {\bibinfo {author} {\bibfnamefont {J.}~\bibnamefont
  {Freudenstein}}, \bibinfo {author} {\bibfnamefont {M.}~\bibnamefont
  {Borsch}}, \bibinfo {author} {\bibfnamefont {M.}~\bibnamefont {Meierhofer}},
  \bibinfo {author} {\bibfnamefont {D.}~\bibnamefont {Afanasiev}}, \bibinfo
  {author} {\bibfnamefont {C.~P.}\ \bibnamefont {Schmid}}, \bibinfo {author}
  {\bibfnamefont {F.}~\bibnamefont {Sandner}}, \bibinfo {author} {\bibfnamefont
  {M.}~\bibnamefont {Liebich}}, \bibinfo {author} {\bibfnamefont
  {A.}~\bibnamefont {Girnghuber}}, \bibinfo {author} {\bibfnamefont
  {M.}~\bibnamefont {Knorr}}, \bibinfo {author} {\bibfnamefont
  {M.}~\bibnamefont {Kira}}, \emph {et~al.},\ }\bibfield  {title} {\bibinfo
  {title} {Attosecond clocking of correlations between {B}loch electrons},\
  }\href@noop {} {\bibfield  {journal} {\bibinfo  {journal} {Nature}\ }\textbf
  {\bibinfo {volume} {610}},\ \bibinfo {pages} {290} (\bibinfo {year}
  {2022})}\BibitemShut {NoStop}%
\bibitem [{\citenamefont {Liu}\ \emph {et~al.}(2024)\citenamefont {Liu},
  \citenamefont {Zhu}, \citenamefont {Jiang}, \citenamefont {Huang},
  \citenamefont {Luo}, \citenamefont {Zhang}, \citenamefont {Yan},
  \citenamefont {Zhang}, \citenamefont {Lu},\ and\ \citenamefont
  {Tao}}]{liu2024dephasing}%
  \BibitemOpen
  \bibfield  {author} {\bibinfo {author} {\bibfnamefont {Y.}~\bibnamefont
  {Liu}}, \bibinfo {author} {\bibfnamefont {B.}~\bibnamefont {Zhu}}, \bibinfo
  {author} {\bibfnamefont {S.}~\bibnamefont {Jiang}}, \bibinfo {author}
  {\bibfnamefont {S.}~\bibnamefont {Huang}}, \bibinfo {author} {\bibfnamefont
  {M.}~\bibnamefont {Luo}}, \bibinfo {author} {\bibfnamefont {S.}~\bibnamefont
  {Zhang}}, \bibinfo {author} {\bibfnamefont {H.}~\bibnamefont {Yan}}, \bibinfo
  {author} {\bibfnamefont {Y.}~\bibnamefont {Zhang}}, \bibinfo {author}
  {\bibfnamefont {R.}~\bibnamefont {Lu}},\ and\ \bibinfo {author}
  {\bibfnamefont {Z.}~\bibnamefont {Tao}},\ }\bibfield  {title} {\bibinfo
  {title} {Dephasing of strong-field-driven excitonic {A}utler-{T}ownes
  doublets revealed by time-and spectrum-resolved quantum-path
  interferometry},\ }\href@noop {} {\bibfield  {journal} {\bibinfo  {journal}
  {Phys. Rev. Lett.}\ }\textbf {\bibinfo {volume} {133}},\ \bibinfo {pages}
  {026901} (\bibinfo {year} {2024})}\BibitemShut {NoStop}%
\bibitem [{\citenamefont {O'Hara}\ \emph {et~al.}(2024)\citenamefont {O'Hara},
  \citenamefont {Costello}, \citenamefont {Wu}, \citenamefont {West},
  \citenamefont {Pfeiffer},\ and\ \citenamefont {Sherwin}}]{o2024bloch}%
  \BibitemOpen
  \bibfield  {author} {\bibinfo {author} {\bibfnamefont {S.~D.}\ \bibnamefont
  {O'Hara}}, \bibinfo {author} {\bibfnamefont {J.~B.}\ \bibnamefont
  {Costello}}, \bibinfo {author} {\bibfnamefont {Q.}~\bibnamefont {Wu}},
  \bibinfo {author} {\bibfnamefont {K.}~\bibnamefont {West}}, \bibinfo {author}
  {\bibfnamefont {L.}~\bibnamefont {Pfeiffer}},\ and\ \bibinfo {author}
  {\bibfnamefont {M.~S.}\ \bibnamefont {Sherwin}},\ }\bibfield  {title}
  {\bibinfo {title} {Bloch-wave interferometry of driven quasiparticles in bulk
  {G}a{A}s},\ }\href@noop {} {\bibfield  {journal} {\bibinfo  {journal} {Phys.
  Rev. B}\ }\textbf {\bibinfo {volume} {109}},\ \bibinfo {pages} {054308}
  (\bibinfo {year} {2024})}\BibitemShut {NoStop}%
\bibitem [{\citenamefont {Costello}\ \emph {et~al.}(2023)\citenamefont
  {Costello}, \citenamefont {O'Hara}, \citenamefont {Wu}, \citenamefont {Jang},
  \citenamefont {Pfeiffer}, \citenamefont {West},\ and\ \citenamefont
  {Sherwin}}]{costello2023breaking}%
  \BibitemOpen
  \bibfield  {author} {\bibinfo {author} {\bibfnamefont {J.~B.}\ \bibnamefont
  {Costello}}, \bibinfo {author} {\bibfnamefont {S.~D.}\ \bibnamefont
  {O'Hara}}, \bibinfo {author} {\bibfnamefont {Q.}~\bibnamefont {Wu}}, \bibinfo
  {author} {\bibfnamefont {M.}~\bibnamefont {Jang}}, \bibinfo {author}
  {\bibfnamefont {L.~N.}\ \bibnamefont {Pfeiffer}}, \bibinfo {author}
  {\bibfnamefont {K.~W.}\ \bibnamefont {West}},\ and\ \bibinfo {author}
  {\bibfnamefont {M.~S.}\ \bibnamefont {Sherwin}},\ }\bibfield  {title}
  {\bibinfo {title} {Breaking a {B}loch-wave interferometer: {Q}uasiparticle
  species-specific temperature-dependent nonequilibrium dephasing},\
  }\href@noop {} {\bibfield  {journal} {\bibinfo  {journal} {Phys. Rev. B}\
  }\textbf {\bibinfo {volume} {108}},\ \bibinfo {pages} {195205} (\bibinfo
  {year} {2023})}\BibitemShut {NoStop}%
\bibitem [{\citenamefont {Luttinger}\ and\ \citenamefont
  {Kohn}(1955)}]{luttinger1955motion}%
  \BibitemOpen
  \bibfield  {author} {\bibinfo {author} {\bibfnamefont {J.~M.}\ \bibnamefont
  {Luttinger}}\ and\ \bibinfo {author} {\bibfnamefont {W.}~\bibnamefont
  {Kohn}},\ }\bibfield  {title} {\bibinfo {title} {Motion of electrons and
  holes in perturbed periodic fields},\ }\href@noop {} {\bibfield  {journal}
  {\bibinfo  {journal} {Phys. Rev.}\ }\textbf {\bibinfo {volume} {97}},\
  \bibinfo {pages} {869} (\bibinfo {year} {1955})}\BibitemShut {NoStop}%
\bibitem [{\citenamefont {Sell}(1972)}]{sell1972resolved}%
  \BibitemOpen
  \bibfield  {author} {\bibinfo {author} {\bibfnamefont {D.~D.}\ \bibnamefont
  {Sell}},\ }\bibfield  {title} {\bibinfo {title} {Resolved free-exciton
  transitions in the optical-absorption spectrum of {G}a{A}s},\ }\href@noop {}
  {\bibfield  {journal} {\bibinfo  {journal} {Phys. Rev. B}\ }\textbf {\bibinfo
  {volume} {6}},\ \bibinfo {pages} {3750} (\bibinfo {year} {1972})}\BibitemShut
  {NoStop}%
\bibitem [{\citenamefont {Wu}\ and\ \citenamefont
  {Sherwin}(2023)}]{wu2023explicit}%
  \BibitemOpen
  \bibfield  {author} {\bibinfo {author} {\bibfnamefont {Q.}~\bibnamefont
  {Wu}}\ and\ \bibinfo {author} {\bibfnamefont {M.~S.}\ \bibnamefont
  {Sherwin}},\ }\bibfield  {title} {\bibinfo {title} {Explicit formula for
  high-order sideband polarization by extreme tailoring of {F}eynman path
  integrals},\ }\href@noop {} {\bibfield  {journal} {\bibinfo  {journal} {Phys.
  Rev. B}\ }\textbf {\bibinfo {volume} {107}},\ \bibinfo {pages} {174308}
  (\bibinfo {year} {2023})}\BibitemShut {NoStop}%
\bibitem [{\citenamefont {Varshni}(1967)}]{varshni1967temperature}%
  \BibitemOpen
  \bibfield  {author} {\bibinfo {author} {\bibfnamefont {Y.~P.}\ \bibnamefont
  {Varshni}},\ }\bibfield  {title} {\bibinfo {title} {Temperature dependence of
  the energy gap in semiconductors},\ }\href@noop {} {\bibfield  {journal}
  {\bibinfo  {journal} {physica}\ }\textbf {\bibinfo {volume} {34}},\ \bibinfo
  {pages} {149} (\bibinfo {year} {1967})}\BibitemShut {NoStop}%
\bibitem [{\citenamefont {Ramachandran}(2001)}]{feenstra2001recent}%
  \BibitemOpen
  \bibfield  {author} {\bibinfo {author} {\bibfnamefont {V.}~\bibnamefont
  {Ramachandran}},\ }\bibfield  {title} {\bibinfo {title} {Recent developments
  in scanning tunneling spectroscopy of semiconductor surfaces},\ }\href@noop
  {} {\bibfield  {journal} {\bibinfo  {journal} {Appl. Phys. A}\ }\textbf
  {\bibinfo {volume} {72}},\ \bibinfo {pages} {S193} (\bibinfo {year}
  {2001})}\BibitemShut {NoStop}%
\bibitem [{\citenamefont {Fr{\"o}hlich}\ \emph {et~al.}(1950)\citenamefont
  {Fr{\"o}hlich}, \citenamefont {Pelzer},\ and\ \citenamefont
  {Zienau}}]{frohlich1950xx}%
  \BibitemOpen
  \bibfield  {author} {\bibinfo {author} {\bibfnamefont {H.}~\bibnamefont
  {Fr{\"o}hlich}}, \bibinfo {author} {\bibfnamefont {H.}~\bibnamefont
  {Pelzer}},\ and\ \bibinfo {author} {\bibfnamefont {S.}~\bibnamefont
  {Zienau}},\ }\bibfield  {title} {\bibinfo {title} {Xx. properties of slow
  electrons in polar materials},\ }\href@noop {} {\bibfield  {journal}
  {\bibinfo  {journal} {The London, Edinburgh, and Dublin Philosophical
  Magazine and Journal of Science}\ }\textbf {\bibinfo {volume} {41}},\
  \bibinfo {pages} {221} (\bibinfo {year} {1950})}\BibitemShut {NoStop}%
\bibitem [{\citenamefont {Lee}\ and\ \citenamefont
  {Pines}(1952)}]{lee1952motion}%
  \BibitemOpen
  \bibfield  {author} {\bibinfo {author} {\bibfnamefont {T.-D.}\ \bibnamefont
  {Lee}}\ and\ \bibinfo {author} {\bibfnamefont {D.}~\bibnamefont {Pines}},\
  }\bibfield  {title} {\bibinfo {title} {The motion of slow electrons in polar
  crystals},\ }\href@noop {} {\bibfield  {journal} {\bibinfo  {journal} {Phys.
  Rev.}\ }\textbf {\bibinfo {volume} {88}},\ \bibinfo {pages} {960} (\bibinfo
  {year} {1952})}\BibitemShut {NoStop}%
\bibitem [{\citenamefont {Yan}(2008)}]{yan2008theory}%
  \BibitemOpen
  \bibfield  {author} {\bibinfo {author} {\bibfnamefont {J.-Y.}\ \bibnamefont
  {Yan}},\ }\bibfield  {title} {\bibinfo {title} {Theory of excitonic
  high-order sideband generation in semiconductors under a strong terahertz
  field},\ }\href@noop {} {\bibfield  {journal} {\bibinfo  {journal} {Phys.
  Rev. B}\ }\textbf {\bibinfo {volume} {78}},\ \bibinfo {pages} {075204}
  (\bibinfo {year} {2008})}\BibitemShut {NoStop}%
\bibitem [{\citenamefont {Xie}\ \emph {et~al.}(2013)\citenamefont {Xie},
  \citenamefont {Zhu},\ and\ \citenamefont {Liu}}]{xie2013effects}%
  \BibitemOpen
  \bibfield  {author} {\bibinfo {author} {\bibfnamefont {X.-T.}\ \bibnamefont
  {Xie}}, \bibinfo {author} {\bibfnamefont {B.-F.}\ \bibnamefont {Zhu}},\ and\
  \bibinfo {author} {\bibfnamefont {R.-B.}\ \bibnamefont {Liu}},\ }\bibfield
  {title} {\bibinfo {title} {Effects of excitation frequency on high-order
  terahertz sideband generation in semiconductors},\ }\href@noop {} {\bibfield
  {journal} {\bibinfo  {journal} {New J. Phys.}\ }\textbf {\bibinfo {volume}
  {15}},\ \bibinfo {pages} {105015} (\bibinfo {year} {2013})}\BibitemShut
  {NoStop}%
\bibitem [{\citenamefont {Soma}\ \emph {et~al.}(1982)\citenamefont {Soma},
  \citenamefont {Satoh},\ and\ \citenamefont {Matsuo}}]{soma1982thermal}%
  \BibitemOpen
  \bibfield  {author} {\bibinfo {author} {\bibfnamefont {T.}~\bibnamefont
  {Soma}}, \bibinfo {author} {\bibfnamefont {J.}~\bibnamefont {Satoh}},\ and\
  \bibinfo {author} {\bibfnamefont {H.}~\bibnamefont {Matsuo}},\ }\bibfield
  {title} {\bibinfo {title} {Thermal expansion coefficient of {G}a{A}s and
  {I}n{P}},\ }\href@noop {} {\bibfield  {journal} {\bibinfo  {journal} {Solid
  State Communications}\ }\textbf {\bibinfo {volume} {42}},\ \bibinfo {pages}
  {889} (\bibinfo {year} {1982})}\BibitemShut {NoStop}%
\bibitem [{\citenamefont {Driscoll}\ \emph {et~al.}()\citenamefont {Driscoll},
  \citenamefont {Willoughby}, \citenamefont {Mullin},\ and\ \citenamefont
  {Straughan}}]{driscoll1975precision}%
  \BibitemOpen
  \bibfield  {author} {\bibinfo {author} {\bibfnamefont {C.~M.~H.}\
  \bibnamefont {Driscoll}}, \bibinfo {author} {\bibfnamefont {A.~F.~W.}\
  \bibnamefont {Willoughby}}, \bibinfo {author} {\bibfnamefont {J.~B.}\
  \bibnamefont {Mullin}},\ and\ \bibinfo {author} {\bibfnamefont {B.~W.}\
  \bibnamefont {Straughan}},\ }\bibfield  {title} {\bibinfo {title} {Precision
  lattice parameter measurements on doped gallium arsenide},\ }in\ \href@noop
  {} {\emph {\bibinfo {booktitle} {Proceedings of the Fifth International
  Symposium on Gallium arsenide and related compounds, Deauville, 1974}}}\
  (\bibinfo  {publisher} {Institute of Physics, London, 1975, Conf. Ser. No.
  24})\ p.\ \bibinfo {pages} {275}\BibitemShut {NoStop}%
\bibitem [{\citenamefont {Ahmed}\ \emph {et~al.}(1992)\citenamefont {Ahmed},
  \citenamefont {Agool}, \citenamefont {Wright}, \citenamefont {Mitchell},
  \citenamefont {Koohian}, \citenamefont {Adams}, \citenamefont {Pidgeon},
  \citenamefont {Cavenett}, \citenamefont {Stanley},\ and\ \citenamefont
  {Kean}}]{ahmed1992far}%
  \BibitemOpen
  \bibfield  {author} {\bibinfo {author} {\bibfnamefont {N.}~\bibnamefont
  {Ahmed}}, \bibinfo {author} {\bibfnamefont {I.}~\bibnamefont {Agool}},
  \bibinfo {author} {\bibfnamefont {M.~G.}\ \bibnamefont {Wright}}, \bibinfo
  {author} {\bibfnamefont {K.}~\bibnamefont {Mitchell}}, \bibinfo {author}
  {\bibfnamefont {A.}~\bibnamefont {Koohian}}, \bibinfo {author} {\bibfnamefont
  {S.~J.~A.}\ \bibnamefont {Adams}}, \bibinfo {author} {\bibfnamefont {C.~R.}\
  \bibnamefont {Pidgeon}}, \bibinfo {author} {\bibfnamefont {B.~C.}\
  \bibnamefont {Cavenett}}, \bibinfo {author} {\bibfnamefont {C.~R.}\
  \bibnamefont {Stanley}},\ and\ \bibinfo {author} {\bibfnamefont {A.~H.}\
  \bibnamefont {Kean}},\ }\bibfield  {title} {\bibinfo {title} {Far-infrared
  optically detected cyclotron resonance in {G}a{A}s layers and low-dimensional
  structures},\ }\href@noop {} {\bibfield  {journal} {\bibinfo  {journal}
  {Semiconductor science and technology}\ }\textbf {\bibinfo {volume} {7}},\
  \bibinfo {pages} {357} (\bibinfo {year} {1992})}\BibitemShut {NoStop}%
\bibitem [{\citenamefont {Skolnick}\ \emph {et~al.}(1976)\citenamefont
  {Skolnick}, \citenamefont {Jain}, \citenamefont {Stradling}, \citenamefont
  {Leotin},\ and\ \citenamefont {Ousset}}]{skolnick1976investigation}%
  \BibitemOpen
  \bibfield  {author} {\bibinfo {author} {\bibfnamefont {M.~S.}\ \bibnamefont
  {Skolnick}}, \bibinfo {author} {\bibfnamefont {A.~K.}\ \bibnamefont {Jain}},
  \bibinfo {author} {\bibfnamefont {R.~A.}\ \bibnamefont {Stradling}}, \bibinfo
  {author} {\bibfnamefont {J.}~\bibnamefont {Leotin}},\ and\ \bibinfo {author}
  {\bibfnamefont {J.~C.}\ \bibnamefont {Ousset}},\ }\bibfield  {title}
  {\bibinfo {title} {An investigation of the anisotropy of the valence band of
  {G}a{A}s by cyclotron resonance},\ }\href@noop {} {\bibfield  {journal}
  {\bibinfo  {journal} {J. Phys. C: Solid State Physics}\ }\textbf {\bibinfo
  {volume} {9}},\ \bibinfo {pages} {2809} (\bibinfo {year} {1976})}\BibitemShut
  {NoStop}%
\bibitem [{\citenamefont {Liu}\ and\ \citenamefont
  {Zhu}(2002)}]{liu2002adiabatic}%
  \BibitemOpen
  \bibfield  {author} {\bibinfo {author} {\bibfnamefont {R.-B.}\ \bibnamefont
  {Liu}}\ and\ \bibinfo {author} {\bibfnamefont {B.-F.}\ \bibnamefont {Zhu}},\
  }\bibfield  {title} {\bibinfo {title} {Adiabatic stabilization of excitons in
  an intense terahertz laser},\ }\href@noop {} {\bibfield  {journal} {\bibinfo
  {journal} {Phys. Rev. B}\ }\textbf {\bibinfo {volume} {66}},\ \bibinfo
  {pages} {033106} (\bibinfo {year} {2002})}\BibitemShut {NoStop}%
\bibitem [{\citenamefont {Tong-Yi}\ and\ \citenamefont
  {Wei}(2008)}]{tong2008excitonic}%
  \BibitemOpen
  \bibfield  {author} {\bibinfo {author} {\bibfnamefont {Z.}~\bibnamefont
  {Tong-Yi}}\ and\ \bibinfo {author} {\bibfnamefont {Z.}~\bibnamefont {Wei}},\
  }\bibfield  {title} {\bibinfo {title} {Excitonic optical absorption in
  semiconductors under intense terahertz radiation},\ }\href@noop {} {\bibfield
   {journal} {\bibinfo  {journal} {Chinese Physics B}\ }\textbf {\bibinfo
  {volume} {17}},\ \bibinfo {pages} {4285} (\bibinfo {year}
  {2008})}\BibitemShut {NoStop}%
\bibitem [{\citenamefont {Sell}\ \emph {et~al.}(1974)\citenamefont {Sell},
  \citenamefont {Casey~Jr},\ and\ \citenamefont
  {Wecht}}]{sell1974concentration}%
  \BibitemOpen
  \bibfield  {author} {\bibinfo {author} {\bibfnamefont {D.~D.}\ \bibnamefont
  {Sell}}, \bibinfo {author} {\bibfnamefont {H.~C.}\ \bibnamefont {Casey~Jr}},\
  and\ \bibinfo {author} {\bibfnamefont {K.~W.}\ \bibnamefont {Wecht}},\
  }\bibfield  {title} {\bibinfo {title} {Concentration dependence of the
  refractive index for n-and p-type {G}a{A}s between 1.2 and 1.8 e{V}},\
  }\href@noop {} {\bibfield  {journal} {\bibinfo  {journal} {J. Appl. Phys.}\
  }\textbf {\bibinfo {volume} {45}},\ \bibinfo {pages} {2650} (\bibinfo {year}
  {1974})}\BibitemShut {NoStop}%
\bibitem [{\citenamefont {Samara}(1983)}]{samara1983temperature}%
  \BibitemOpen
  \bibfield  {author} {\bibinfo {author} {\bibfnamefont {G.~A.}\ \bibnamefont
  {Samara}},\ }\bibfield  {title} {\bibinfo {title} {Temperature and pressure
  dependences of the dielectric constants of semiconductors},\ }\href@noop {}
  {\bibfield  {journal} {\bibinfo  {journal} {Phys. Rev. B}\ }\textbf {\bibinfo
  {volume} {27}},\ \bibinfo {pages} {3494} (\bibinfo {year}
  {1983})}\BibitemShut {NoStop}%
\bibitem [{\citenamefont {Irmer}\ \emph {et~al.}(1996)\citenamefont {Irmer},
  \citenamefont {Wenzel},\ and\ \citenamefont
  {Monecke}}]{irmer1996temperature}%
  \BibitemOpen
  \bibfield  {author} {\bibinfo {author} {\bibfnamefont {G.}~\bibnamefont
  {Irmer}}, \bibinfo {author} {\bibfnamefont {M.}~\bibnamefont {Wenzel}},\ and\
  \bibinfo {author} {\bibfnamefont {J.}~\bibnamefont {Monecke}},\ }\bibfield
  {title} {\bibinfo {title} {The temperature dependence of the {LO}
  ({$\Gamma$}) and {TO} ({$\Gamma$}) phonons in {G}a{A}s and {I}n{P}},\
  }\href@noop {} {\bibfield  {journal} {\bibinfo  {journal} {physica status
  solidi (b)}\ }\textbf {\bibinfo {volume} {195}},\ \bibinfo {pages} {85}
  (\bibinfo {year} {1996})}\BibitemShut {NoStop}%
\bibitem [{\citenamefont {Uesugi}\ \emph {et~al.}(2000)\citenamefont {Uesugi},
  \citenamefont {Suemune}, \citenamefont {Hasegawa}, \citenamefont
  {Akutagawa},\ and\ \citenamefont {Nakamura}}]{uesugi2000temperature}%
  \BibitemOpen
  \bibfield  {author} {\bibinfo {author} {\bibfnamefont {K.}~\bibnamefont
  {Uesugi}}, \bibinfo {author} {\bibfnamefont {I.}~\bibnamefont {Suemune}},
  \bibinfo {author} {\bibfnamefont {T.}~\bibnamefont {Hasegawa}}, \bibinfo
  {author} {\bibfnamefont {T.}~\bibnamefont {Akutagawa}},\ and\ \bibinfo
  {author} {\bibfnamefont {T.}~\bibnamefont {Nakamura}},\ }\bibfield  {title}
  {\bibinfo {title} {Temperature dependence of band gap energies of {G}a{A}s{N}
  alloys},\ }\href@noop {} {\bibfield  {journal} {\bibinfo  {journal} {Appl.
  Phys. Lett.}\ }\textbf {\bibinfo {volume} {76}},\ \bibinfo {pages} {1285}
  (\bibinfo {year} {2000})}\BibitemShut {NoStop}%
\bibitem [{\citenamefont {Haug}\ and\ \citenamefont
  {Jauho}(2008)}]{haug2008quantum}%
  \BibitemOpen
  \bibfield  {author} {\bibinfo {author} {\bibfnamefont {H.}~\bibnamefont
  {Haug}}\ and\ \bibinfo {author} {\bibfnamefont {A.-P.}\ \bibnamefont
  {Jauho}},\ }\href@noop {} {\emph {\bibinfo {title} {Quantum kinetics in
  transport and optics of semiconductors}}}\ (\bibinfo  {publisher}
  {Springer},\ \bibinfo {year} {2008})\ pp.\ \bibinfo {pages}
  {115--156}\BibitemShut {NoStop}%
\bibitem [{\citenamefont {Liu}(2002)}]{renbao2002electron}%
  \BibitemOpen
  \bibfield  {author} {\bibinfo {author} {\bibfnamefont {R.-B.}\ \bibnamefont
  {Liu}},\ }\href@noop {} {\bibinfo {title} {private communication}} (\bibinfo
  {year} {2002})\BibitemShut {NoStop}%
\bibitem [{\citenamefont {Moore}\ and\ \citenamefont
  {Holm}(1996)}]{moore1996infrared}%
  \BibitemOpen
  \bibfield  {author} {\bibinfo {author} {\bibfnamefont {W.~J.}\ \bibnamefont
  {Moore}}\ and\ \bibinfo {author} {\bibfnamefont {R.~T.}\ \bibnamefont
  {Holm}},\ }\bibfield  {title} {\bibinfo {title} {Infrared dielectric constant
  of gallium arsenide},\ }\href@noop {} {\bibfield  {journal} {\bibinfo
  {journal} {J. Appl. Phys.}\ }\textbf {\bibinfo {volume} {80}},\ \bibinfo
  {pages} {6939} (\bibinfo {year} {1996})}\BibitemShut {NoStop}%
\bibitem [{\citenamefont {Scholz}(1995)}]{scholz1995hole}%
  \BibitemOpen
  \bibfield  {author} {\bibinfo {author} {\bibfnamefont {R.}~\bibnamefont
  {Scholz}},\ }\bibfield  {title} {\bibinfo {title} {Hole--phonon scattering
  rates in gallium arsenide},\ }\href@noop {} {\bibfield  {journal} {\bibinfo
  {journal} {J. Appl. Phys.}\ }\textbf {\bibinfo {volume} {77}},\ \bibinfo
  {pages} {3219} (\bibinfo {year} {1995})}\BibitemShut {NoStop}%
\bibitem [{\citenamefont {Wu}(2026)}]{qile2025electron}%
  \BibitemOpen
  \bibfield  {author} {\bibinfo {author} {\bibfnamefont {Q.}~\bibnamefont
  {Wu}},\ }\href@noop {} {\bibinfo {title} {Electron-hole propagators from
  high-order sideband polarimetry in bulk {G}a{A}s and effective-{H}amiltonian
  parameters from {M}onte {C}arlo simulations}},\ \bibinfo {howpublished}
  {Dataset on Zenodo. https://doi.org/10.5281/zenodo.18225853} (\bibinfo {year}
  {2026})\BibitemShut {NoStop}%
\bibitem [{\citenamefont {Born}\ and\ \citenamefont
  {Wolf}(2013)}]{born2013principles}%
  \BibitemOpen
  \bibfield  {author} {\bibinfo {author} {\bibfnamefont {M.}~\bibnamefont
  {Born}}\ and\ \bibinfo {author} {\bibfnamefont {E.}~\bibnamefont {Wolf}},\
  }\href@noop {} {\emph {\bibinfo {title} {Principles of optics:
  electromagnetic theory of propagation, interference and diffraction of
  light}}}\ (\bibinfo  {publisher} {Pergamon Press},\ \bibinfo {year} {2013})\
  pp.\ \bibinfo {pages} {51--60}\BibitemShut {NoStop}%
\bibitem [{\citenamefont {Vurgaftman}\ \emph {et~al.}(2001)\citenamefont
  {Vurgaftman}, \citenamefont {Meyer},\ and\ \citenamefont
  {Ram-Mohan}}]{vurgaftman2001band}%
  \BibitemOpen
  \bibfield  {author} {\bibinfo {author} {\bibfnamefont {I.}~\bibnamefont
  {Vurgaftman}}, \bibinfo {author} {\bibfnamefont {J.~R.}\ \bibnamefont
  {Meyer}},\ and\ \bibinfo {author} {\bibfnamefont {L.~R.}\ \bibnamefont
  {Ram-Mohan}},\ }\bibfield  {title} {\bibinfo {title} {Band parameters for
  {III}--{V} compound semiconductors and their alloys},\ }\href@noop {}
  {\bibfield  {journal} {\bibinfo  {journal} {J. Appl. Phys.}\ }\textbf
  {\bibinfo {volume} {89}},\ \bibinfo {pages} {5815} (\bibinfo {year}
  {2001})}\BibitemShut {NoStop}%
\bibitem [{\citenamefont {Schilp}\ \emph {et~al.}(1994)\citenamefont {Schilp},
  \citenamefont {Kuhn},\ and\ \citenamefont {Mahler}}]{schilp1994electron}%
  \BibitemOpen
  \bibfield  {author} {\bibinfo {author} {\bibfnamefont {J.}~\bibnamefont
  {Schilp}}, \bibinfo {author} {\bibfnamefont {T.}~\bibnamefont {Kuhn}},\ and\
  \bibinfo {author} {\bibfnamefont {G.}~\bibnamefont {Mahler}},\ }\bibfield
  {title} {\bibinfo {title} {Electron-phonon quantum kinetics in pulse-excited
  semiconductors: {M}emory and renormalization effects},\ }\href@noop {}
  {\bibfield  {journal} {\bibinfo  {journal} {Phys. Rev. B}\ }\textbf {\bibinfo
  {volume} {50}},\ \bibinfo {pages} {5435} (\bibinfo {year}
  {1994})}\BibitemShut {NoStop}%
\end{thebibliography}
%

\end{document}